\documentclass[aps,prd,preprint,groupedaddress,showpacs]{revtex4}

\usepackage{epsfig}
\usepackage{graphics}
\usepackage{slashed}
\usepackage{color}
\begin{document}

\leftmargin -2cm
\def\choosen{\atopwithdelims..}

\boldmath
\title{Large-$p_T$ production of $D$ mesons at the LHCb in the parton Reggeization approach}
\unboldmath
\author{\firstname{A.V.}\surname{Karpishkov}} \email{karpishkov@ssau.ru}
\affiliation{Samara National Research University, Moscow Highway,
34, 443086, Samara, Russia} \affiliation{{II.} Institut f\"ur
Theoretische Physik, Universit\" at Hamburg, Luruper Chaussee 149,
22761 Hamburg, Germany}
\author{\firstname{V.A.}\surname{Saleev}} \email{saleev@samsu.ru}
\author{\firstname{A.V.}
\surname{Shipilova}} \email{alexshipilova@samsu.ru}
\affiliation{Samara National Research University, Moscow Highway,
34, 443086, Samara, Russia}

\begin{abstract}
The production of $D$ mesons in proton-proton collisions at the LHCb
detector is studied. We consider the single production of $D^0$,
$D^+$, $D^{\star +}$, and $D_s^+$ mesons and correlation spectra in
the production of $D\bar D$ and $DD$ pairs at the $\sqrt{S}=7$~TeV
and $\sqrt{S}=13$~TeV. In case of the single $D$-meson production we
calculate differential cross sections over transverse momentum $p_T$
while in the pair $D\bar D,~DD$-meson production the cross sections
are calculated over the azimuthal angle  difference $\Delta\varphi$,
rapidity difference $\Delta y$, invariant mass of the pair $M$  and
over the $p_T$ of the one meson from a pair. The cross sections are
obtained at the leading order of the parton Reggeization approach
using Kimber-Martin-Ryskin unintegrated parton distribution
functions in a proton. To describe the $D$-meson production we use
universal scale-dependent $c$-quark and gluon fragmentation
functions fitted to $e^+e^-$ annihilation data from CERN LEP1. Our
predictions find a good  agreement with the LHCb Collaboration data
within uncertainties and without free parameters.
\end{abstract}
\pacs{12.38.-t,12.40.Nn,13.85.Ni,14.40.Lb}

\maketitle

\section{Introduction}

The open charm production in collisions of high energy hadrons is an appropriate class of processes to test the perturbative
 quantum chromodynamics (QCD) \cite{pQCD}. A large mass of charm quark $m_c$ as a lower bound of hard energy scale $\mu\geq m_c$ leads to
  the small value of the strong coupling constant $\alpha_S(\mu)$ due to the condition $m_c\gg\Lambda_{QCD}$, where $\Lambda_{QCD}$
  is the asymptotic scale parameter of QCD. Production of charm
  quark  with large transverse momentum ($p_T>>m_c$) is a typical
  multiscale hard process for which the fixed-order QCD calculations
  should be corrected by the fragmentation contribution to resum large logarithmic terms  $\sim \alpha_S \log p_T/m_c$,
  where $\alpha_S=g^2_S/4\pi$ is the strong coupling constant.

Nowadays, the theoretical study of inclusive $D$-meson
hadroproduction were performed at the leading order (LO) for double
production ($D\bar D, DD$) and the next-to-leading order (NLO) for
single production in the collinear parton model (CPM) of QCD within
the two approaches: the general-mass variable-flavor-number (GM-VFN)
scheme~\cite{GMVFN}, and the so-called fixed order scheme improved
with next-to-leading logarithms (FONLL scheme)~\cite{FONLL}. In the
first one, realized in the Refs.~\cite{KniehlKramer, KKSS,
KKSS2012}, the large transverse momenta collinear logarithms are
resummed through the evolution of the fragmentation functions (FFs),
satisfying the Dokshitzer-Gribov-Lipatov-Altarelli-Parisi (DGLAP)
evolution equations ~\cite{DGLAP}. The $D$-meson FFs were extracted
both at the LO and NLO in the fixed factorization scheme from the
fit of $e^+e^-$ data taken by the OPAL Collaboration at CERN
LEP1~\cite{LEP1}. Opposite, in the FONLL approach, the NLO $D$-meson
production cross sections are calculated with a non-perturbative
$c$-quark FF, that is not a subject to DGLAP evolution. The FONLL
scheme was implemented in the Refs.~\cite{Nason,newNLO} and its main
ingredients are the following: the NLO fixed order calculation  with
resummation of large transverse momentum logarithms at the
next-to-leading level for heavy quark production. The important
difference of these two fragmentation approaches is in following: in
the case of scale-depended GM-VFN scheme $D$-mesons are produced
both from $c$-quarks and from gluons, in case of scale-independent
FONLL scheme $D$-mesons are produced only from $c$-quarks.

At the high-energy Regge limit $\sqrt{S}>>p_T>>m_c$, where
$\sqrt{S}$ is the invariant collision energy, one has new dynamic
regime of particle production in the multi-Regge kinematics (MRK) or
in the quasi-multi-Regge kinematics (QMRK), when one particle or
a group of particles are produced in the rapidity region, being
strongly separated in rapidity from other particles. Radiative
corrections in this regime are dominated by the production of
additional hard jets. The only one way in CPM to treat these
processes is to calculate higher-order corrections, which is a
challenging task for some processes even at the NLO level, such as
for relevant to our study process of double charm production in the
gluon fusion, $gg\to cc\bar c\bar c$.

Alternatively, to solve this problem we should change a
factorization scheme from the collinear approximation of the PM to
the high-energy or $k_T$-factorization
\cite{KTCollins,KTGribov,KTCatani} and take into account a large part
of the higher-order corrections which are incorporated in the
transverse-momentum-dependent parton distribution functions (PDFs) of
off-shell initial partons. Recently, the studies of single and
double $D$-meson hadroproduction at the LHC were performed in the
$k_T$-factorization approach with the scale-independent Peterson
$c$-quark FF in Refs.~\cite{shurek1,shurek2} and with the
scale-dependent $c$-quark and gluon FFs from \cite{KKSS2012} in
Refs.~\cite{SingleD,DoubleD}. The latter are based on the parton
Reggeization approach (PRA), which is a combination of
$k_T$-factorization framework with the fully gauge-invariant
amplitudes with Reggeized partons in the initial state.

In our previous study of the single $D$-meson production at the
Tevatron and the LHC~\cite{SingleD} in the central rapidity region
of produced mesons ($|y|<1$ and $|y|<0.5$, respectively) in the PRA
we obtained quite a good agreement between theoretical calculations
and the experimental data. In this work we continue the
investigation and extend it into the region of large rapidity for
single and double $D$-meson production. Not so long ago the LHCb
Collaboration has provided the forward $D$-meson production
data~\cite{LHCb1}. The differential cross sections on $D$-meson
transverse momentum $d\sigma/dp_T$ were measured for $D^0$, $D^+$,
$D^{\star+}$, and $D_s^+$ mesons in proton-proton collisions at the
CERN LHC at a center-of-mass energy of $\sqrt{S}=7$~TeV and with
rapidities in the range of $2.0<y<4.5$. Nowadays, the recent data
come from the LHC Run II and the newest measurements of $D$-meson
production were presented by the LHCb Collaboration at
$\sqrt{S}=13$~TeV~\cite{LHCb2}. They are presented as $p_T$
distributions of $D$ mesons at the same rapidity region and in a
wider $p_T$ range then in the LHC Run I. The data are presented
in three kinds: absolute double differential cross sections
$d^2\sigma/(dp_Tdy)$, ratios of $D$-meson production cross sections
between different center-of-mass energies $R_{13/7}$, and ratios of
the cross sections for different mesons.


Not only the single but also a pair production of $D$ mesons is
under consideration. The first observations of double charm
production were performed at the $\sqrt{S}=7$ TeV by the LHCb
Collaboration~\cite{LHCb_Pair}. The published results include the
cross sections differential in azimuthal angle difference,
transverse momentum, rapidity difference and invariant mass of
various combinations of $D$-meson pairs. The spectra were measured
at the region of large rapidities $2.0<y<4.0$ at the collision
energy $\sqrt{S}=7$~TeV. We continue our recent study of double
$D$-meson production in PRA~\cite{DoubleD} testing the new
contributions from Reggeized quark -- Reggeized antiquark
annihilation processes and studying different combinations of
$D$-mesons in the measured pair production spectra.


Recently, PRA was successfully applied for the analysis of inclusive
production of single jet \cite{KSSY}, pair of jets \cite{PRA},
prompt-photon \cite{tevatronY,heraY}, photon plus jet \cite{KNS14},
Drell-Yan lepton pairs \cite{NNS_DY}, bottom-flavored jets
\cite{bbTEV,bbLHC}, charmonium and bottomonium production
\cite{KniehlSaleevVasin1,KniehlSaleevVasin2,PRD2003,NSS_charm,NSS_bot,NKS16}
at the Tevatron and LHC. This study  has demonstrated the advantages
of the high-energy factorization scheme based on PRA in the
descriptions of data comparing to the CPM calculations.

This paper is organized as follows. In Sec.~\ref{sec:two} we present the
relevant Reggeized amplitudes as they are obtained from the Lipatov's
effective high-energy action in QCD. In Sec.~\ref{sec:three} the
formalism of our calculation in the PRA and the fragmentation model
are discussed. In Secs.~\ref{sec:four} and ~\ref{sec:five} our
results for single and double $D$-meson production, respectively, are
presented in comparison with the experimental data and discussed. In
Sec.~\ref{sec:six} we summarize our conclusions.

\boldmath
\section{Reggeized amplitudes}
\unboldmath
\label{sec:two}


The Reggeization of amplitudes at the high energy is a natural
property of gauge-invariant quantum field theories, such as quantum
electrodynamics~\cite{QED} and QCD~\cite{IFL,BFKL}. At large $\sqrt
S$ the dominant contribution to cross sections of QCD processes
give MRK or QMRK parton scattering processes with the gluon or
quark $t$-channel exchange. Due to the Reggeization of quarks and
gluons, an important role is dedicated to the vertices of
Reggeon-particle interactions. Nowadays, they can be straightforwardly
derived from the non-Abelian gauge-invariant effective action for
the interactions of the Reggeized partons with the usual QCD
partons, which was firstly introduced in Ref.~\cite{KTLipatov} for
Reggeized gluons only, and then extended by inclusion of Reggeized
quark fields in the Ref.~\cite{LipVyaz}. The full set of the induced
and effective vertices together with Feynman rules can be found in
Refs.~\cite{LipVyaz,KTAntonov}.

In our recent paper~\cite{SingleD} we have shown that gluon and
$c$-quark fragmentation give the leading contribution to the
$D$-meson production in the PRA. The same result was also obtained
in the Ref. \cite{KKSS} in NLO of the CPM, as well as the light
quark fragmentation turns out to be negligible. According to this
statement in case of the forward production of charmed mesons in the
framework of the PRA we will also consider the gluon and $c$-quark
fragmentation. The lowest order in $\alpha_S$ parton subprocesses of
PRA  which give a contribution to a single $D$-meson production via
gluon fragmentation are the following
\begin{eqnarray}
\mathcal {R} + \mathcal {R} \to g,\label{eq:RRg}
\end{eqnarray}
and
\begin{eqnarray}
\mathcal {Q} + \bar{\mathcal{Q}} \to g.\label{eq:QQg}
\end{eqnarray}
The corresponding lowest order parton subprocesses which contribute to a
single $D$-meson production via $c$-quark fragmentation are charm
quark-antiquark pair production in the gluon-gluon fusion
\begin{eqnarray}
\mathcal {R} + \mathcal {R} \to c + \bar c, \label{eq:RRqq}
\end{eqnarray}
and in the quark-antiquark annihilation
\begin{eqnarray}
\mathcal {Q} + \bar{\mathcal {Q}} \to c + \bar c, \label{eq:QQqq}
\end{eqnarray}
where $\mathcal {R}$ are the Reggeized gluons and $\mathcal {Q}$
denotes Reggeized $u$, $d$ and $s$ quarks.

The double $D \bar D$-meson production at the lowest order of PRA is
described by the parton subprocesses (\ref{eq:RRqq}) and
(\ref{eq:QQqq}) of $c\bar c$-pair production  as well as by the
parton subprocesses of $gg$-pair production
\begin{eqnarray}
\mathcal {R} + \mathcal {R} \to g + g, \label{eq:RRgg}
\end{eqnarray}
\begin{eqnarray}
\mathcal {Q} + \bar{\mathcal {Q}} \to g + g. \label{eq:QQgg}
\end{eqnarray}

In the case of $DD$-meson production we should consider
contributions from six parton subprocesses. At first, these are
(\ref{eq:RRgg}) and (\ref{eq:QQgg}). There are also $ 2\to 3$ and
$2\to 4$ subprocesses
\begin{eqnarray}
\mathcal {R} + \mathcal {R} \to c + \bar c + g, \label{eq:RRqqg}
\end{eqnarray}
\begin{eqnarray}
\mathcal {Q} + \bar{\mathcal {Q}} \to c + \bar c +g,
\label{eq:QQqqg}
\end{eqnarray}
\begin{eqnarray}
\mathcal {R} + \mathcal {R} \to c + \bar c + c + \bar c,
\label{eq:RRqqqq}
\end{eqnarray}
\begin{eqnarray}
\mathcal {Q} + \bar{\mathcal {Q}} \to c + \bar c + c + \bar c.
\label{eq:QQqqqq}
\end{eqnarray}
Formally, the matrix elements of the $2\to 2$, $2\to 3$ and $2\to 4$
subprocesses are of different order in the strong coupling constant
$\alpha_S$, but their contributions can be of the same order because
the fragmentation probabilities for gluon and $c$-quark
fragmentation to a $D$-meson are related as $P_{g\to D}\sim \alpha_S
P_{c\to D}$. Taking into account our previous study
\cite{SingleD,DoubleD} and the results of Ref.~\cite{shurek1}, we
conclude that {gluon fragmentation dominates over the $c$-quark
fragmentation} in the $DD$-pair production at the LHC energy, so we
will consider the contributions of (\ref{eq:RRgg}) and
(\ref{eq:QQgg}) subprocesses only.

Let us define four-vectors $(n^+)^\mu=P_2^\mu/E_1$ and
$(n^-)^\mu=P_1^\mu/E_2$, where $P_{1,2}^\mu$ are the four-momenta of
the colliding protons, and $E_{1,2}$ are their energies. We have
$(n^\pm)^2=0$, $n^+\cdot n^-=2$, and $S=(P_1+P_2)^2=4E_1E_2$. For
any four-momentum $k^\mu$, we define $k^\pm=k\cdot n^\pm$.  It is
easy to see that the four-momenta of the Reggeized gluons can be
represented as
\begin{eqnarray}
&&q_1^\mu = \frac{q_1^+}{2}(n^-)^\mu+q_{1T}^\mu\mbox{,}\nonumber\\
&&q_2^\mu = \frac{q_2^-}{2}(n^+)^\mu+q_{2T}^\mu\mbox{,}
\end{eqnarray}
where $q_{T}=(0,{\bf q}_{T},0)$.
Then the amplitude of gluon production in a fusion of two Reggeized gluons
can be presented as a convolution of the Fadin-Kuraev-Lipatov effective
PRR vertex $C_{\mathcal{RR}}^{g,\mu}(q_1,q_2)$ and polarization
four-vector of final gluon $\varepsilon_\mu(k)$:
\begin{equation}
{\cal M}(\mathcal{R}+\mathcal{R}\to
g)=C_{\mathcal{RR}}^{g,\mu}(q_1,q_2)\varepsilon_\mu(k),
\end{equation}
where
\begin{equation}
C_{\mathcal{RR}}^{g,\mu}(q_1,q_2)=-\sqrt{4\pi\alpha_s}f^{abc}
\frac{q_1^+q_2^-}{2\sqrt{t_1t_2}} \left[\left(q_1-q_2\right)^\mu+
\frac{(n^+)^\mu}{q_1^+}\left(q_2^2+q_1^+q_2^- \right)-
\frac{(n^-)^\mu}{q_2^-}\left(q_1^2+q_1^+q_2^-\right)\right],
\label{amp:RRg}
\end{equation}
$a$ and $b$ are the color indices of the Reggeized gluons with
incoming four-momenta $q_1$ and $q_2$, and $f^{abc}$ with
$a=1,...,N_c^2-1$ is the antisymmetric structure constant of color
gauge group $SU_C(3)$. Similarly, we can write down the amplitudes
for subprocess (\ref{eq:QQg}) with massless spinors of Reggeized
quark $\mathcal {Q}$ and antiquark $\bar {\mathcal {Q}}$,
$U(x_1P_1)$ and $V(x_2P_2)$ respectively:
\begin{eqnarray}
{\cal M}(\mathcal{Q}+\bar{\mathcal{Q}}\to g)&=&\varepsilon_\mu(k)\bar{V}(x_2P_2)C_{\mathcal{Q}\bar{\mathcal{Q}}}^{g,\mu}(q_1,q_2)U(x_1P_1).
\end{eqnarray}
The Fadin-Sherman effective vertex~\cite{FadinSherman} is given by
\begin{eqnarray}
C_{\mathcal{Q}\bar{\mathcal{Q}}}^{g,\mu}(q_1,q_2)&=&-i\sqrt{4\pi\alpha_s}T^a
\left[\gamma^\mu-\hat{q}_1\frac{\left(n^-\right)^{\mu}}{q_1^-+q_2^-}-\hat{q}_2\frac{\left(n^+\right)^\mu}{q_1^++q_2^+}\right].
\label{amp:QQg}
\end{eqnarray}
where  $T^a$ are the generators of the color gauge group $SU_C(3)$,
$k=q_1+q_2$ and
\begin{equation}
\gamma^{(-)\mu}(q_1,q_2)=\gamma^{\mu}+\hat{q}_1\frac{\left(n^-\right)^{\mu}}{q_2^-}.
\end{equation}
Then the squared amplitudes have a simple form and read:
\begin{eqnarray}
\overline{|{\cal M}(\mathcal{R}+\mathcal{R}\to g)|^2}&=&\frac{3}{2}\pi
\alpha_s \mathbf{k}_T^2,\label{sqamp:RRg} \\
\overline{|{\cal M}(\mathcal {Q} + \bar{\mathcal{Q}}\to g)|^2}&=&\frac{16}{3}\pi
\alpha_s(t_1+t_2), \label{sqamp:QQg}.
\end{eqnarray}
where $t_1 = - q_1^2 = |{\bf q}_{1T}|^2$, $t_2 = - q_2^2
= |{\bf q}_{2T}|^2$.

The amplitudes for the subprocesses (\ref{eq:RRqq}),
(\ref{eq:QQqq}), (\ref{eq:RRgg}), and (\ref{eq:QQgg}), can be
written as a convolution of corresponding effective vertices with
final gluon polarization four-vectors $\varepsilon_\mu(k_i)$ in case
of gluon production or with spinors $U(x_1P_1)$ and $V(x_2P_2)$ in
case of initial-state Reggeized quark $Q$ and antiquark~$\bar Q$:
\begin{eqnarray}
{\cal M}(\mathcal{R}+\mathcal{R}\to
\bar c+c)&=&\varepsilon_\mu(k_1)\varepsilon_\nu(k_2)C_{\mathcal{RR}}^{c\bar c}(q_1,q_2,k_1,k_2),\\
{\cal M}(\mathcal{Q}+\bar{\mathcal{Q}}\to
c+\bar c)&=&\bar{V}(x_2P_2)C_{\mathcal{Q}\bar{\mathcal{Q}}}^{c\bar c}(q_1,q_2,k_1,k_2)U(x_1P_1),\\
{\cal M}(\mathcal{R}+\mathcal{R}\to
g+g)&=&\varepsilon_\mu(k_1)\varepsilon_\nu(k_2)C_{\mathcal{RR}}^{gg,\mu\nu}(q_1,q_2,k_1,k_2),\\
{\cal M}(\mathcal{Q}+\bar{\mathcal{Q}}\to
g+g)&=&\varepsilon_\mu(k_1)\varepsilon_\nu(k_2)\bar{V}(x_2P_2)C_{\mathcal{Q}\bar{\mathcal{Q}}}^{gg,\mu\nu}(q_1,q_2,k_1,k_2)U(x_1P_1),
\end{eqnarray}
where $C_{\mathcal{RR},\mathcal{Q}\bar{\mathcal{Q}}}$ are the
effective vertices of the transition of two Reggeized partons into
the gluon pair or charm quark-antiquark pair. The general form of
the squared amplitudes is:
\begin{equation}
\overline{|{\mathcal M}|^2}=\pi^2\alpha_{s}^2A\sum_{n=0}^{4}w_nS^n,
\end{equation}
where $A$ and $w_n$ are functions which depend on variables $s$,
$t$, $u$, $a_1$, $a_2$, $b_1$, $b_2$, $S$. The corresponding
effective vertices and matrix elements squared are too large to be
presented here and can be found in an explicit form in
Ref.~\cite{PRA} in the case of the massless final charm
quarks(antiquarks).  We perform our analysis in the region of $\sqrt
S, p_T\gg m_c$, that allows us to use ZM VFNS scheme, where the
masses of the charm  quarks in the hard-scattering amplitude are
neglected.


For all the squared amplitudes of $2\to 2$ subprocesses we checked
that in the collinear limit, i.e. $\textbf{q}_{(1,2)T}=0$, the
Reggeized squared amplitudes transform to the squared amplitudes of
the corresponding parton subprocesses in CPM. It is evident that the
squared amplitudes for  $2\to 1$ subprocesses are vanishing in the
collinear limit.

\section{Hadronic cross sections}
\unboldmath \label{sec:three}

In the conventional CPM the factorization theorem allows to present
the cross sections of processes which take place in high-energy
hadron-hadron collisions and have a one hard scale $\mu^2$ as a
convolution of scale-dependent parton (quark or gluon)
\textit{collinear} distributions $f(x,\mu^2)$ and squared amplitude
of the hard parton scattering. These distributions represent the
density of partons carrying the longitudinal momentum fraction $x$
of the proton, integrated over the parton transverse momentum $k_T$
up to $k_T=\mu$. The evolution of this density from some scale
$\mu_0$, fixing a non-perturbative regime, to the typical scale
$\mu$ is described by DGLAP~\cite{DGLAP} evolution equations where
the large logarithms of type $\log(\mu^2/\Lambda_{QCD}^2)$
(collinear logarithms) are summed. The typical scale $\mu$ of the
hard-scattering processes is usually of order of the transverse
momentum  of the produced particle,  ${p}_T$.

The PRA gives the description of QCD parton scattering amplitudes in
the region of large $S$ and fixed momentum transfer $t$, $S \gg |t|$
(Regge region), with various color states in the $t$-channel. In the
Regge region we should keep the transverse momenta of the initial
partons and their virtualities.
 It becomes possible introducing the unintegrated over transverse momenta parton distribution functions
 (unPDFs) $\Phi(x,t,\mu^2)$, which depend on parton transverse momentum
${\bf q}_T$ while its virtuality is $t=-|{\bf q}_T|^2$.  The unPDFs
are defined to be related with collinear ones through the equation:
\begin{eqnarray}
xf(x,\mu^2)=\int^{\mu^2}dt \Phi(x,t,\mu^2)\label{eq:norm}.
\end{eqnarray}
The factorization formula in the PRA reads:
\begin{equation}\label{eq:KTtheorem}
d\sigma=\sum_{i,j}\int\frac{dx_1}{x_1}\int\frac{d^2\textbf{q}_{1T}}{\pi}\Phi_i(x_1,t_1,\mu_F^2)
\int\frac{dx_2}{x_2}\int\frac{d^2\textbf{q}_{1T}}{\pi}\Phi_j(x_2,t_2,\mu_F^2)d\hat{\sigma}_{ij}(q_1,q_2,\mu_F,\mu_R),
\end{equation}
where $t_1=-|{\bf q}_{1T}|^2$, $t_2=-|{\bf q}_{2T}|^2$, $\mu_F$ and
$\mu_R$ are the factorization and renormalization scales
respectively, $\Phi_i(x,t,\mu_F^2)$ is the unPDF of an $i$-parton in
the initial state and $d\hat{\sigma}_{ij}$ is the cross section of
the hard off-shell partonic subprocess. The cross section
(\ref{eq:KTtheorem}) is normalized to be in accordance with the CPM
at the collinear limit, when $\textbf{q}_{(1,2)T}=0$.

According to the factorization formula (\ref{eq:KTtheorem}) the
partonic cross section of the $2\to 1$ subprocess (as an example we
take (\ref{eq:RRg})) is being a convolution of gluon production
squared amplitude
 (\ref{sqamp:RRg}) with unPDFs, and it
 can be simplified to the following formula:
\begin{eqnarray}
\frac{d \sigma}{dy dk_T}(p + p \to g + X)=  \frac{1}{k_T^3} \int
d\phi_1 \int dt_1 \Phi(x_1,t_1,\mu^2) \Phi(x_2,t_2,\mu^2)
\overline{|{\cal M}(\mathcal{R} + \mathcal{R} \to g)|^2} \mbox{,}
\label{eq:QMRKg}
\end{eqnarray}
where $\phi_1$ is the azimuthal angle between ${\bf k}_T$ and ${\bf
q}_{1T}$.

The hadronic cross section of any $2 \to 2$ subprocess, here we show the case of heavy-quark pair production
 through the process (\ref{eq:RRqq}), can be written as follows:
\begin{eqnarray}
&&\frac{d\sigma}{dy_1dy_2dk_{1T}dk_{2T}}(p + p \to c(k_1)+\bar
c(k_2) + X)= \frac{k_{1T}k_{2T}}{16 \pi^3} \int d\phi_1 \int
d\Delta\phi \int dt_1\times \nonumber \\
&& \times \Phi(x_1,t_1,\mu^2) \Phi(x_2,t_2,\mu^2)
\frac{\overline{|{\cal M}(\mathcal{R} + \mathcal{R} \to c + \bar
c)|^2}}{(x_1x_2 S)^2}\mbox{,} \label{eq:QMRKqq}
\end{eqnarray}
where $x_1=q_1^+/P_1^+$, $x_2=q_2^-/P_2^-$, $\Delta\phi$ is the
azimuthal angle between ${\bf k}_{1T}$ and ${\bf k}_{2T}$, the
rapidity of the final parton with four-momentum $k$
 can be presented as $\displaystyle{y=\frac{1}{2}\ln (\frac{k^+}{k^-})}$.
Again, we have checked a fact that in the limit of $t_{1,2}\to 0$,
we reproduce the conventional factorization formula of the collinear
parton model from (\ref {eq:QMRKg}) and (\ref{eq:QMRKqq}).

The important ingredient of the our calculation is unPDF, which we
take as one proposed by Kimber,
  Martin and Ryskin (KMR) \cite{KMR}. These distributions are obtained introducing a single-scale auxiliary function which
  satisfies the unified BFKL/DGLAP evolution equation, where the leading BFKL logarithms $\alpha_S\log (1/x)$ are fully resummed
  and even a major (kinematical) part of the subleading BFKL effects are taken into account. The dependence on the second scale, $t$,
  is implemented at the last step of the evolution. This procedure to obtain unPDFs requires less computational efforts than
  the precise solution of two-scale evolution equations such as, for instance, Ciafaloni-Catani-Fiorani-Marchesini equation~\cite{CCFM},
  but we found it to be suitable and adequate
  to the physics of processes under study. The set of Feynman diagrams corresponding to the contributions proportional to $\log (1/x)$
  included into unPDFs appear if we consider the Reggeization of given parton. That makes unPDFs fully compatible with Reggeized amplitudes.
  In our previous study~\cite{SingleD} devoted to the similar processes of
  $D$-meson production we proved that they give the best description of
  $p_T-$spectra measured at the central region of rapidity at the LHC.

To describe the hadronization stage we use the fragmentation model~\cite{MeleNason}.
So, the transition from the produced gluon or $c$-quark to the $D$ meson is described by corresponding fragmentation
function $D_i(z,\mu^2)$.
According the factorization theorem in QCD, in the fragmentation model the basic formula
for the single $D$-meson production cross section reads:
\begin{eqnarray}
\frac{d\sigma(p+p\to D+ X)}{dp_{DT} dy}= \sum_i \int_0^1
\frac{dz}{z} D_{i\to D}(z,\mu^2) \frac{d\sigma(p+p\to i(k_i=p_D/z)+
X)}{dk_{iT}dy_i},\label{eq:frag}
\end{eqnarray}
where $D_{i\to D}(z,\mu^2)$ is the fragmentation function for the
parton $i$ splitting into $D$-meson at the scale $\mu^2$, $z$ is the longitudinal momentum fraction of a fragmenting particle carried by the $D$-meson.
In the zero-mass approximation the fragmentation parameter $z$ can be defined
as follows $p_D^\mu=zk_i^\mu$, $p_D$ and $k_i$ are the $D$-meson and
$i$-parton  four-momenta, and $y_D=y_i$.

In the case of pair $D$-meson production an additional integral over the momentum fraction of second parton appears.
The fragmentation formula now has the following form:
\begin{eqnarray}
&&\frac{d\sigma(p+p\to D+\overline{D} + X)}{dp_{DT} dy_D dp_{\overline{D}T} dy_{\overline{D}}}= \sum_{ij} \int_0^1
\frac{dz_1}{z_1} \int_0^1\frac{dz_2}{z_2} D_{i\to D}(z_1,\mu^2) D_{j\to \overline{D}}(z_2,\mu^2)\times \nonumber \\
&&\times\frac{d\sigma(p+p\to i(k_i=p_D/z_1)+j(k_j=p_{\overline{D} }/z_2)+
X)}{dk_{iT}dy_i dk_{jT}dy_j},\label{eq:fragDD}
\end{eqnarray}
where $D_{i\to D}(z_1,\mu^2)$ and $D_{j\to \overline{D}}(z_2,\mu^2)$
are the fragmentation functions for the transitions of parton
$i=g,c,\bar c$ into $D$ meson with momentum fraction $z_1$ and of
parton $j=g,c,\bar c$ into $\overline{D}$ meson with momentum
fraction $z_2$, respectively, at the scale $\mu^2$. In our
calculations we use the LO FFs from Ref.~\cite{KKSS}, where the fits
of non-perturbative $D^0$, $D^+$, $D^{\star +}$, and $D_s^+$ FF's to
OPAL data from LEP1~\cite{LEP1} were performed. These FFs satisfy
two desirable properties: at first, their scaling violations are
ruled by DGLAP evolution equations; at second, they are universal.

As the contribution of gluon fragmentation at $\mu>\mu_0$ is
initiated by the perturbative transition of gluons to $c\bar
c$-pairs encountered by DGLAP evolution equations, the part of
$c$-quarks produced in the subprocess (\ref{eq:RRqq}) with their
subsequent transition to $D$-mesons are already taken into account
considering $D$-meson production via gluon fragmentation in the
subprocess (\ref{eq:RRg}).
 Such a way, to avoid double counting, we must subtract this contribution, that can be effectively done by the lower cut of the amplitude
  in formula (\ref{eq:QMRKqq}) as $\hat s>4m_c^2$ , i.e. at the threshold of the production of $c\bar c$-pair.

\section{Single $D$-meson production}

\label{sec:four}

\begin{figure}[ph]
\begin{center}
\includegraphics[width=0.5\textwidth, angle=-90,origin=c, clip=]{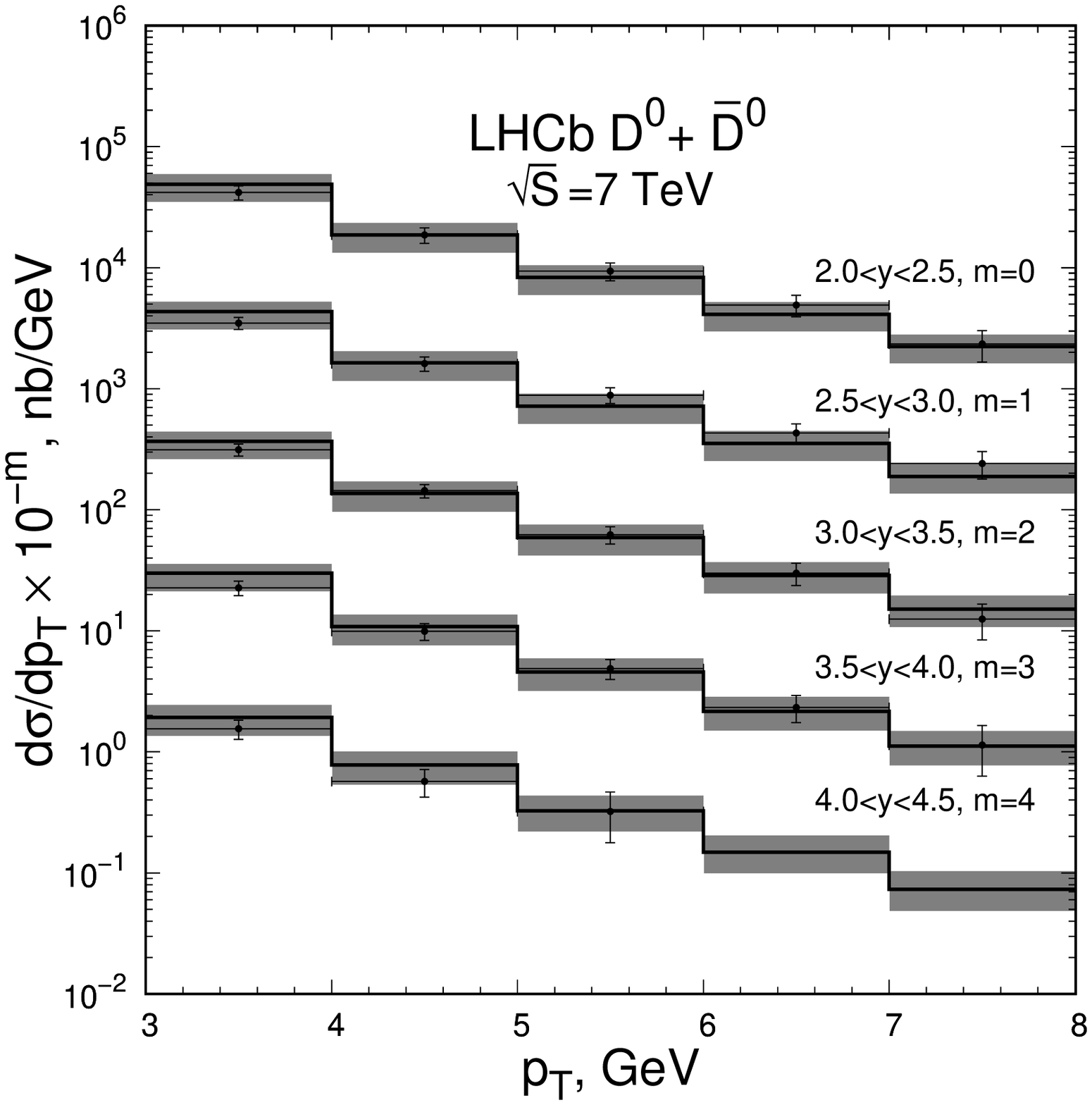}\includegraphics[width=0.5\textwidth, angle=-90,origin=c, clip=]{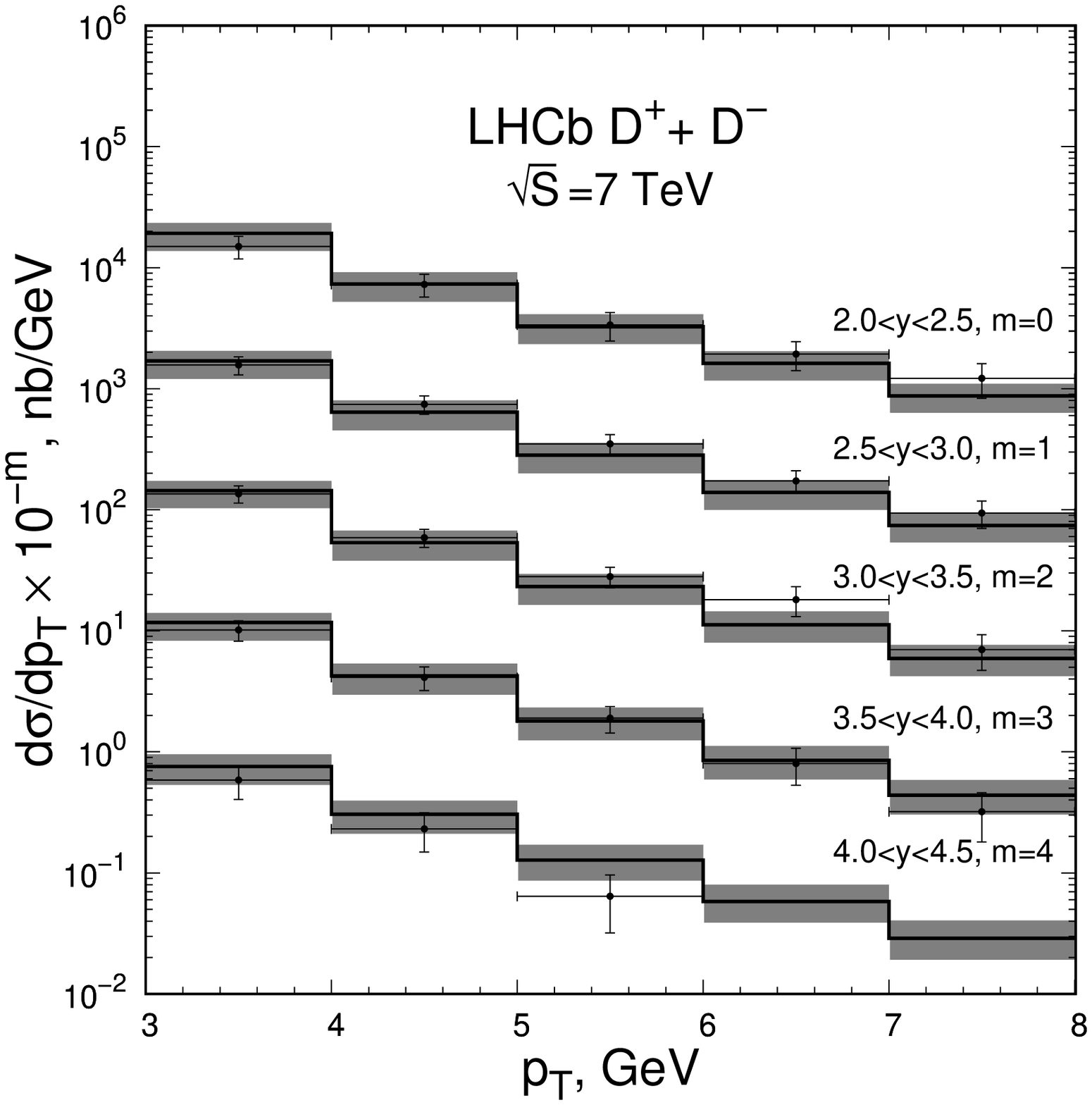}
\includegraphics[width=0.5\textwidth, angle=-90,origin=c, clip=]{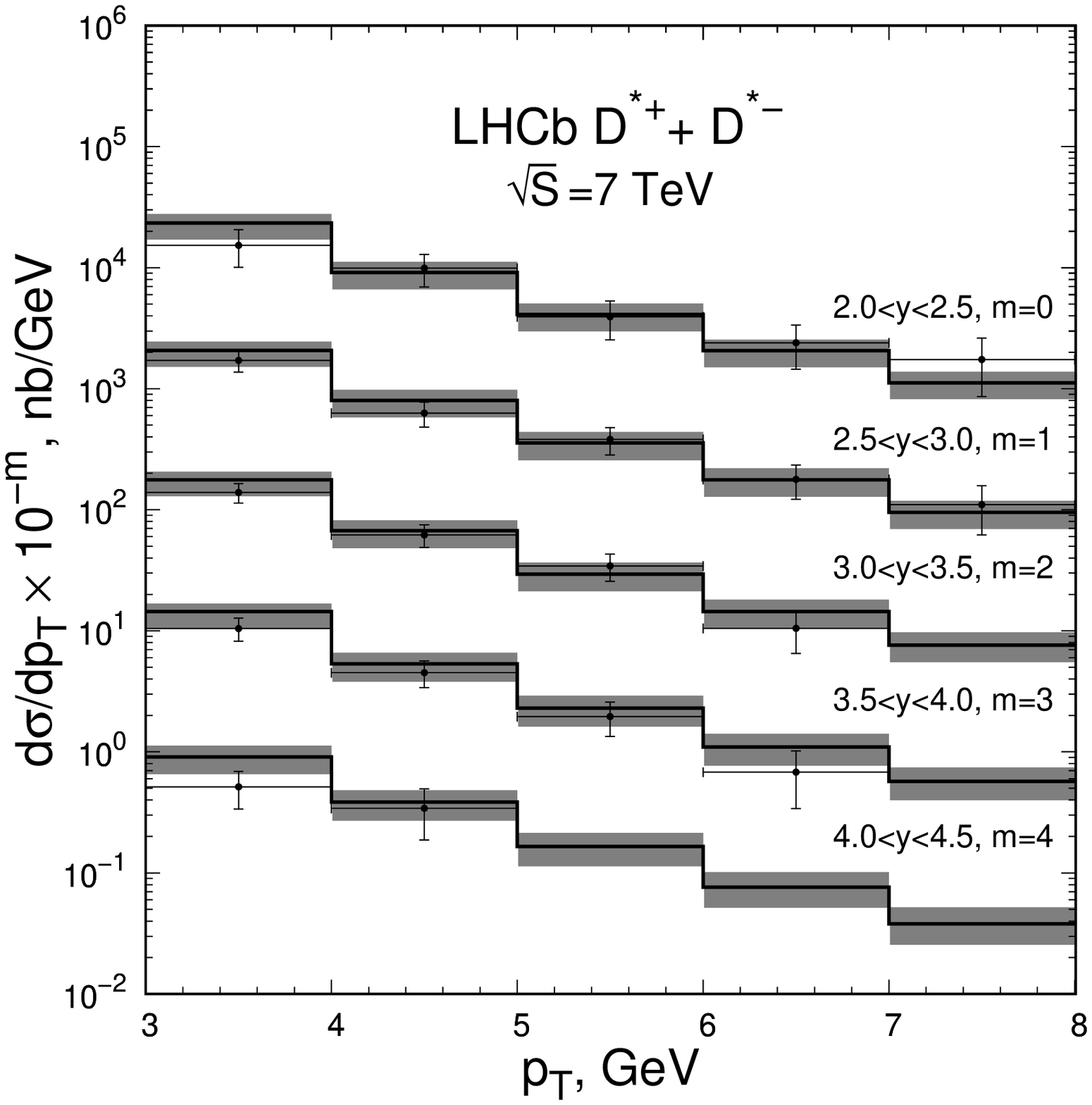}\includegraphics[width=0.5\textwidth, angle=-90,origin=c, clip=]{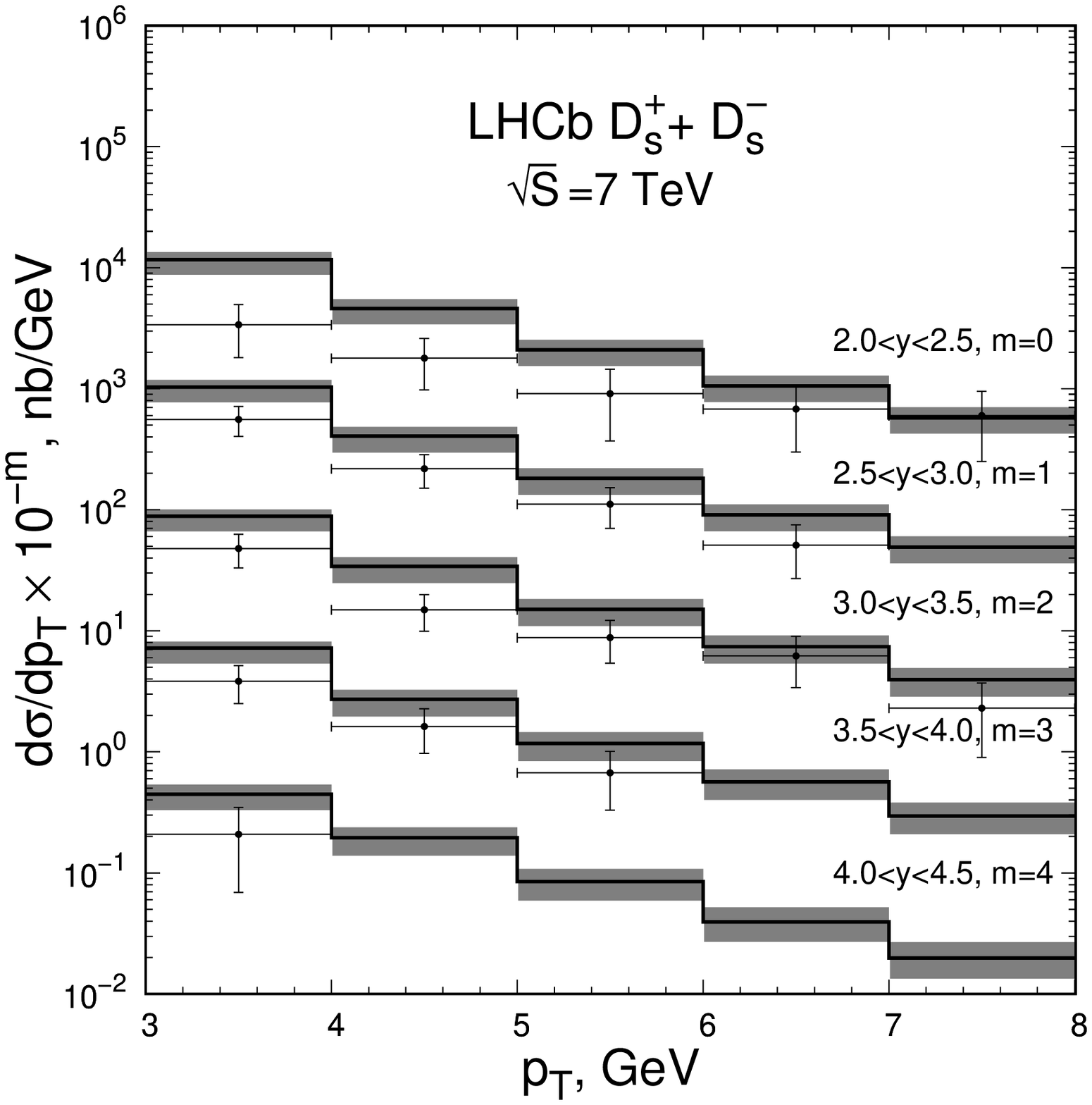}
\caption{Transverse momentum distributions of $D^0$, $D^+$,
$D^{\star+}$, and $D_s^+$ mesons in different rapidity regions at
the energy $\sqrt S=7$~TeV. The experimental data from the LHCb
Collaboration ~\cite{LHCb1}.\label{fig:1}}
\end{center}
\end{figure}

The forward rapidity region in $pp$ collisions at the LHC became
available due to the specially designed LHCb detector where the
measurements of differential cross sections of $D^0$, $D^+$,
$D^{\star+}$, and $D_s^+$ mesons with $2.0<y<4.5$ at the $\sqrt
S=7$~TeV were performed~\cite{LHCb1,LHCb_Pair}. The observed data
divided into 5 rapidity regions
 are presented together with our results obtained in the LO of PRA in the Fig.~\ref{fig:1}.
The sum of all contributions from subprocesses
(\ref{eq:RRg})-(\ref{eq:RRqq}) is shown as solid line. We estimated a
theoretical uncertainty arising from uncertainty of definition of
factorization and renormalization scales ($\mu=\mu_R=\mu_F$) by
varying them between $1/2 p_T$ and $2 p_T$ around their central
value of $p_T$, the transverse momentum of fragmenting parton. The
resulting uncertainty is depicted in the figures by shaded bands. We
find a good agreement between our predictions and experimental data
in the whole $p_T$ interval of $D$-meson transverse momenta within
experimental and theoretical uncertainties only with one exclusion
for $D_s$-mesons, when our results overestimate data by a factor 2. It
may be connected with the bad quality of gluon or $c$-quark FF
extracted from the $e^+e^-$ data, see \cite{KKSS} for details.

In the calculations of Ref.~\cite{shurek1}, which were done in the
similar approach with KMR unPDFs \cite{KMR} and using scale-dependent FFs KKKS08~ \cite{KKSS}, the underestimation of
experimental data from LHCb Collaboration for $D^0-$meson transverse
momentum spectra by a factor 2 or more was found. But, in
the Ref.~\cite{shurek1} the only $c-$quark fragmentation into $D-$mesons
was applied. We correct this results taking into account gluon to
$D-$meson fragmentation and obtain good agreement with the data.

\begin{figure}[ph]
\begin{center}
\includegraphics[width=0.5\textwidth, angle=-90,origin=c, clip=]{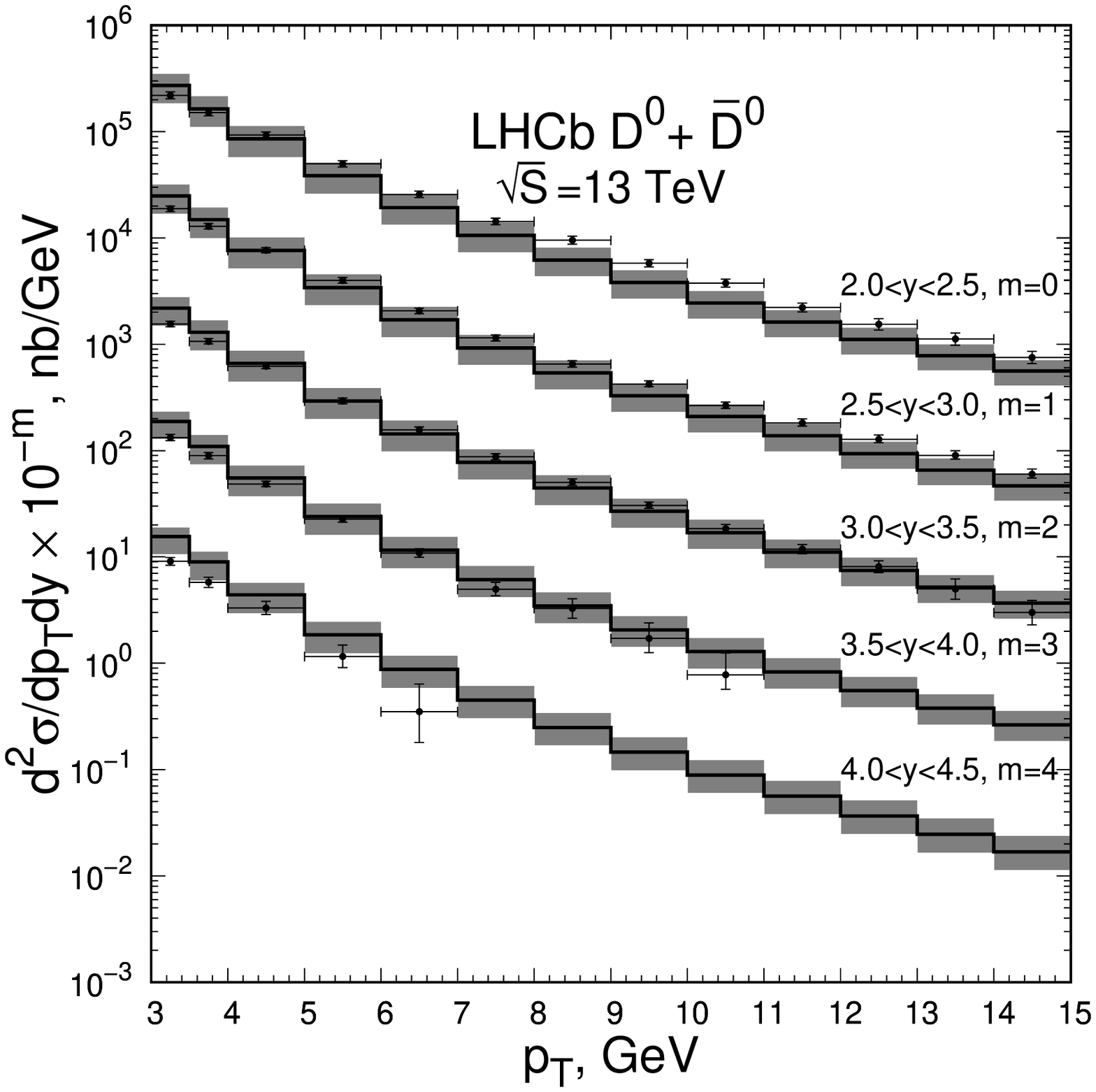}\includegraphics[width=0.5\textwidth, angle=-90,origin=c, clip=]{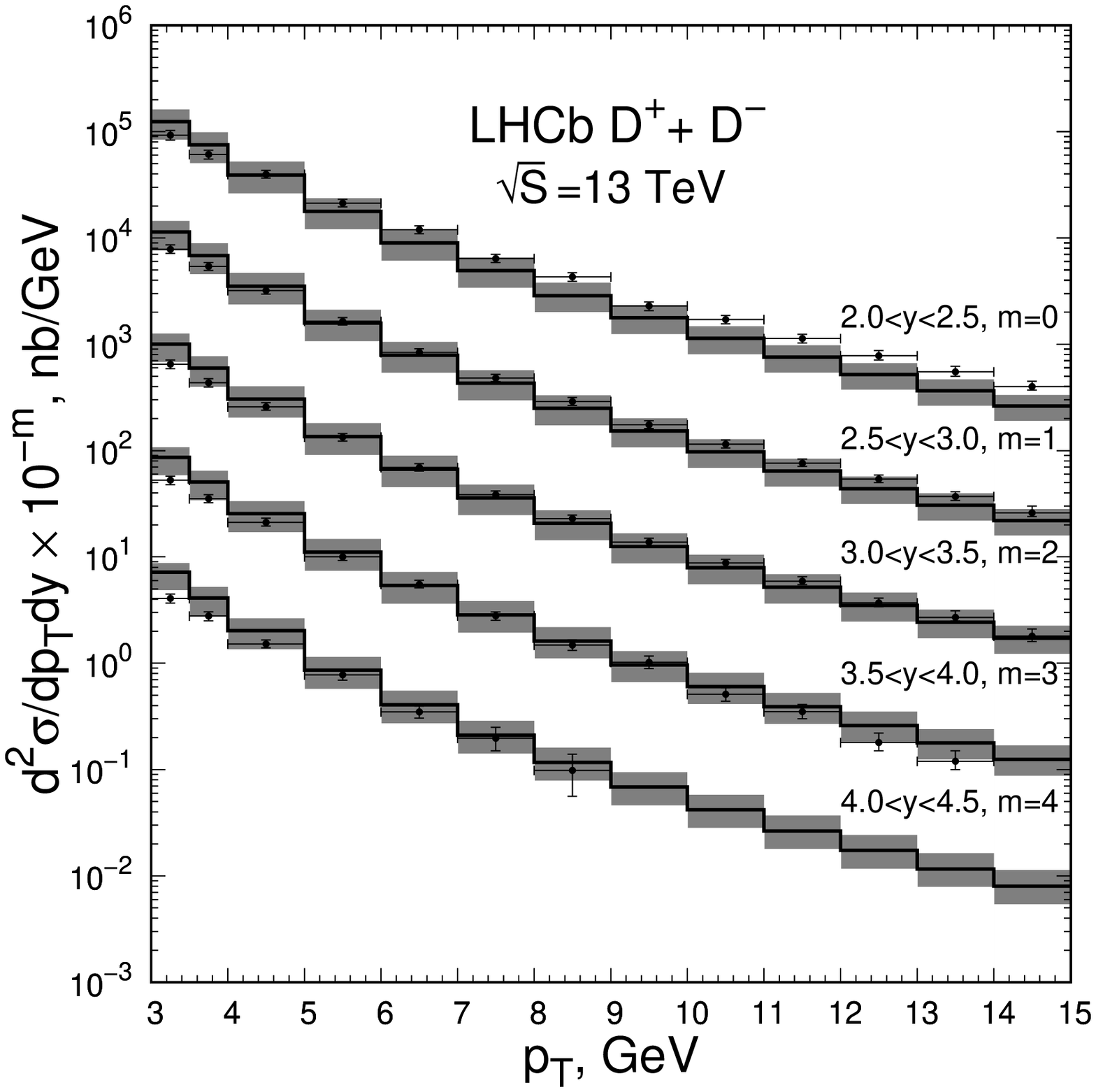}
\includegraphics[width=0.5\textwidth, angle=-90,origin=c, clip=]{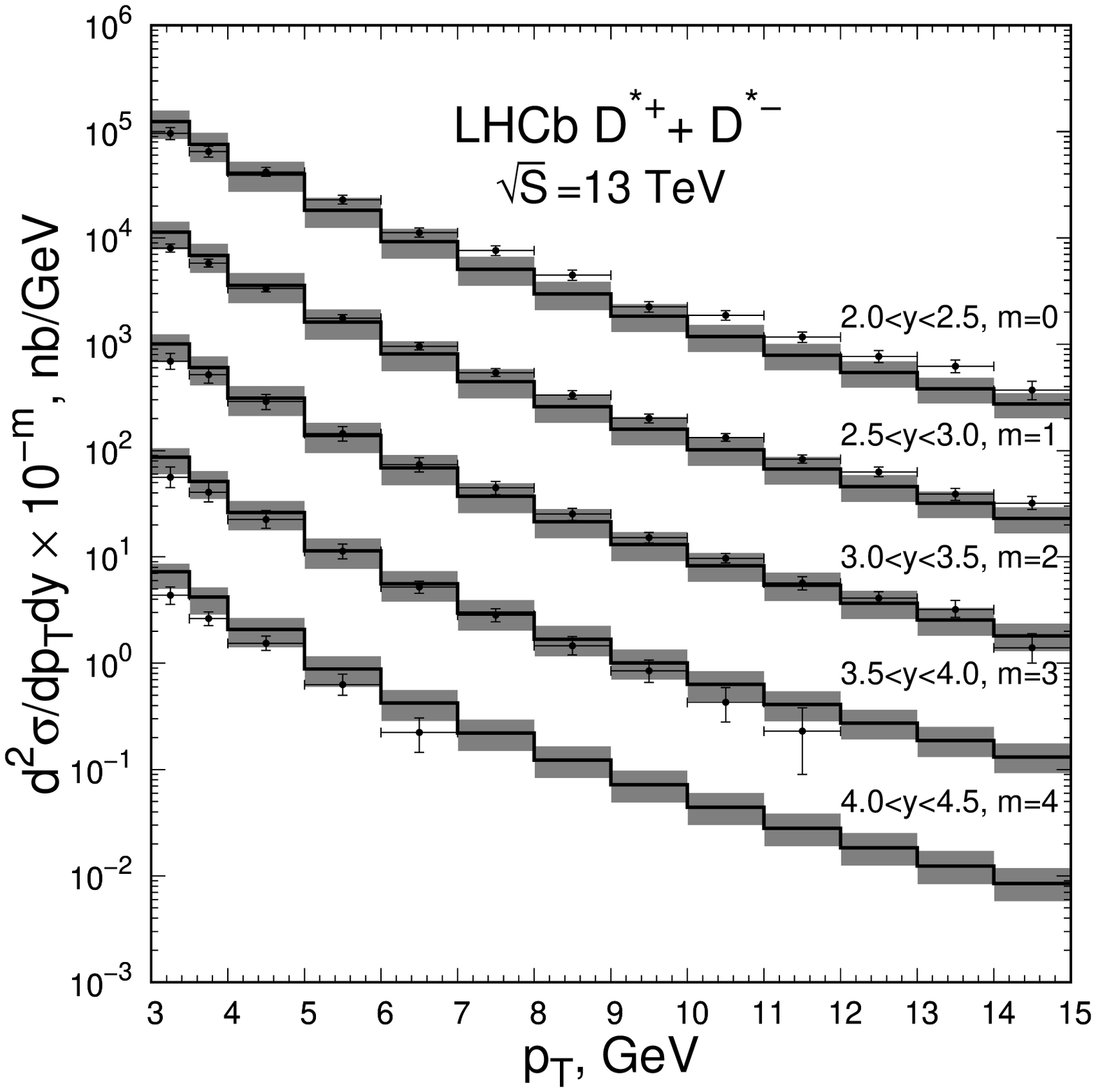}\includegraphics[width=0.5\textwidth, angle=-90,origin=c, clip=]{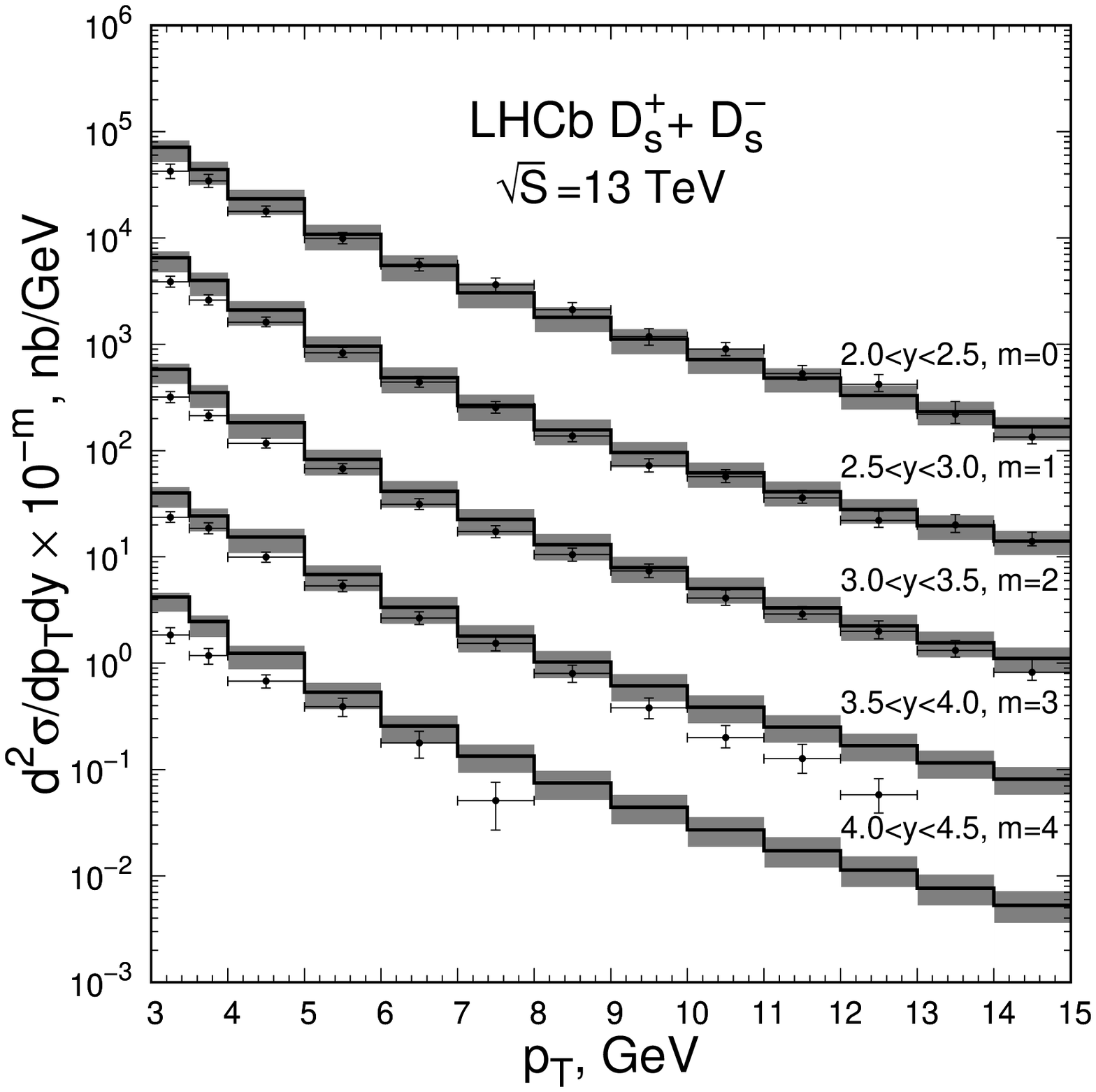}
\caption{Transverse momentum distributions of $D^0$, $D^+$,
$D^{\star+}$, and $D_s^+$ mesons in different rapidity regions at
the energy $\sqrt S=13$~TeV. The experimental data from the LHCb
Collaboration ~\cite{LHCb2}.}\label{fig:2}
\end{center}
\end{figure}

\begin{figure}[ph]
\begin{center}
\includegraphics[width=0.5\textwidth, angle=-90,origin=c, clip=]{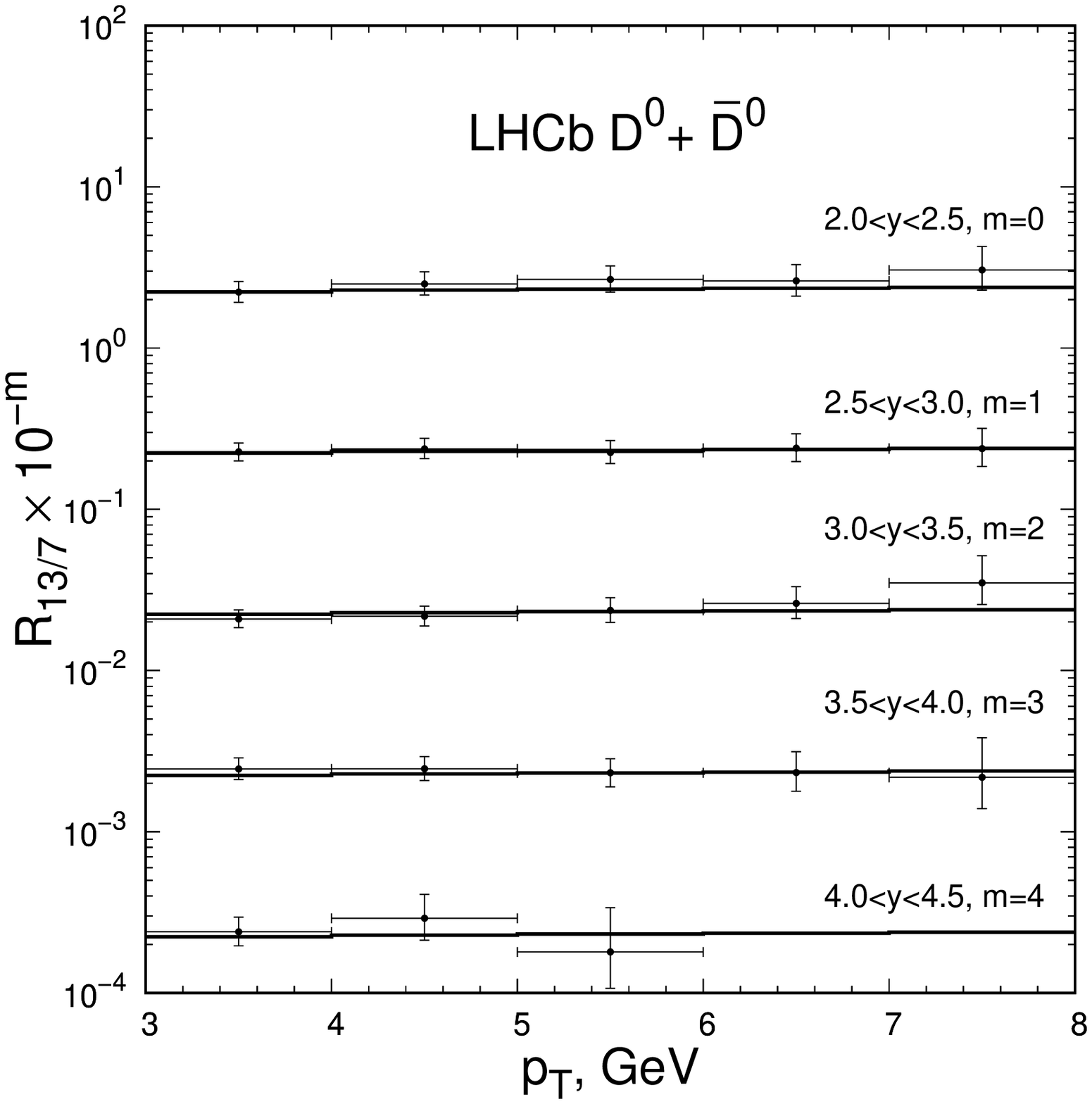}\includegraphics[width=0.5\textwidth, angle=-90,origin=c, clip=]{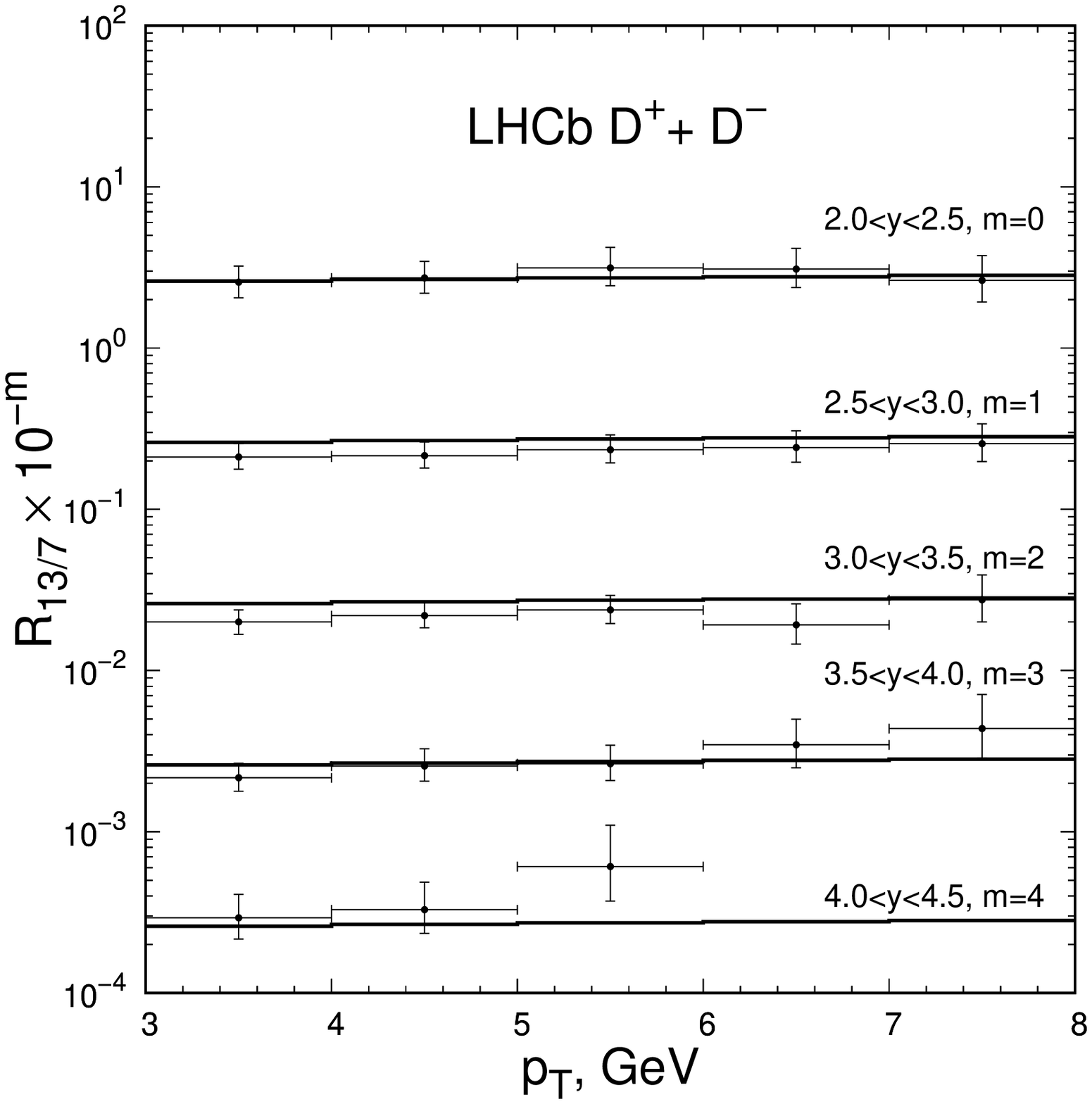}
\includegraphics[width=0.5\textwidth, angle=-90,origin=c, clip=]{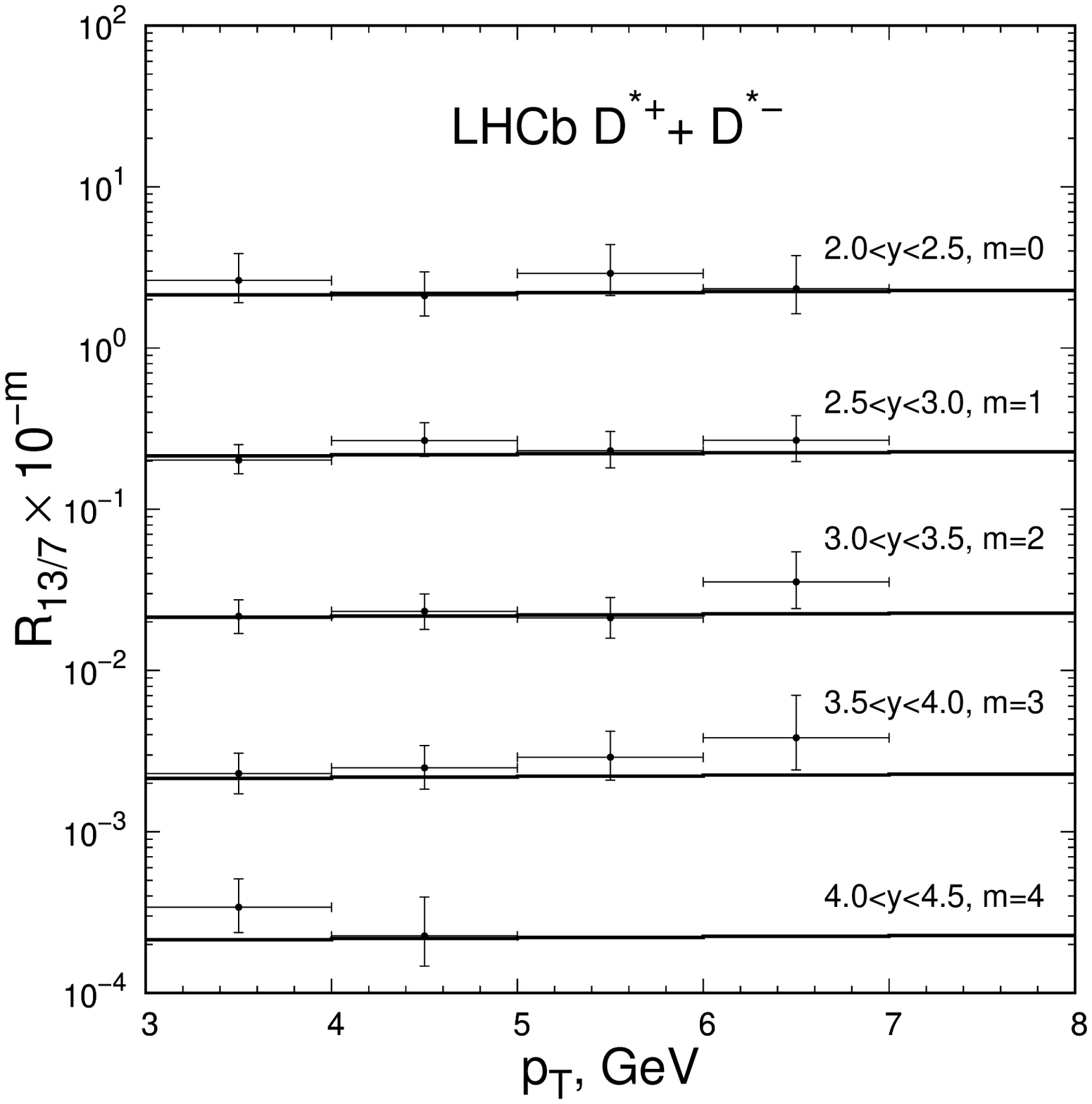}\includegraphics[width=0.5\textwidth, angle=-90,origin=c, clip=]{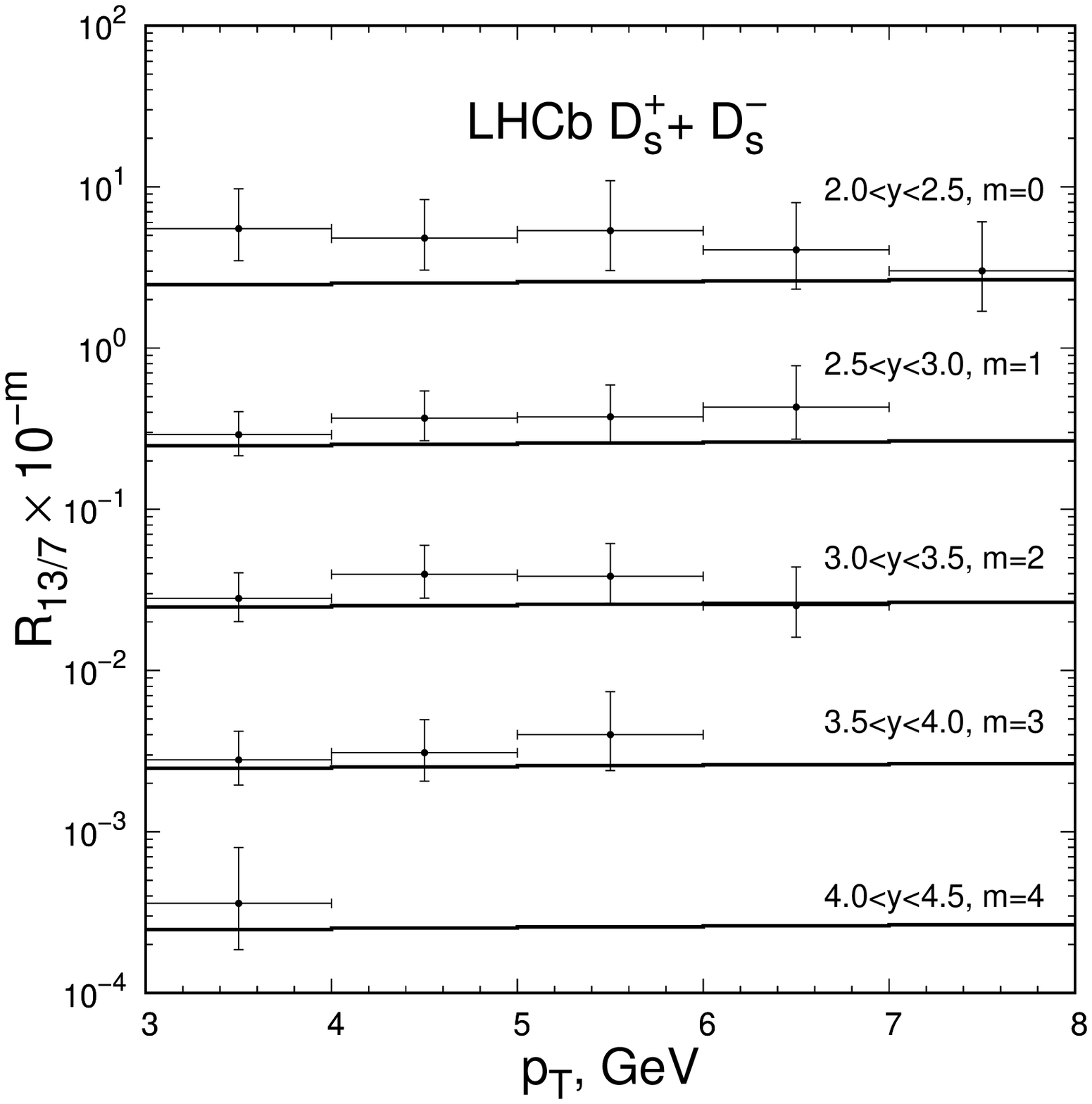}
\caption{{The ratios of prompt $D^0$, $D^+$, $D^{\star+}$, and
$D_s^+$ cross section at $\sqrt{S}=13$~TeV to the same at
$\sqrt{S}=7$~TeV for different regions of rapidity.} The
experimental data from the LHCb Collaboration
~\cite{LHCb2}.}\label{fig:3}
\end{center}
\end{figure}

We also provide theoretical calculations for $D-$production within
the LHCb acceptance at the energy $\sqrt S=13$~TeV. We compare them
with the most recent data from the LHCb~\cite{LHCb2}. The results
are presented as double differential distributions
$d^{2}\sigma/(dp_Tdy)$ for $D^0$, $D^+$, $D^{\star+}$, and $D_s^+$
mesons in the same rapidity region as at the $\sqrt S=7$~TeV in the
Fig.~\ref{fig:2}. Our predictions for $D_s$ mesons became better at
the $\sqrt{S}=13$ TeV than at the smaller energy and they are very
close to data. Again we estimate a theoretical uncertainty varying
the factorization and renormalization scales up and down by the
factor $2$ around the central value and depict it by the shaded
bands at the plot. In addition to the differential cross sections
the LHCb Collaboration presented ratios between differential cross
sections at 13 and 7~TeV in the transverse momentum region
$3<p_T<8$~GeV.
\begin{equation}\label{eq:Ratio}
    R_{13/7}(p_T,y)=\frac{d^2\sigma_{13}}{dp_Tdy}/\frac{d^2\sigma_7}{dp_Tdy}.
\end{equation}
 We present our calculations of the ratio in the Fig.~\ref{fig:3}.
We obtain a good agreement of our predictions with the experimental
data for the $D^0$, $D^+$, and  $D^*$ mesons at the whole ranges of
their transverse momenta and rapidities. In case of $D_s$ mesons,
the agreement is not so good because of the overestimation of data by theoretical predictions at the $\sqrt{S}=7$ TeV.

\section{Pair production of $D$ mesons}
\label{sec:five}

A comparative study of correlation observables in $D\bar D$- and $DD$-pair production is an important test of $D-$meson production
mechanism. As it was found earlier in Ref.~\cite{shurek1} in case of
particle plus antiparticle production ($D\bar D$) the data from LHCb
Collaboration \cite{LHCb_Pair} can be described well in the
$k_T-$factorization approach with inclusion of subprocess
(\ref{eq:RRqq}) only and using the scale-independent
$c-$quark Peterson FF. Working in the same manner, to describe $DD$ pair
production via $c$-quark fragmentation into $D$-meson, we should
include contribution from $2\to 4$ subprocess (\ref{eq:RRqqqq}) as
it was done in Ref.~\cite{shurek2}. It was obtained that predictions
based on the model of $D$-meson production via $c$-quark
fragmentation can not describe data for correlation observables in
$DD$-pair production absolutely and inclusion of double parton
scattering (DPS) production mechanism is needed. The inclusion of
a gluon fragmentation channel, which is the dominant one in the
$D$-meson production at the LHC energy \cite{SingleD}, allows us to
describe data for $DD$-pair production well without hypothesis about
DPS production mechanism \cite{DoubleD}.

\begin{figure}[ph]
\begin{center}
\includegraphics[width=0.5\textwidth, angle=-90,origin=c, clip=]{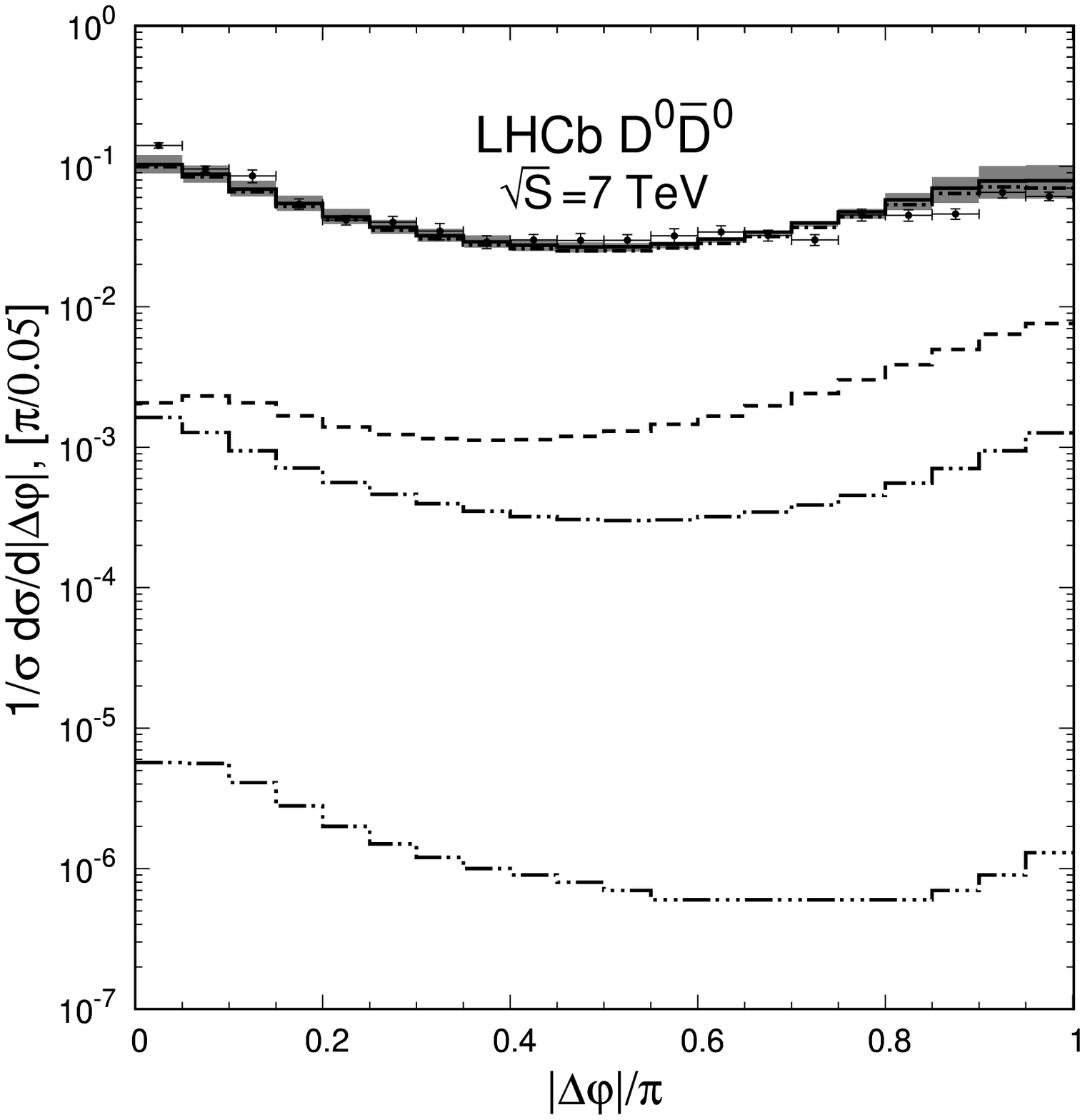}\includegraphics[width=0.5\textwidth, angle=-90,origin=c, clip=]{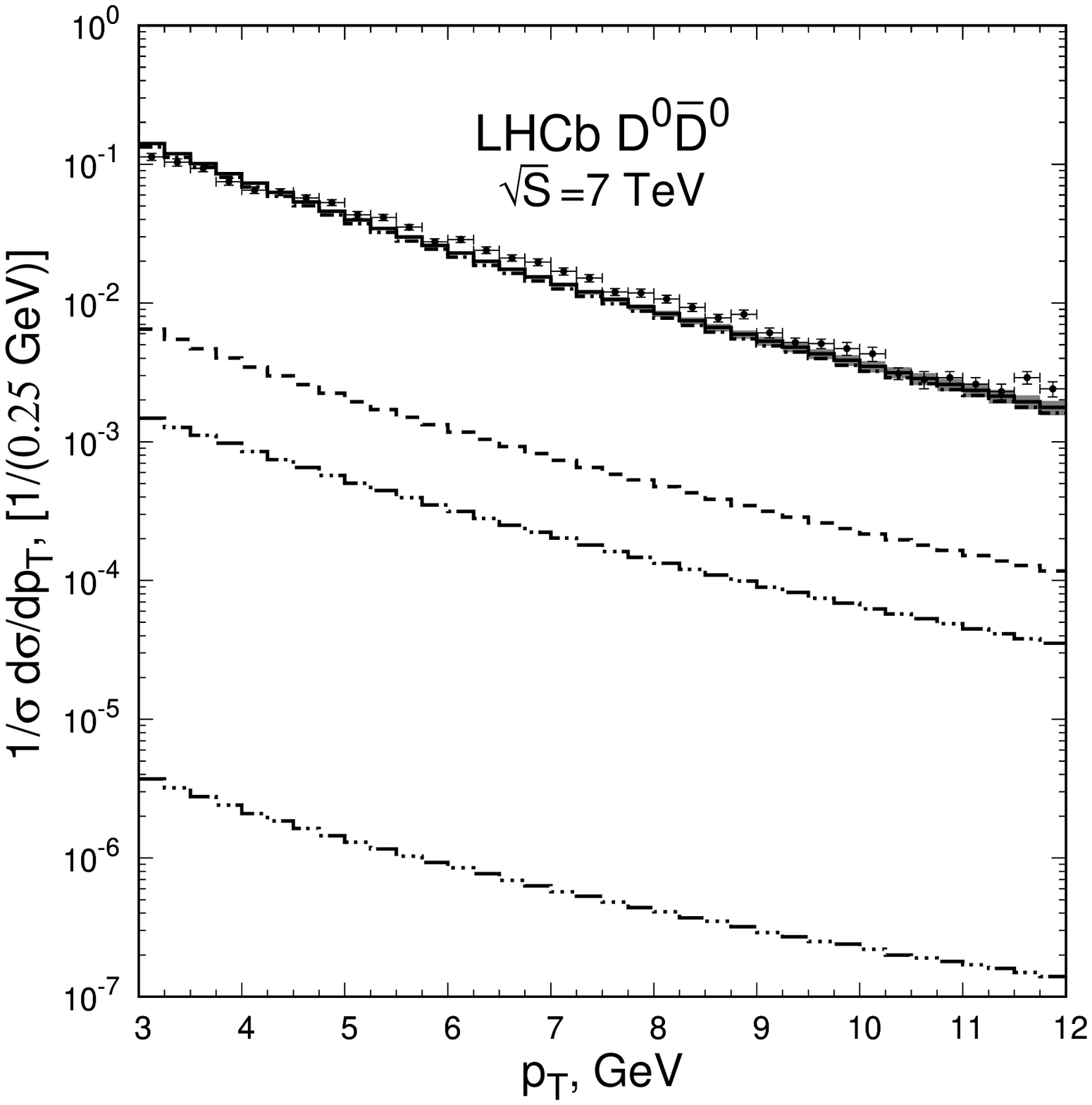}
\includegraphics[width=0.5\textwidth, angle=-90,origin=c, clip=]{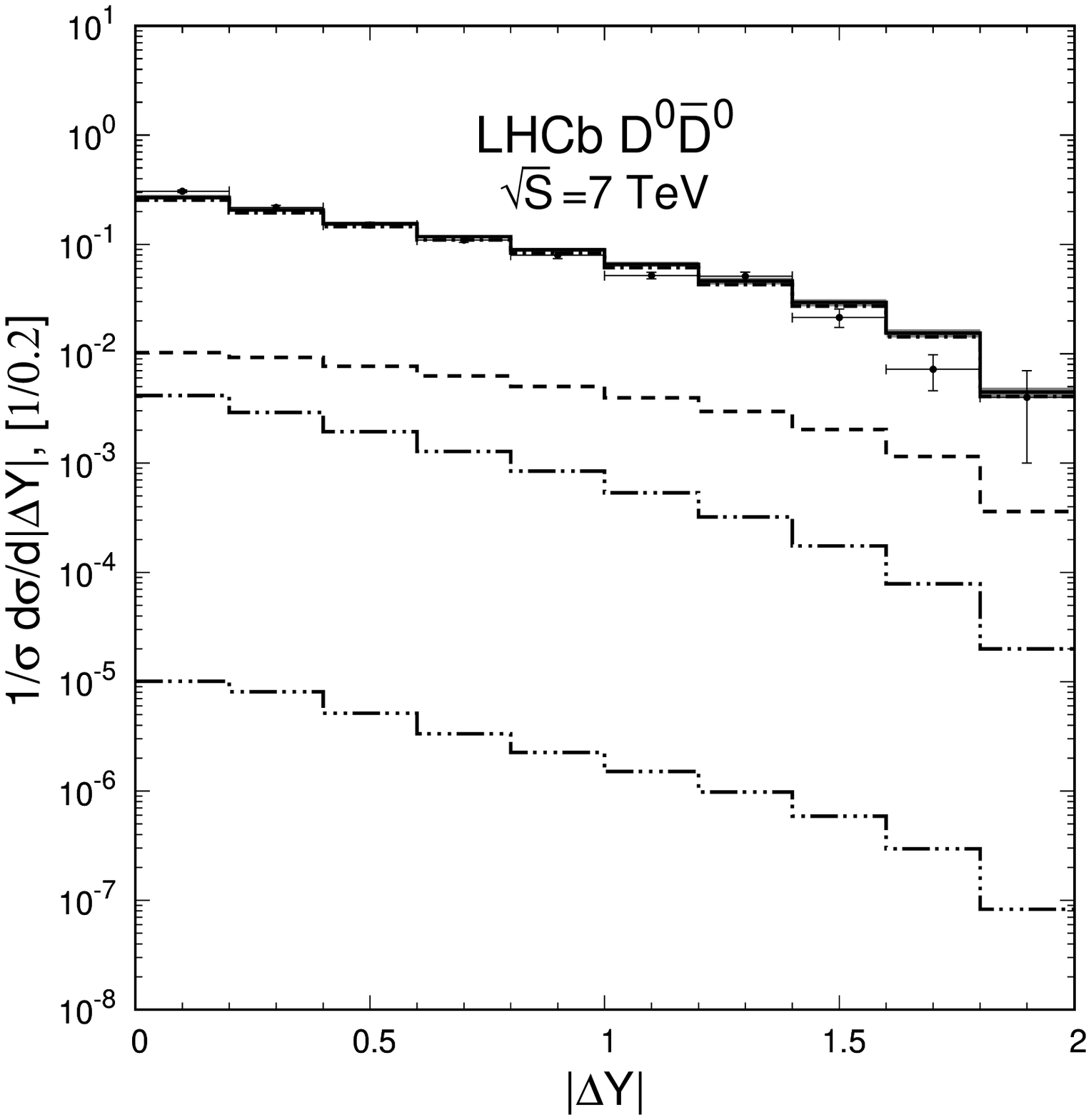}\includegraphics[width=0.5\textwidth, angle=-90,origin=c, clip=]{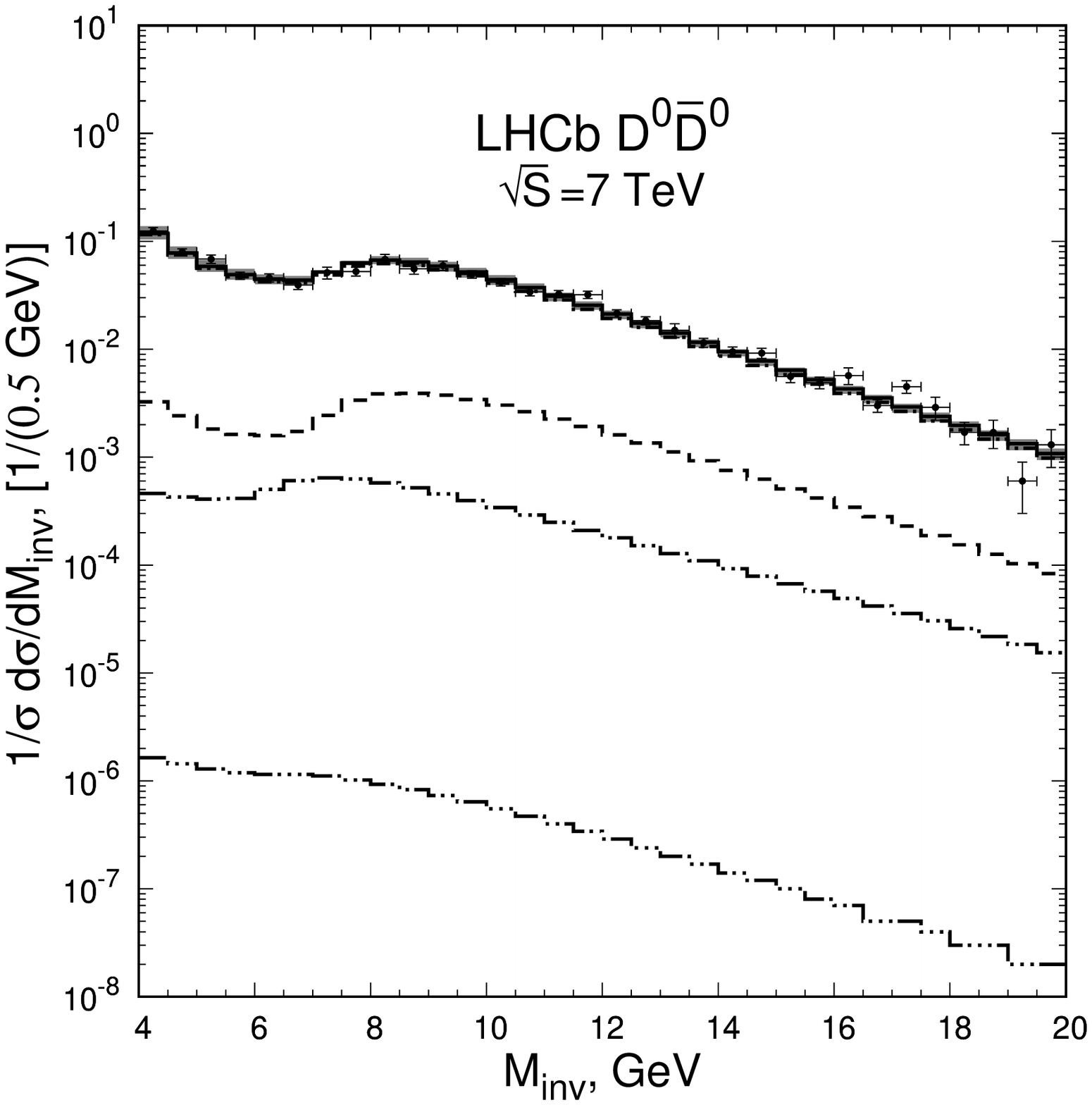}
\caption{The spectra of $D^0\bar D^0$ pairs differential in azimuthal
angle difference (left, top), transverse momentum (right, top),
rapidity distance (left, bottom) and invariant mass of the pair
(right, bottom) at the $2<y<4$ and $\sqrt{S}=7$~TeV. The LHCb data
at LHC are from the Ref.~\cite{LHCb_Pair}. Dashed line represents
the contribution of gluon fragmentation in gluon-gluon fusion,
dash-dotted line -- the $c$-quark fragmentation contribution in
gluon-gluon fusion, double-dot-dashed line is the $c$-quark
fragmentation contribution in quark-antiquark annihilation (sum of
$u-$, $d-$ and $s$-quark contributions), and triple-dot-dashed -- the
same for gluon fragmentation, solid line is their sum.\label{fig:4}}
\end{center}
\end{figure}

\begin{figure}[ph]
\begin{center}
\includegraphics[width=0.5\textwidth, angle=-90,origin=c, clip=]{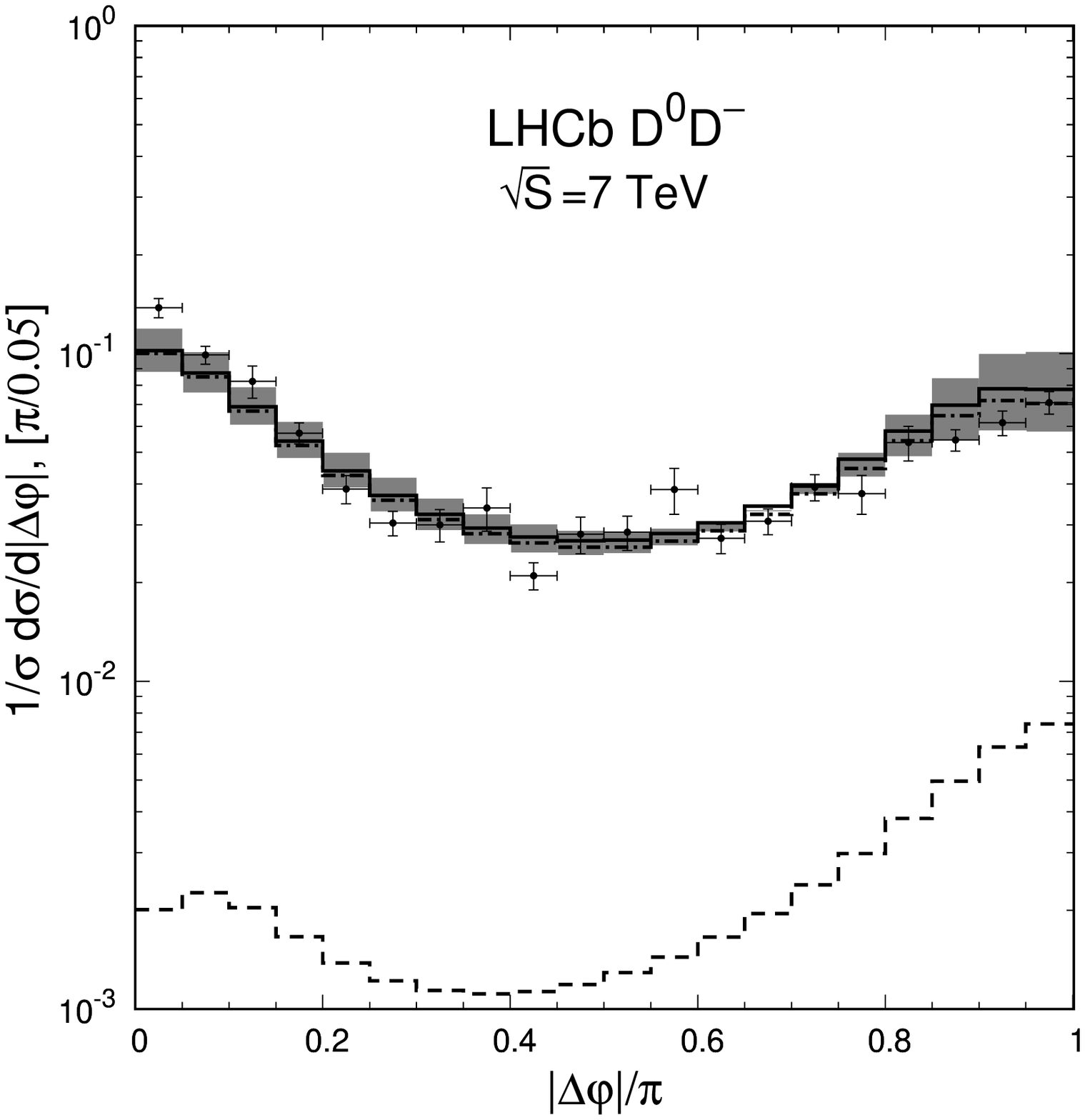}\includegraphics[width=0.5\textwidth, angle=-90,origin=c, clip=]{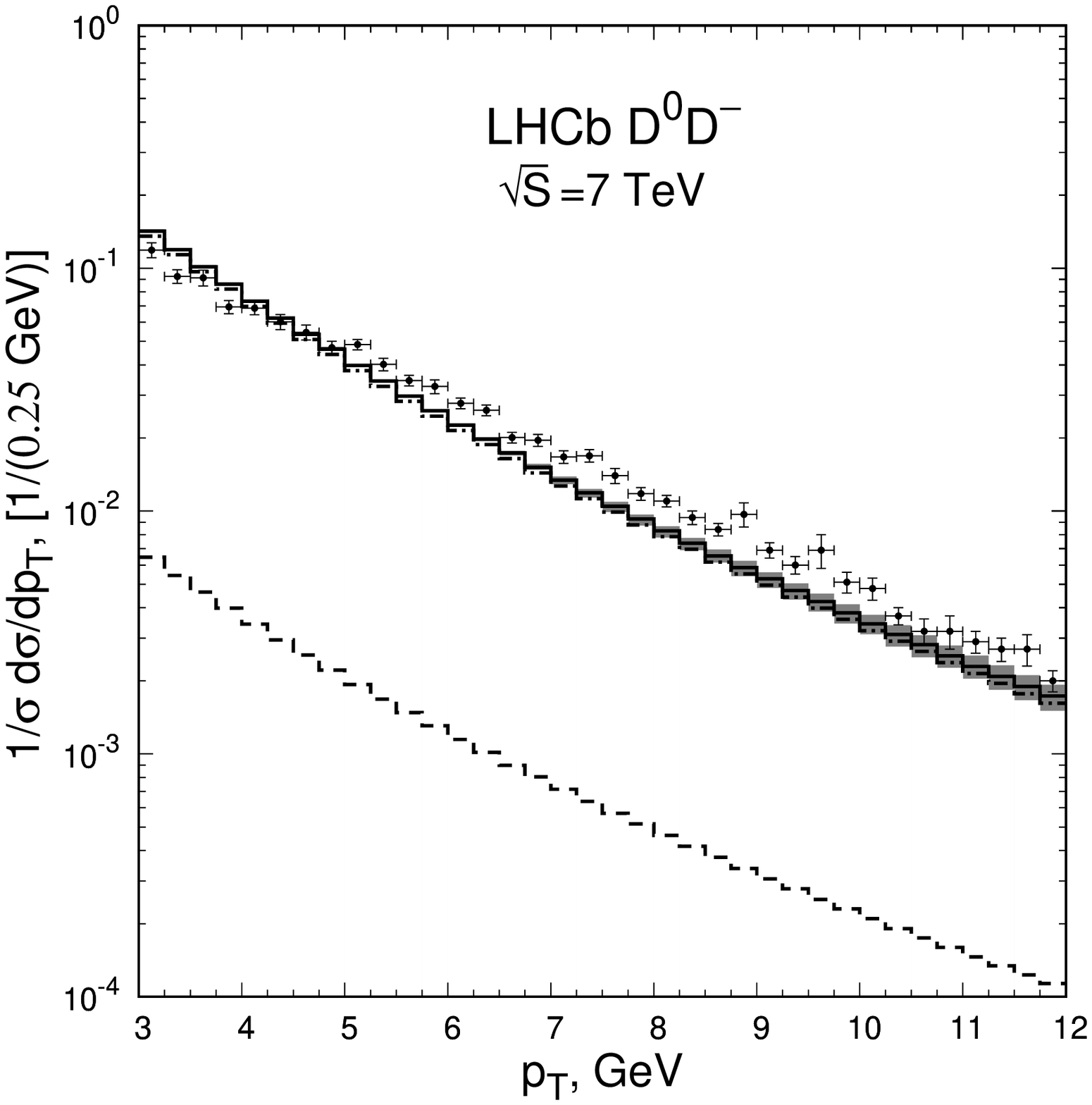}
\includegraphics[width=0.5\textwidth, angle=-90,origin=c, clip=]{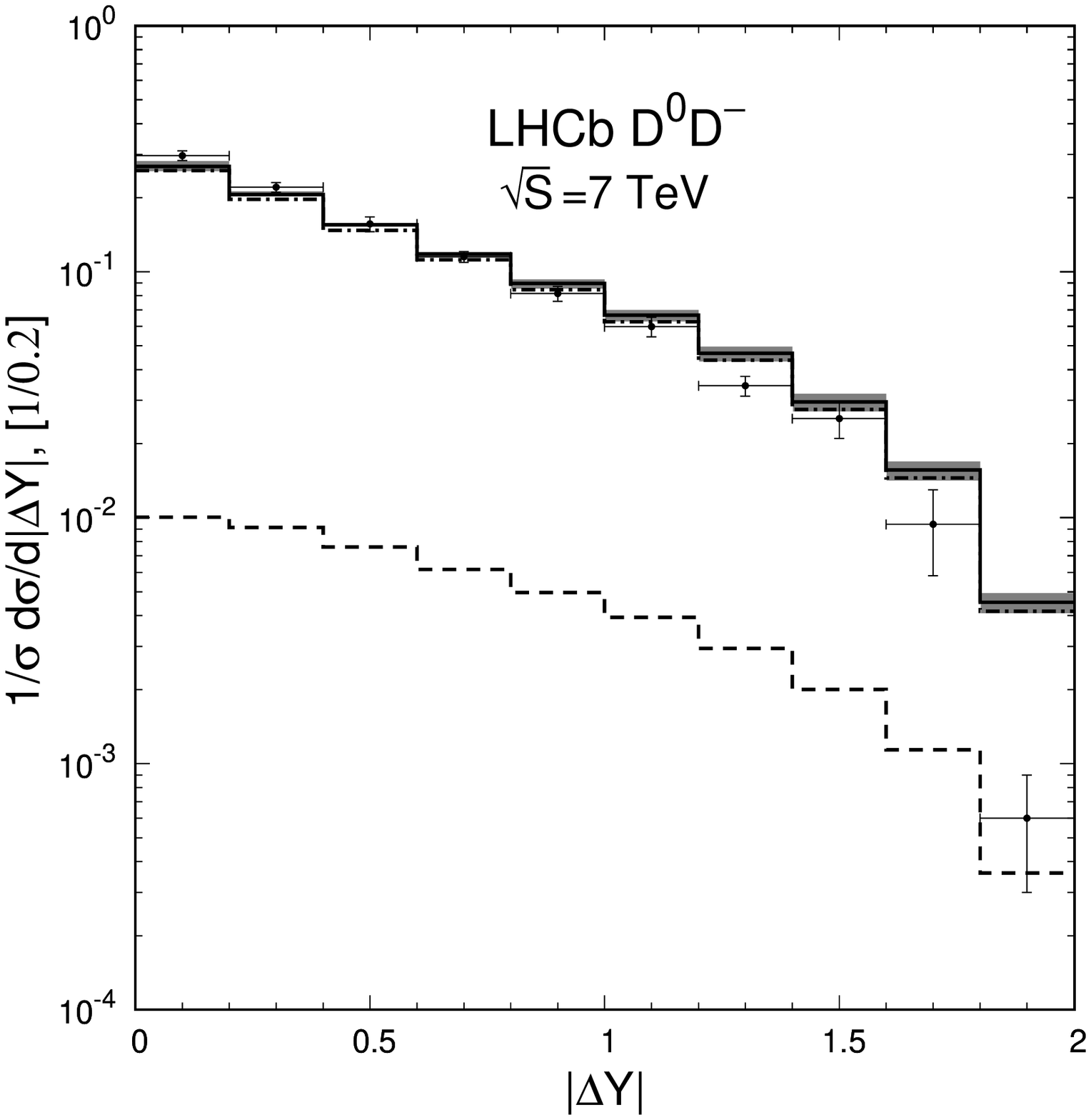}\includegraphics[width=0.5\textwidth, angle=-90,origin=c, clip=]{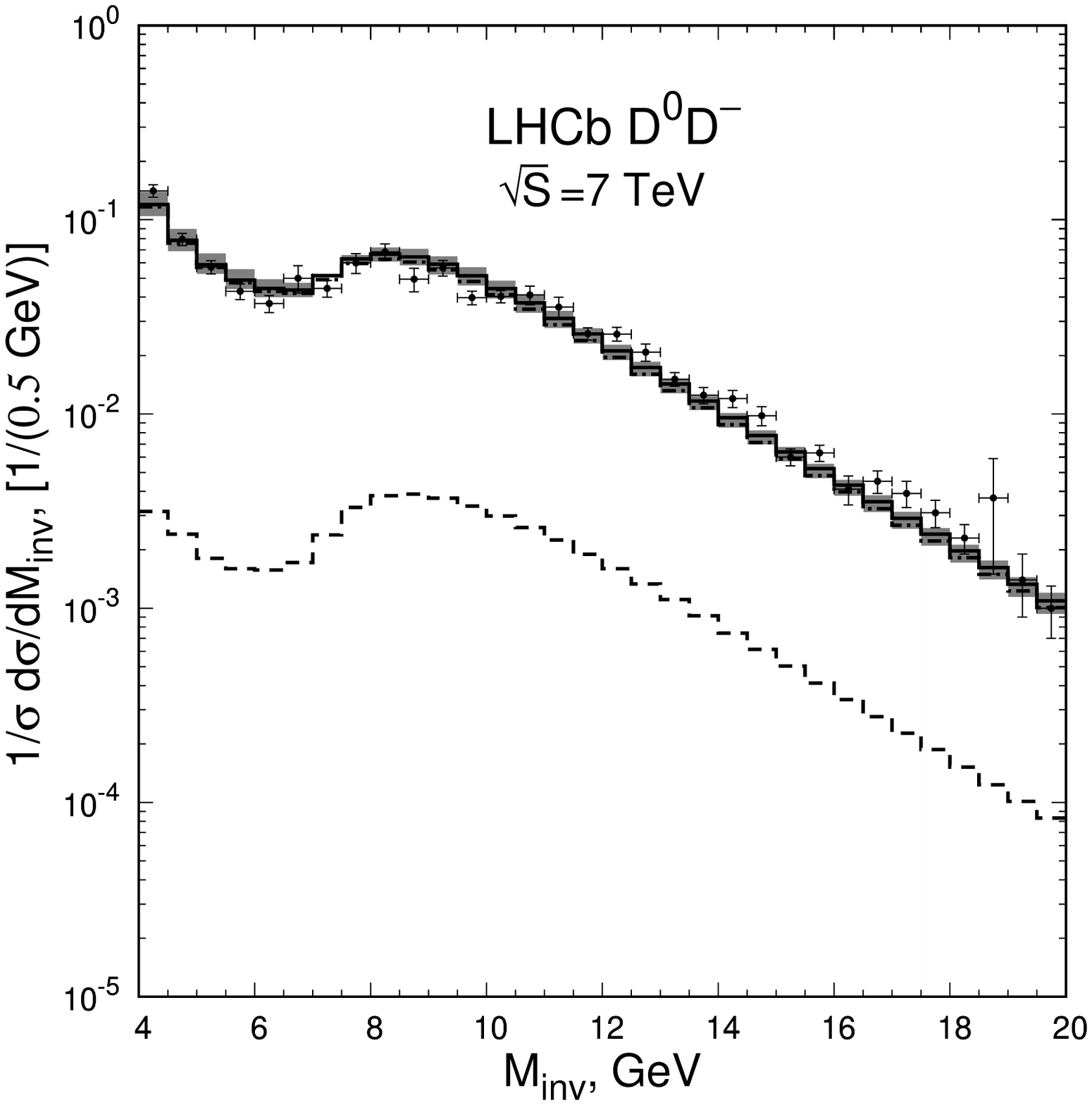}
\caption{The spectra of $D^0D^-$ pairs differential in azimuthal
angle difference (left, top), transverse momentum (right, top),
rapidity distance (left, bottom) and invariant mass of the pair
(right, bottom) at the $2<y<4$ and $\sqrt{S}=7$~TeV. The LHCb data
at LHC are from the Ref.~\cite{LHCb_Pair}. Dashed line represents
the contribution of gluon fragmentation in gluon-gluon fusion,
dash-dotted line -- the $c$-quark fragmentation contribution in
gluon-gluon fusion, solid line is their sum.\label{fig:5}}
\end{center}
\end{figure}

\begin{figure}[ph]
\begin{center}
\includegraphics[width=0.5\textwidth, angle=-90,origin=c, clip=]{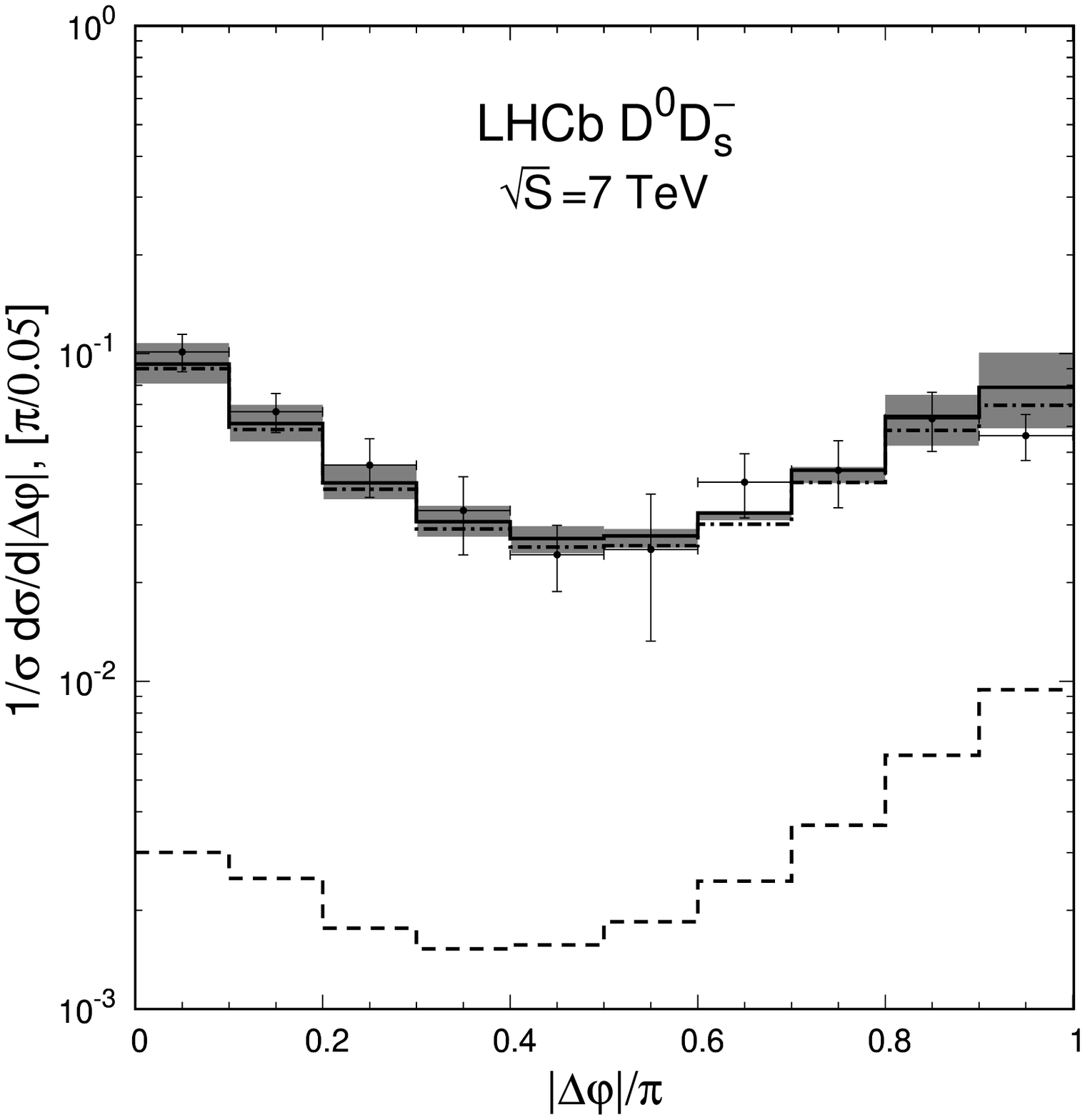}\includegraphics[width=0.5\textwidth, angle=-90,origin=c, clip=]{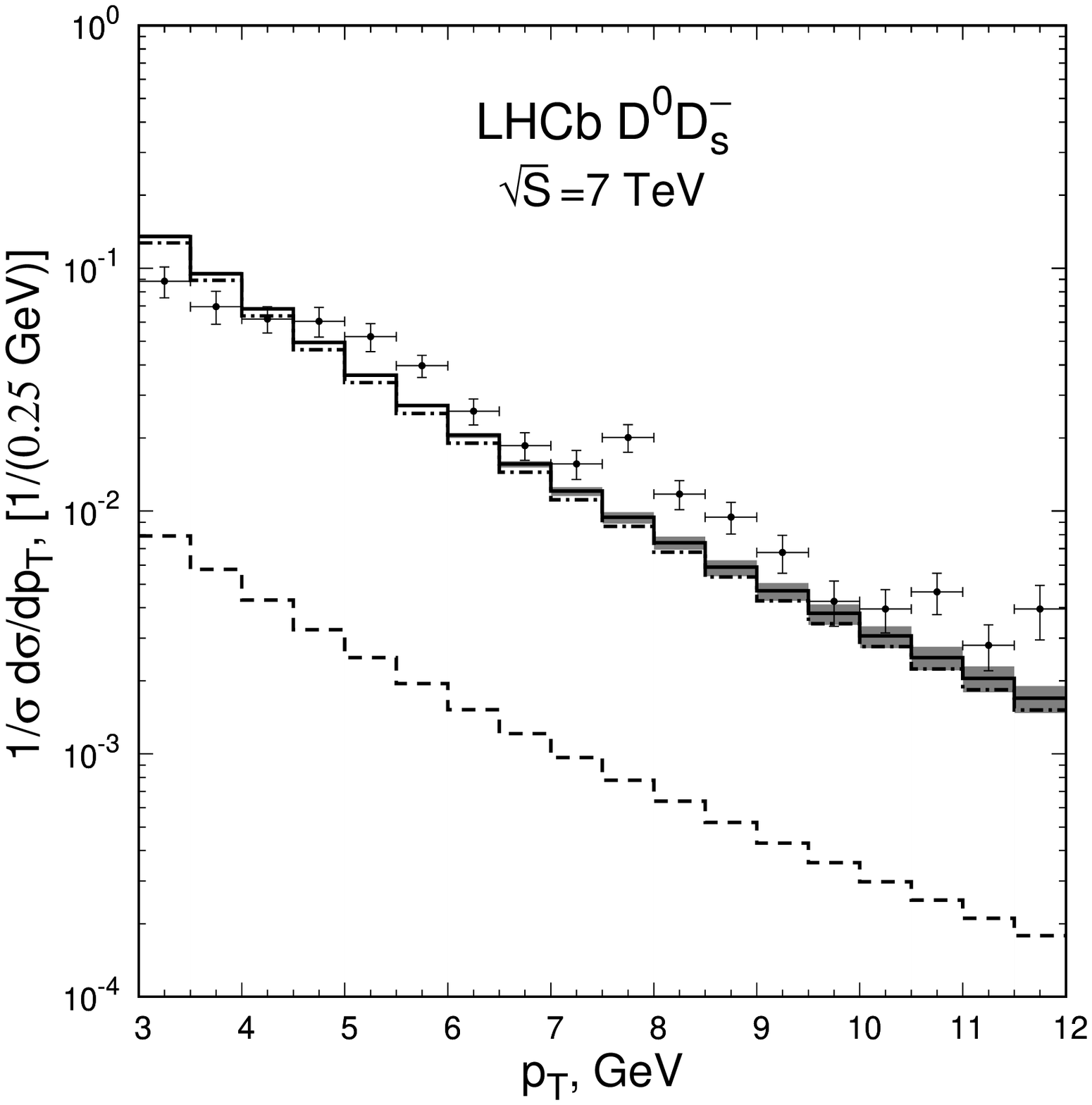}
\includegraphics[width=0.5\textwidth, angle=-90,origin=c, clip=]{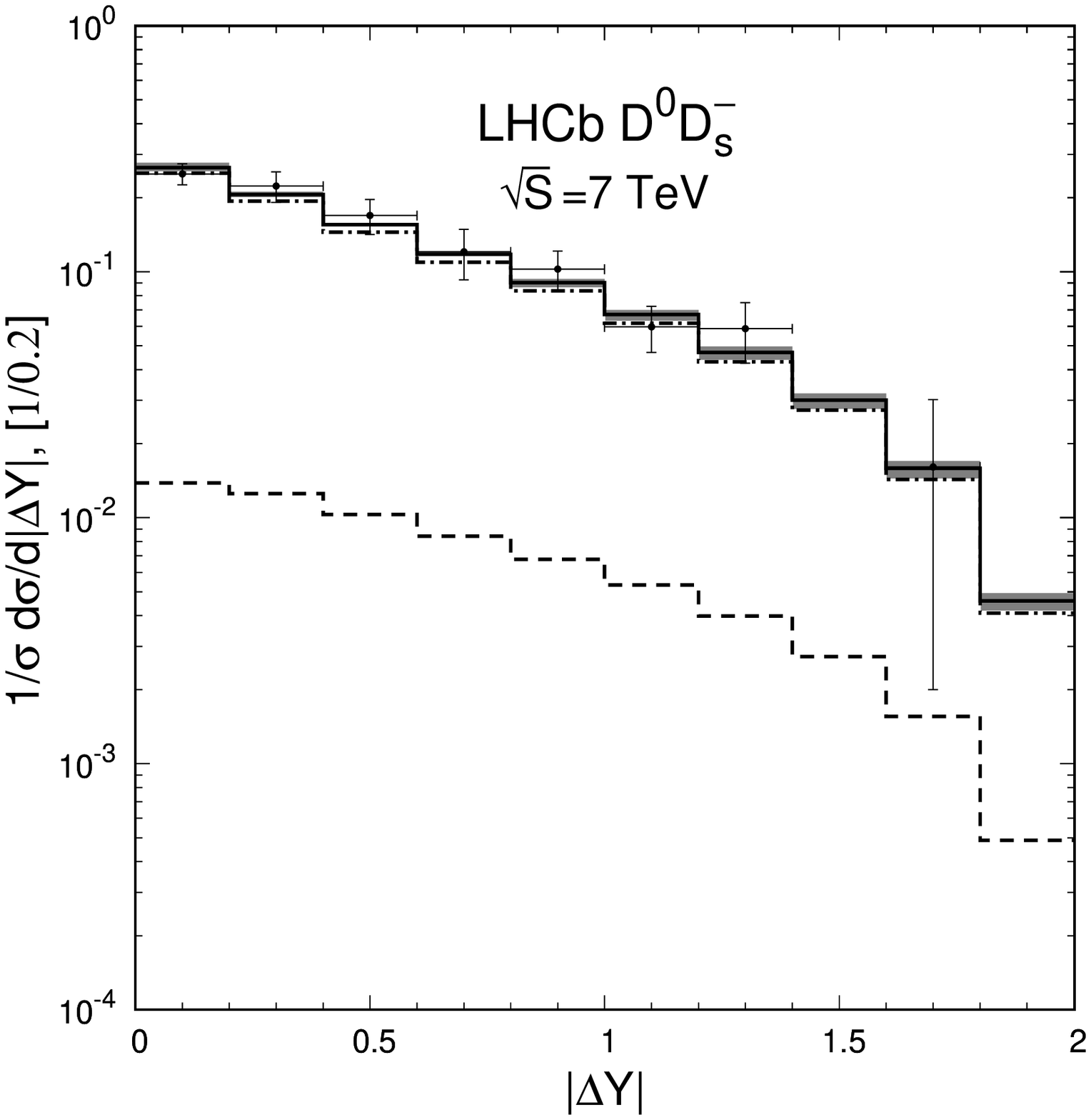}\includegraphics[width=0.5\textwidth, angle=-90,origin=c, clip=]{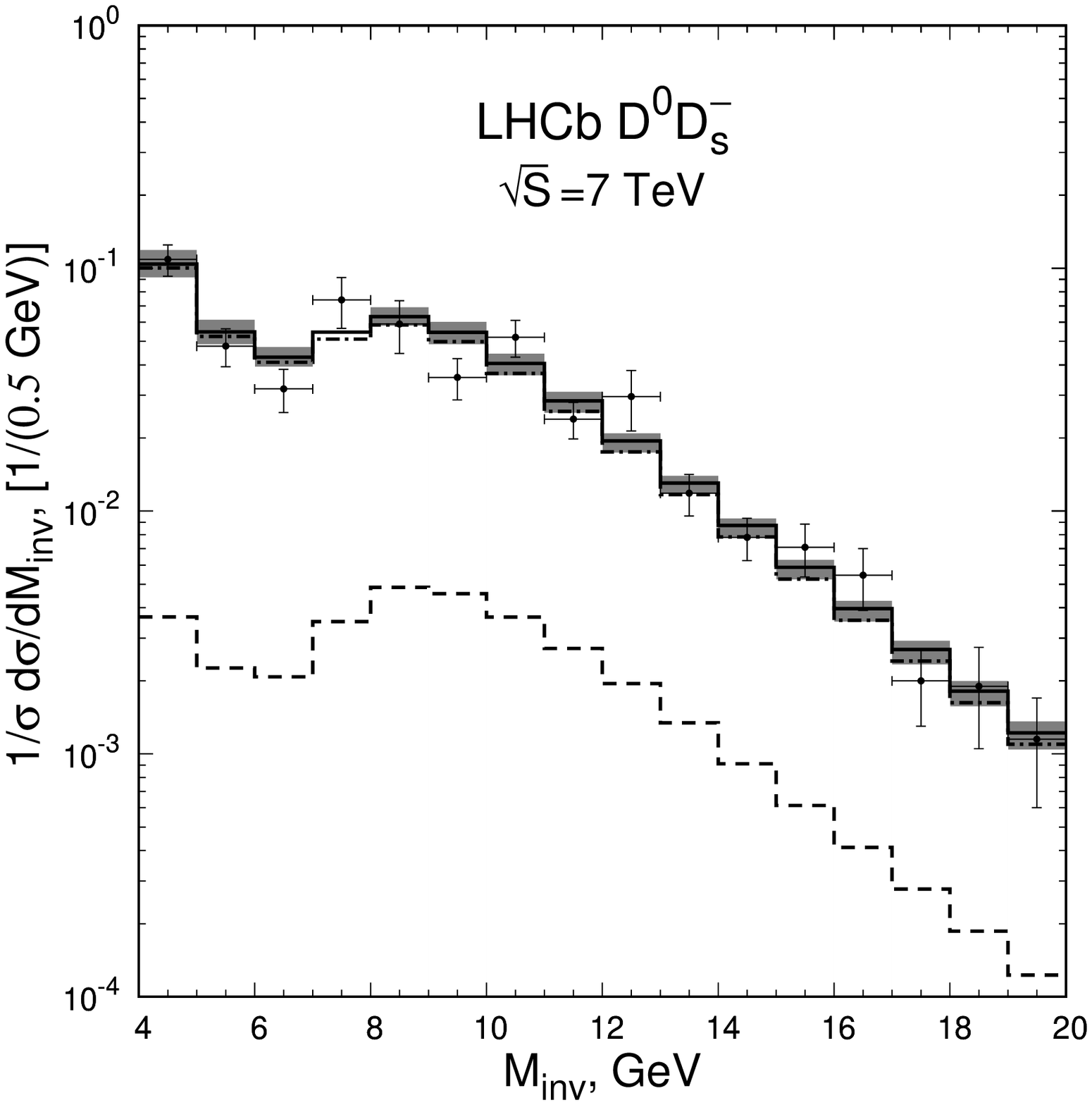}
\caption{The spectra of $D^0D_s^-$ pairs. The notations as in the
Fig.~\ref{fig:5} .\label{fig:6}}
\end{center}
\end{figure}

\begin{figure}[ph]
\begin{center}
\includegraphics[width=0.5\textwidth, angle=-90,origin=c, clip=]{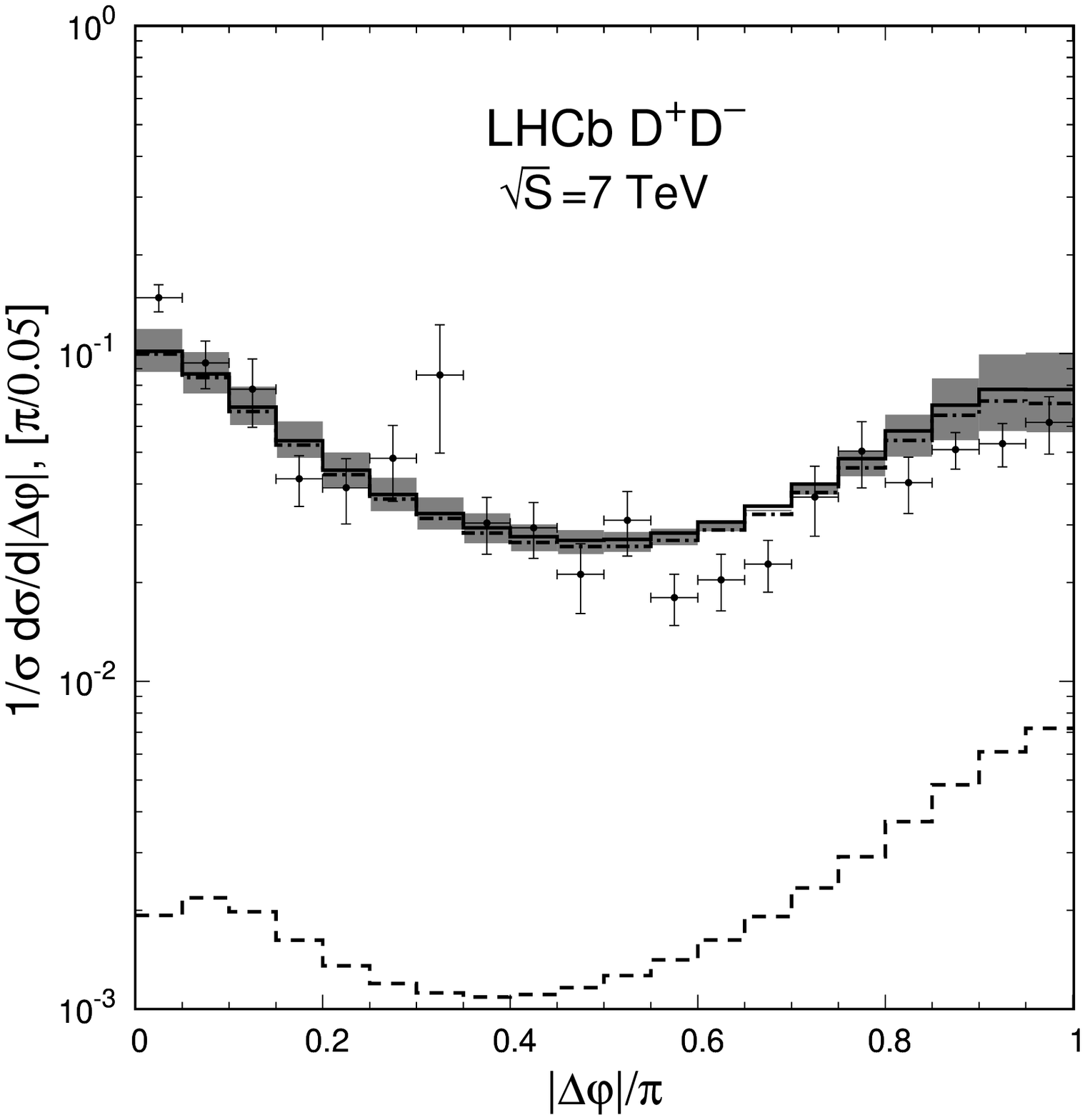}\includegraphics[width=0.5\textwidth, angle=-90,origin=c, clip=]{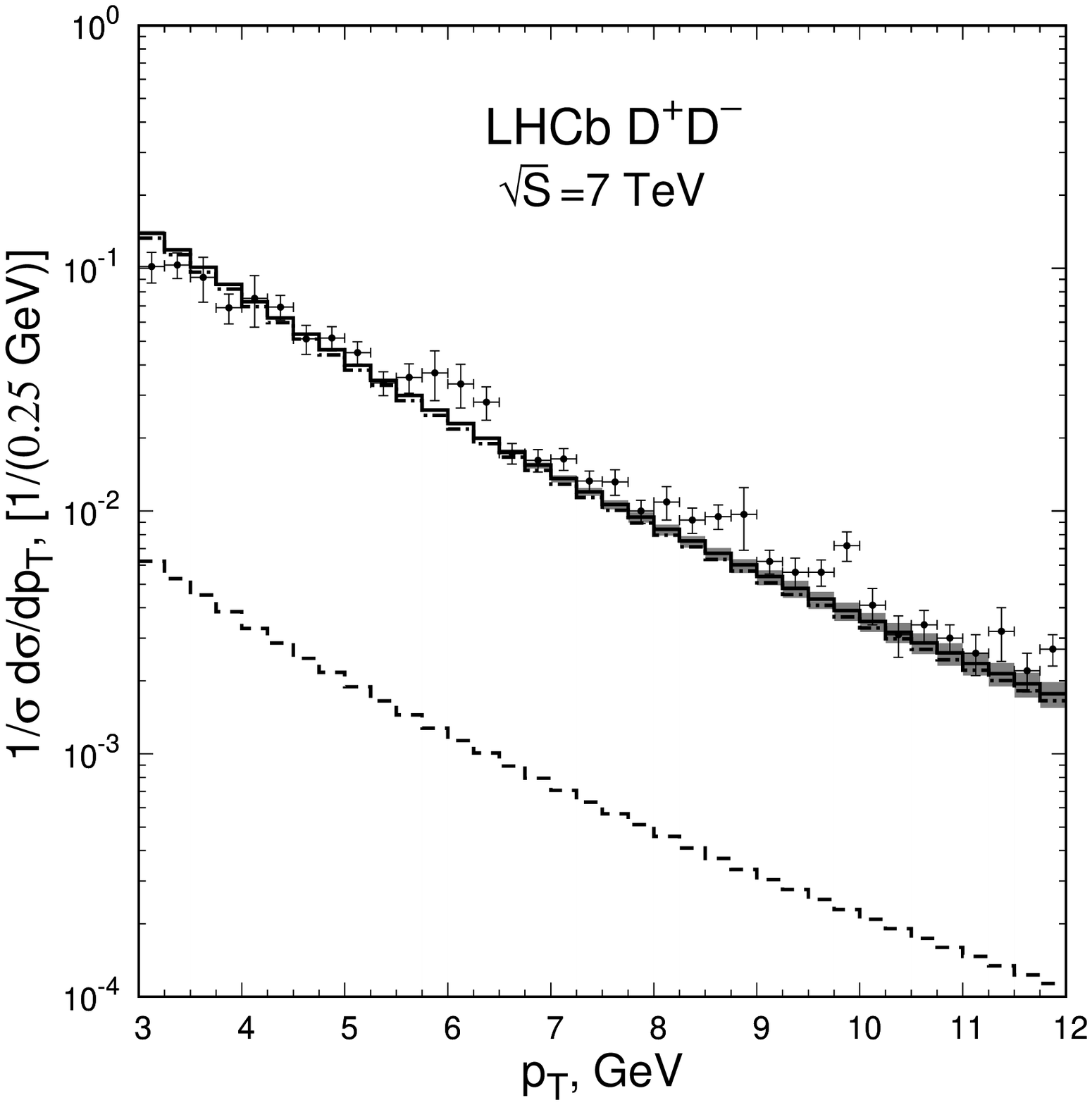}
\includegraphics[width=0.5\textwidth, angle=-90,origin=c, clip=]{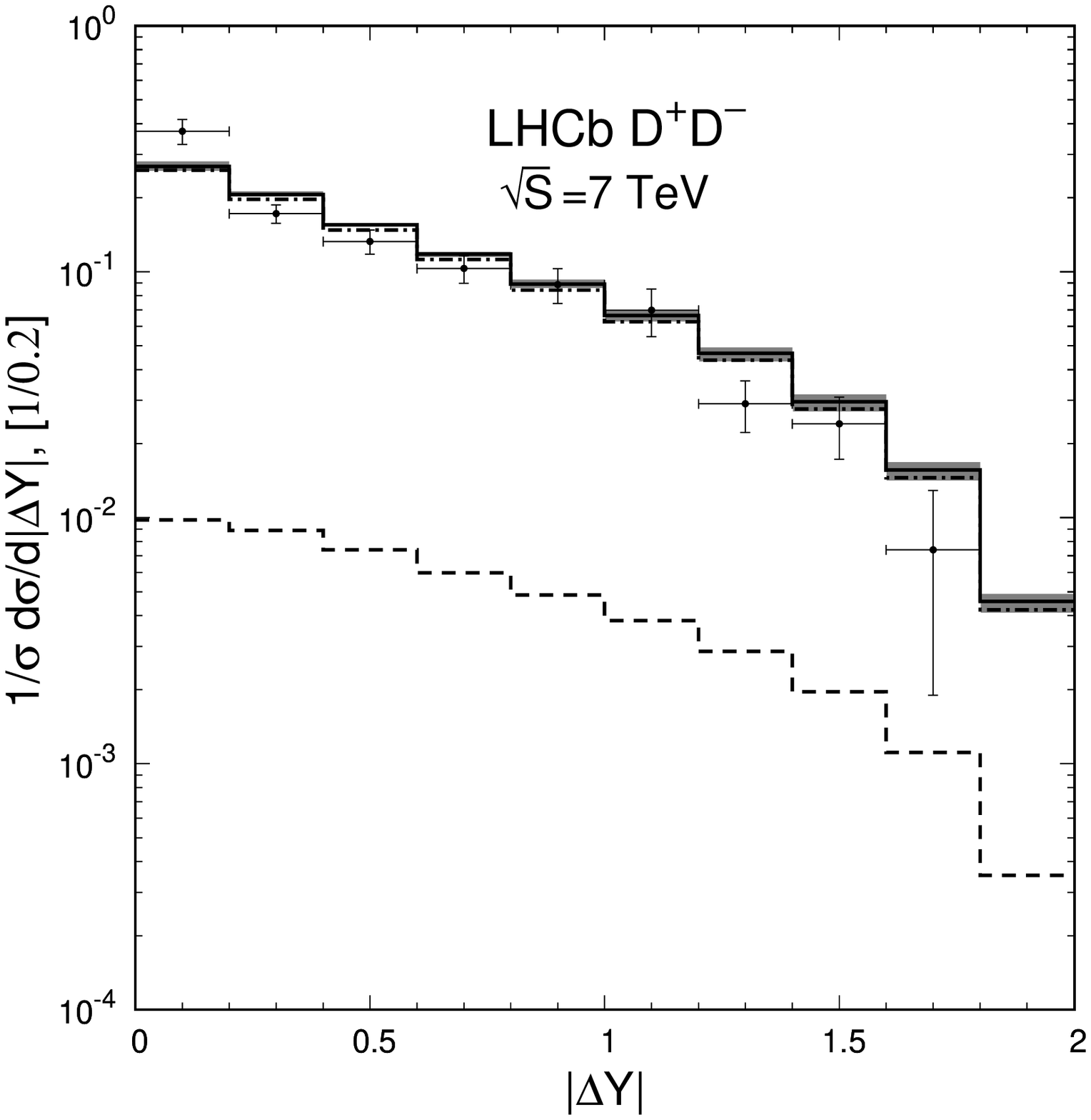}\includegraphics[width=0.5\textwidth, angle=-90,origin=c, clip=]{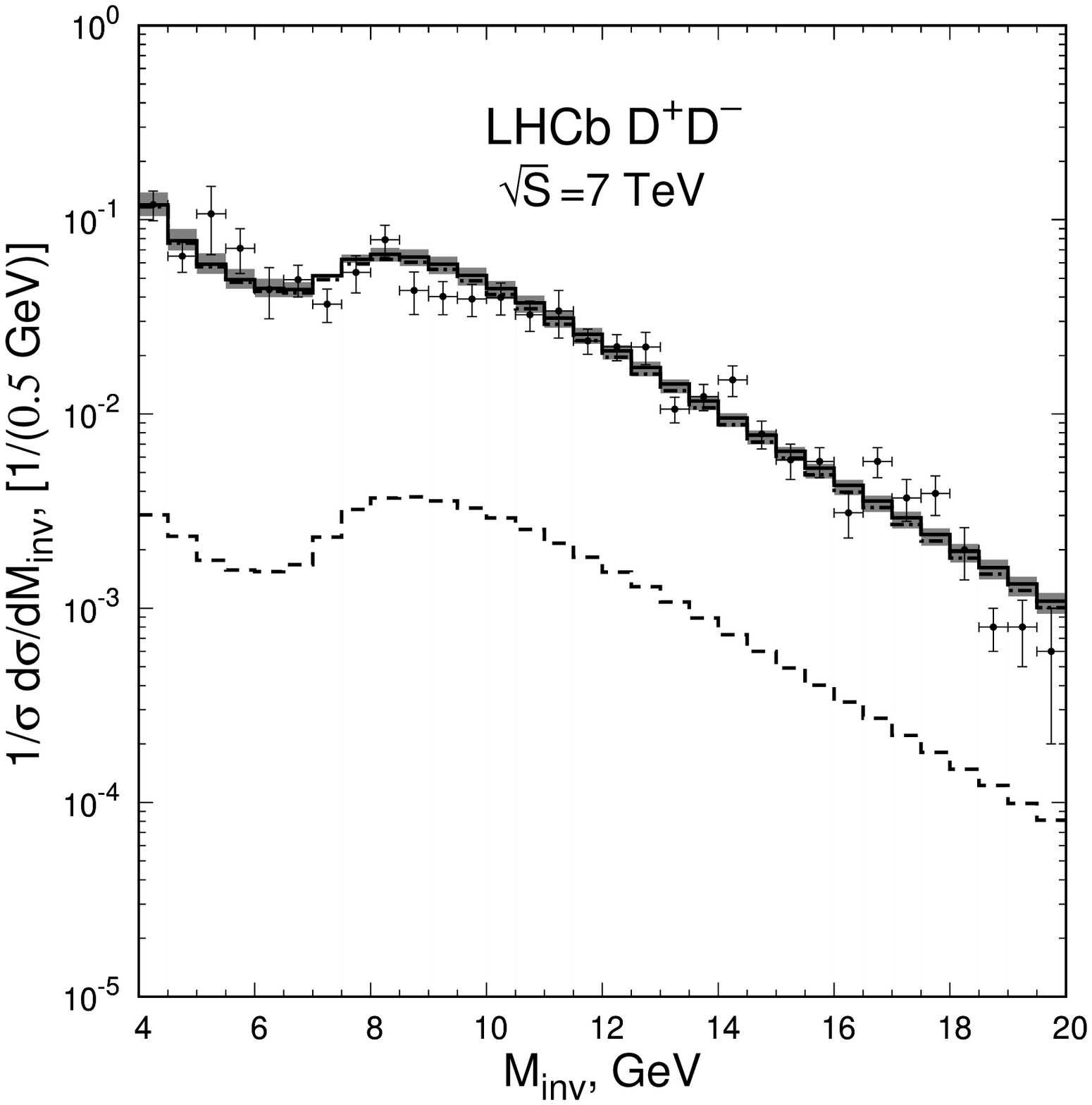}
\caption{The spectra of  $D^+D^-$ pairs. The notations as in the
Fig.~\ref{fig:5} .\label{fig:7}}
\end{center}
\end{figure}

\begin{figure}[ph]
\begin{center}
\includegraphics[width=0.5\textwidth, angle=-90,origin=c, clip=]{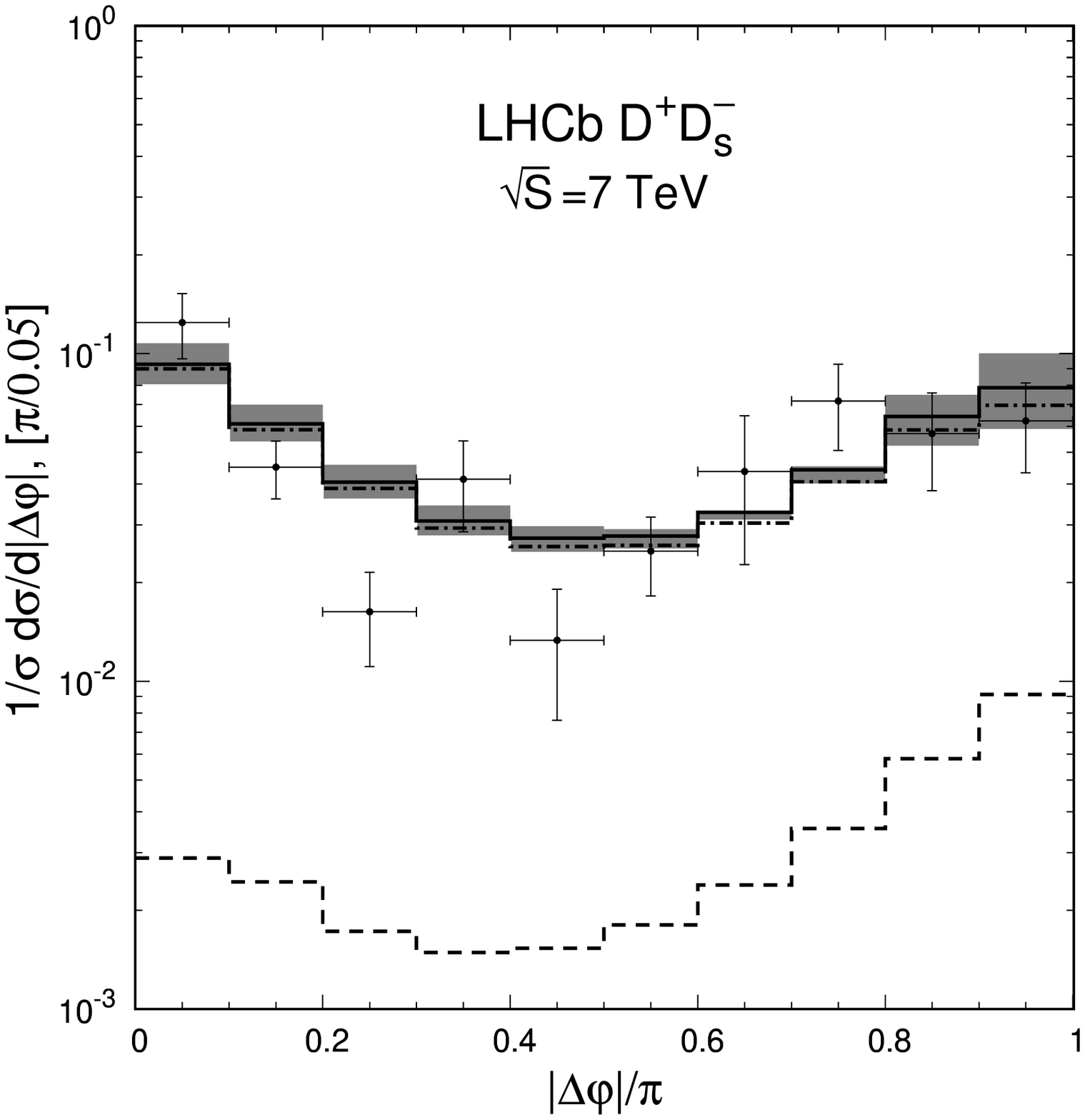}\includegraphics[width=0.5\textwidth, angle=-90,origin=c, clip=]{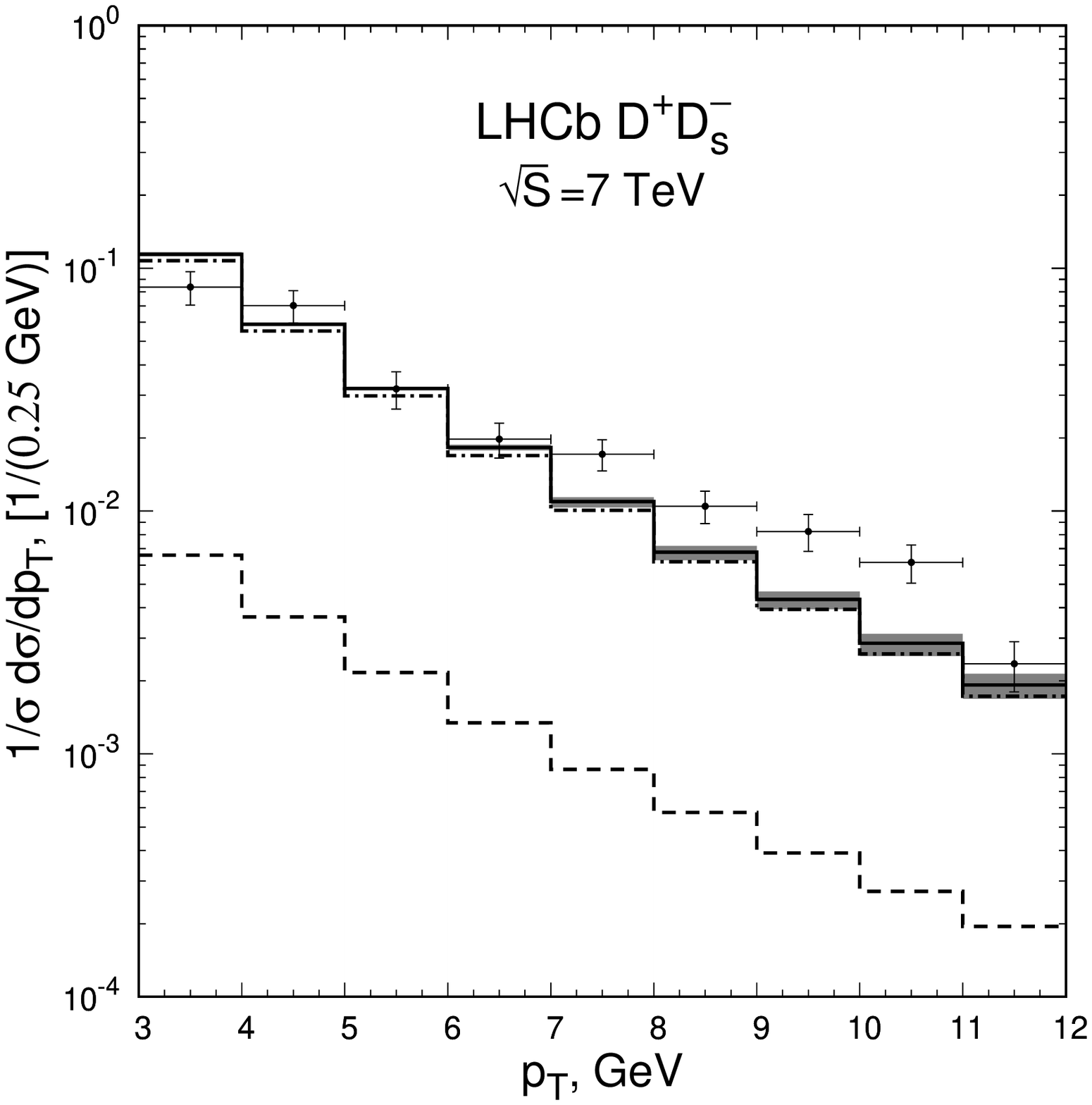}
\includegraphics[width=0.5\textwidth, angle=-90,origin=c, clip=]{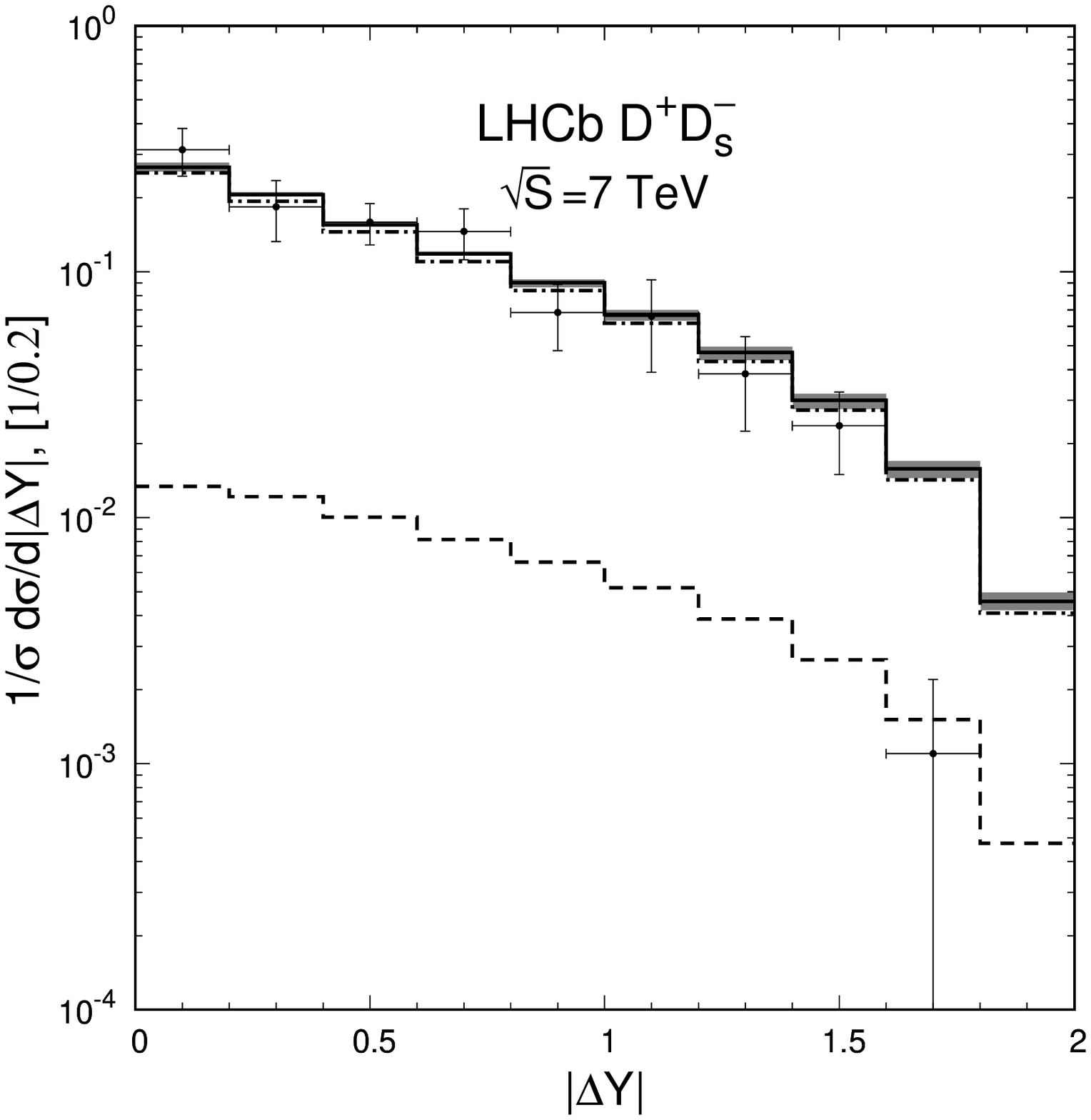}\includegraphics[width=0.5\textwidth, angle=-90,origin=c, clip=]{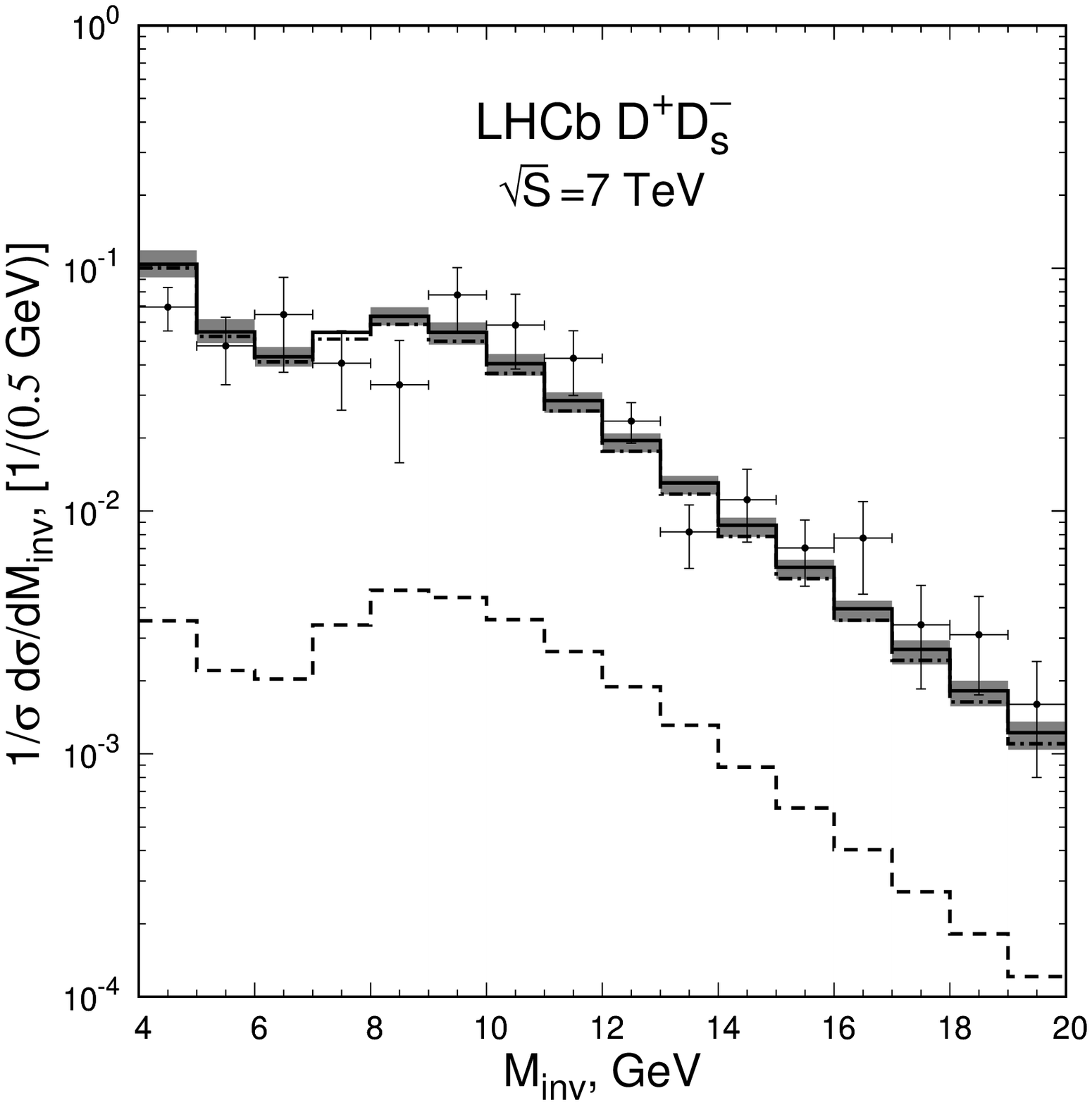}
\caption{The spectra of $D^+D_s^-$ pairs. The notations as in the
Fig.~\ref{fig:5} .\label{fig:8}}
\end{center}
\end{figure}

Now we start with discussion of $D\bar D$-pair production at the
LHCb. In Fig. \ref{fig:4}, we compare our predictions with the LHCb
data at the $\sqrt{S}=7$ TeV for $D^0\bar D^0$-pair production
normalized spectra over the azimuthal angle difference $\Delta
\phi$, $D$-meson transverse momenta $p_T$, rapidity distance $\Delta
Y$ and invariant mass of $D\bar D$ pair $M$. As it is estimated, the
dominating contribution is gluon-gluon fusion into $c\bar c$ pair
with $c-$quark fragmentation into the $D^0$ meson and with $\bar
c-$quark fragmentation into the $\bar D^0$ meson (dash-dotted line).
The contribution from gluon-gluon fusion into two gluons when the
first gluon fragments into $D^0$ meson and second gluon fragments
into $\bar D^0$ meson (dashed line) is lying below the leading
contribution by the order of magnitude. Two next by value
contributions from quark-antiquark annihilation into $c\bar c$ pair
(double-dot-dashed line) or into two gluons (triple-dot-dashed line)
are even smaller and we will ignore these ones below. In the Figs.
(\ref{fig:5})-(\ref{fig:8}), we demonstrate that our predictions for
correlation spectra of $D^0D^-$, $D^0D_s$, $D^+D^-$ and $D^+D^-_s$
are in a good agreement with the experimental data.

\begin{figure}[ph]
\begin{center}
\includegraphics[width=0.5\textwidth, angle=-90,origin=c, clip=]{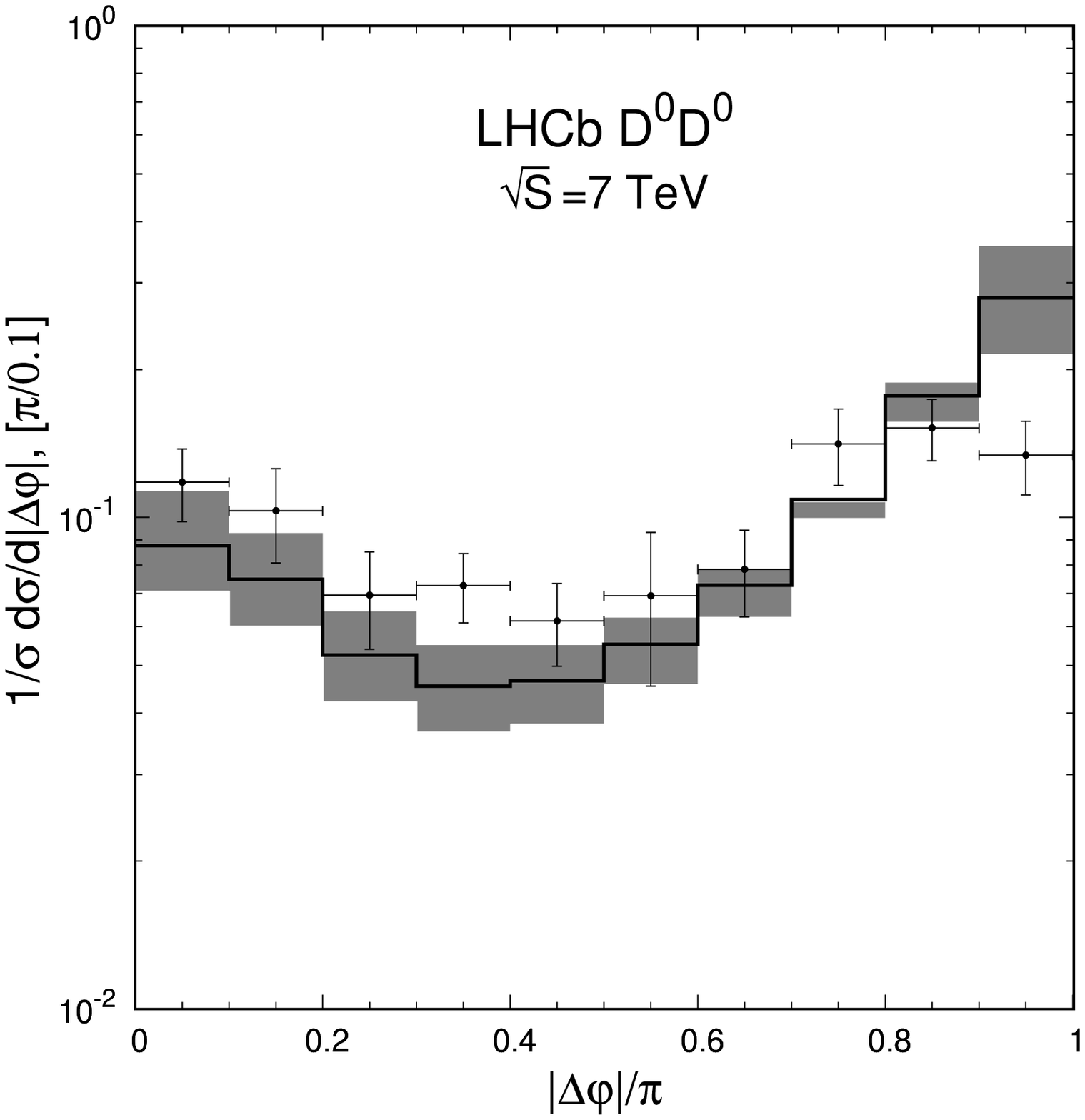}\includegraphics[width=0.5\textwidth, angle=-90,origin=c, clip=]{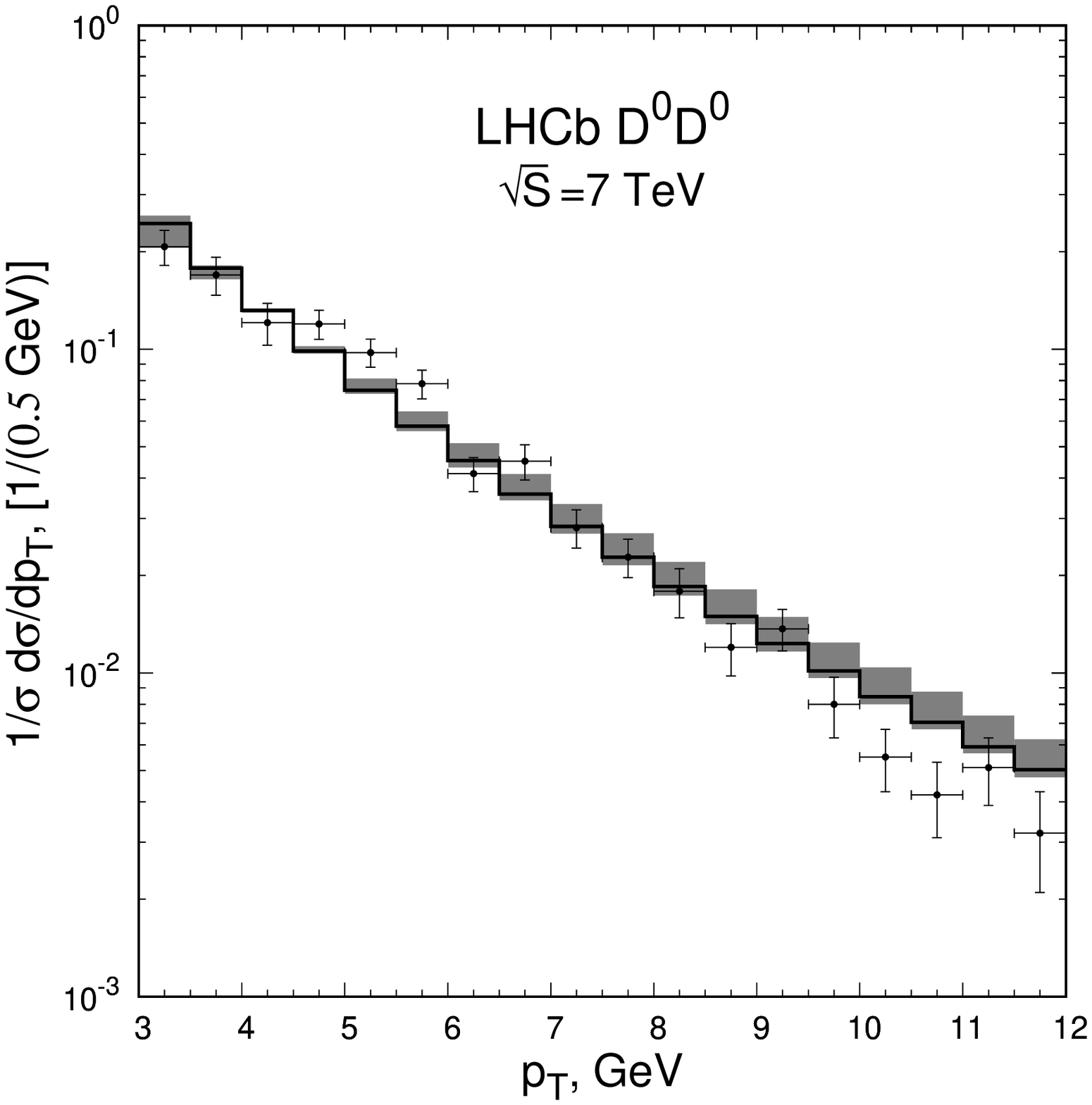}
\includegraphics[width=0.5\textwidth, angle=-90,origin=c, clip=]{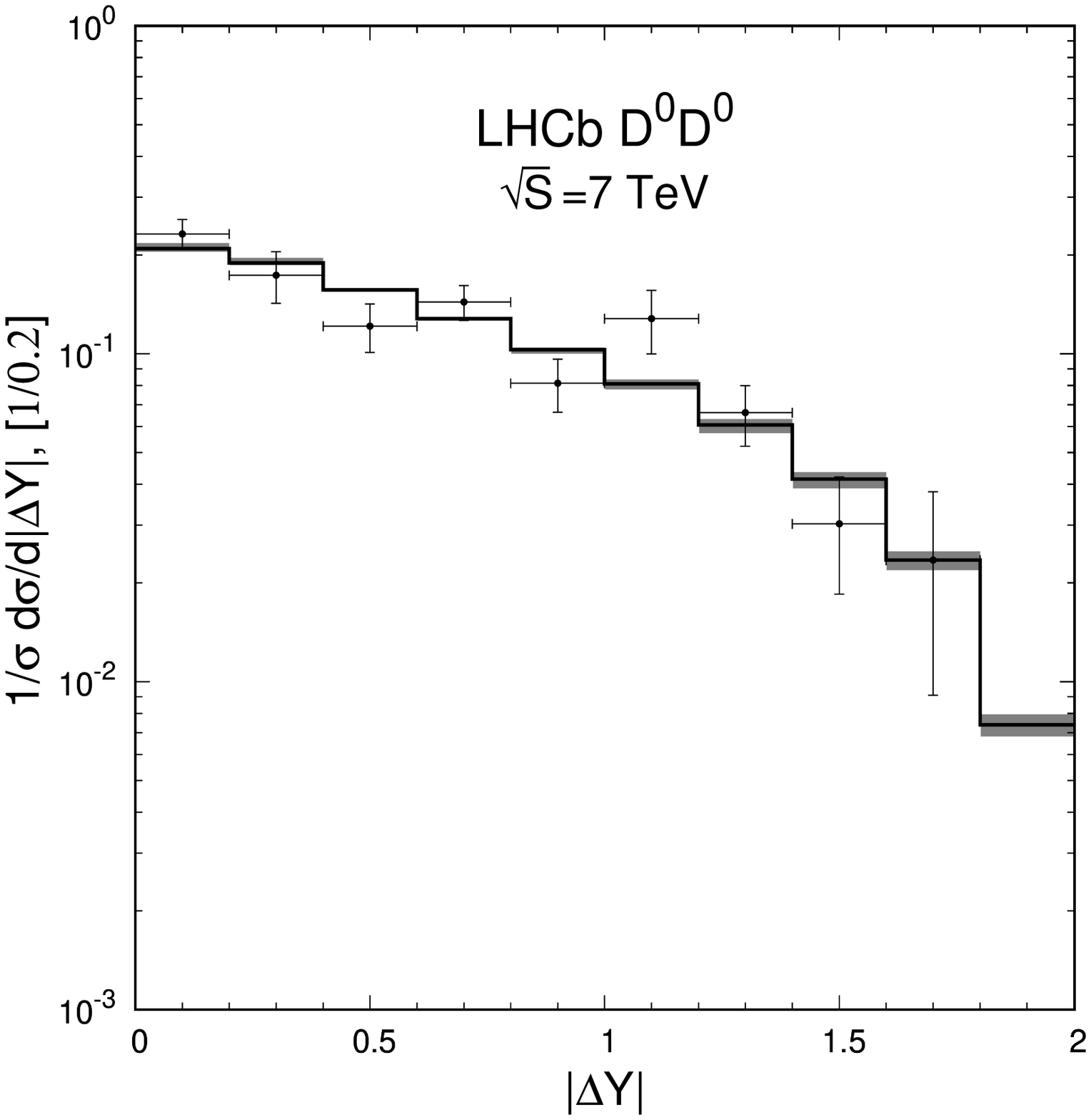}\includegraphics[width=0.5\textwidth, angle=-90,origin=c, clip=]{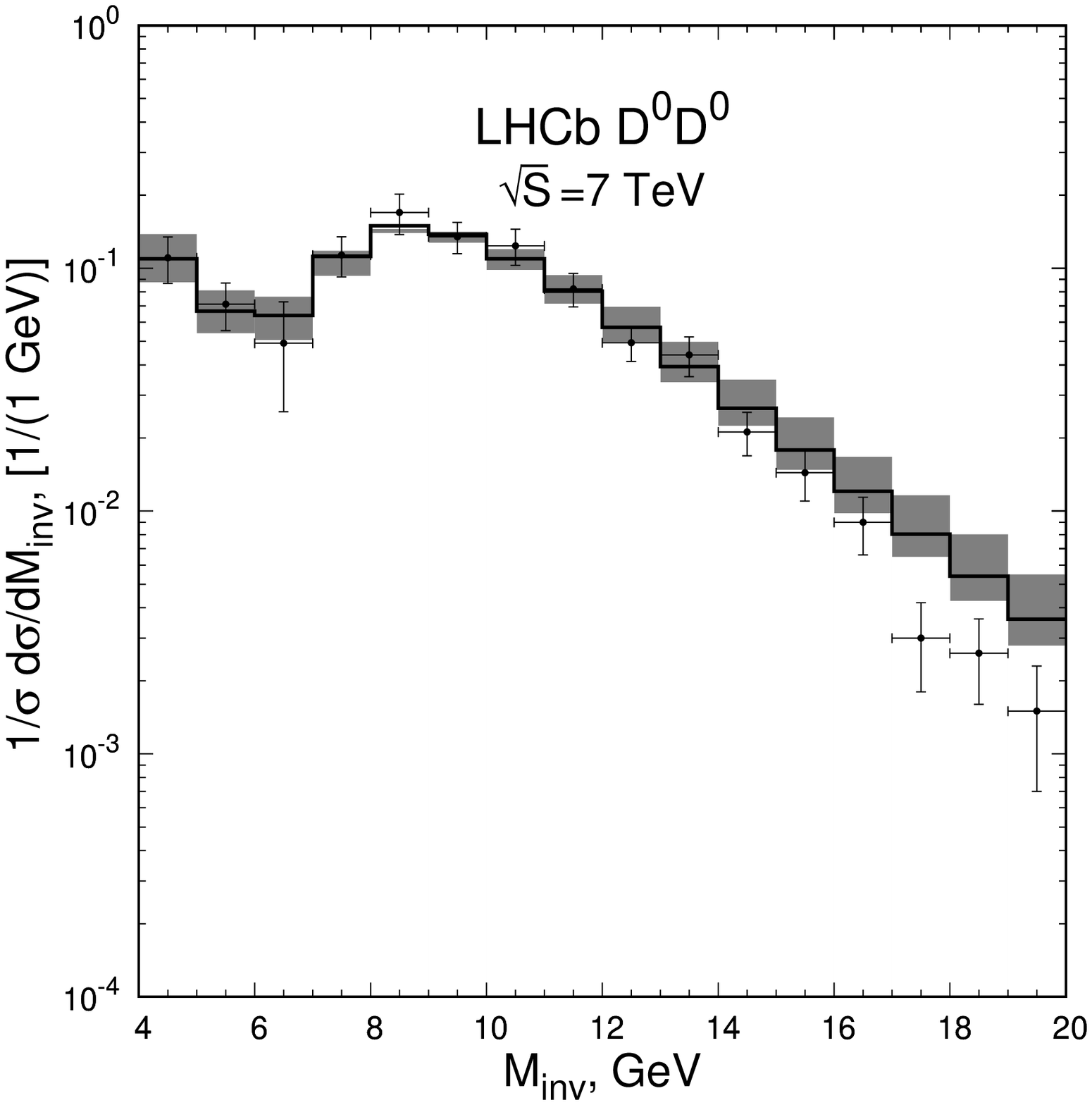}
\caption{The spectra of $D^0D^0$ pairs differential in azimuthal
angle difference (left, top), transverse momentum (right, top),
rapidity distance (left, bottom) and invariant mass of the pair
(right, bottom) at the $2<y<4$ and $\sqrt{S}=7$~TeV. The LHCb data
at LHC are from the Ref.~\cite{LHCb_Pair}. Solid line represents the
leading contribution of gluon fragmentation in gluon-gluon
fusion.\label{fig:9}}
\end{center}
\end{figure}

\begin{figure}[ph]
\begin{center}
\includegraphics[width=0.5\textwidth, angle=-90,origin=c, clip=]{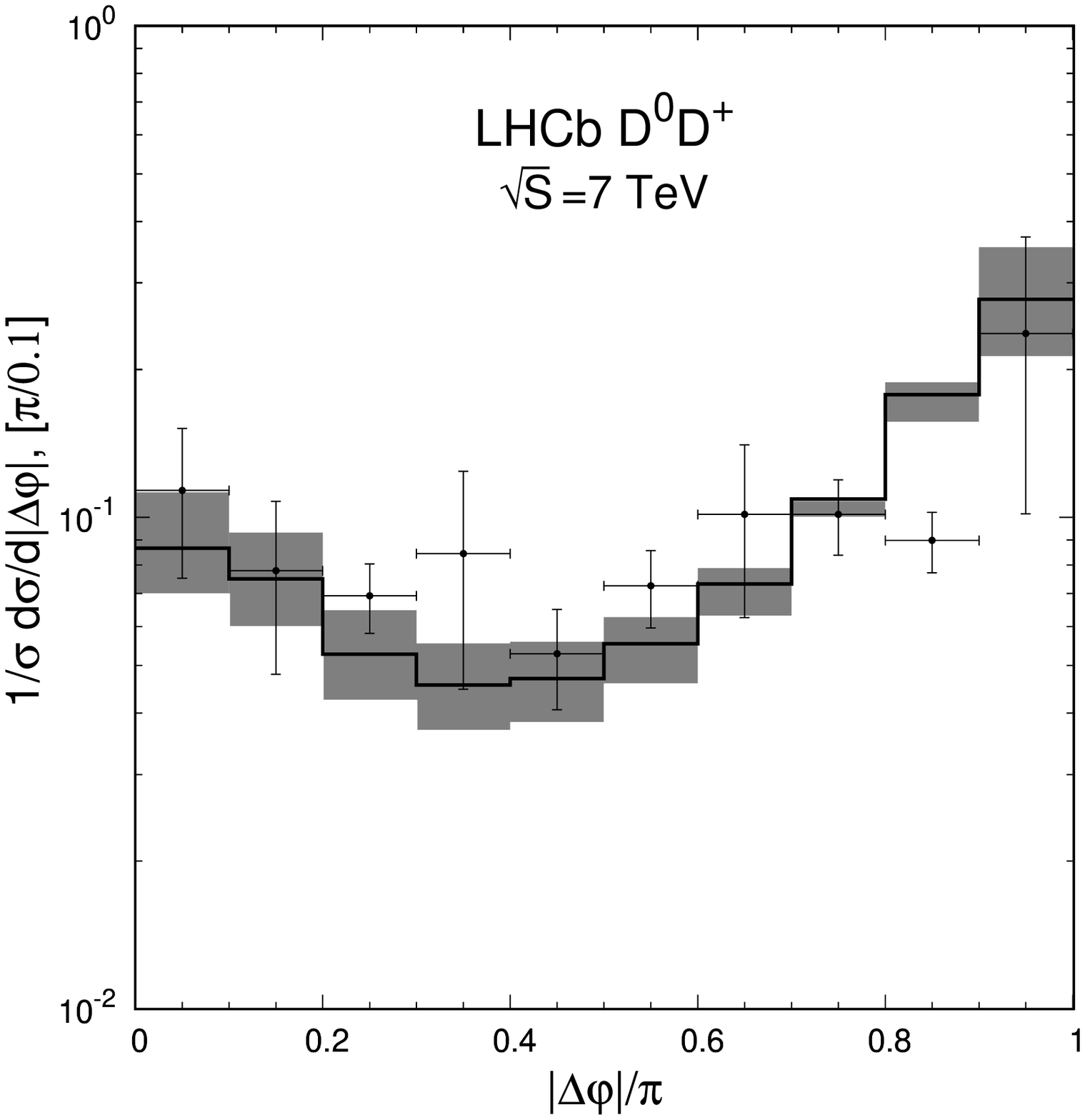}\includegraphics[width=0.5\textwidth, angle=-90,origin=c, clip=]{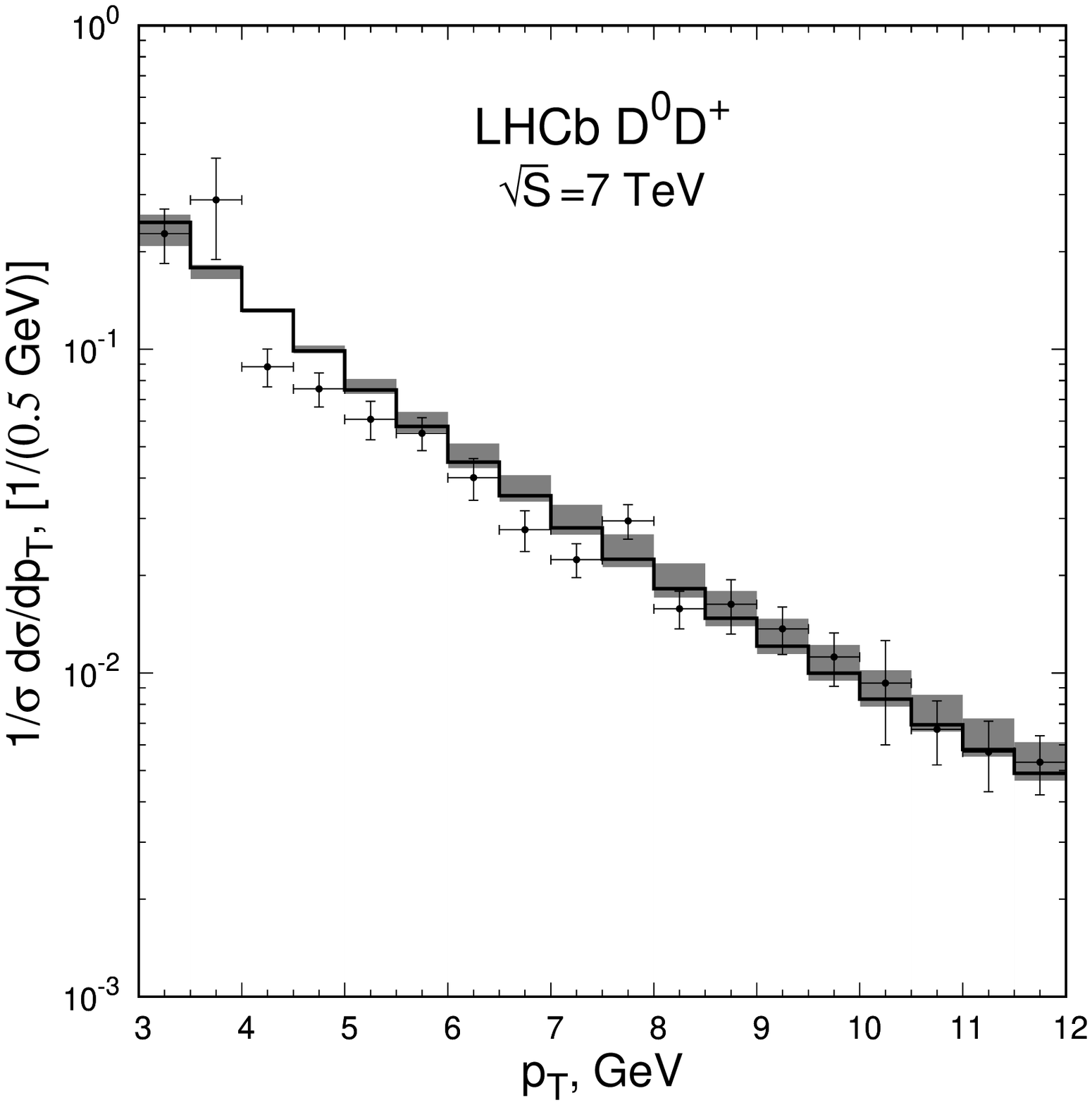}
\includegraphics[width=0.5\textwidth, angle=-90,origin=c, clip=]{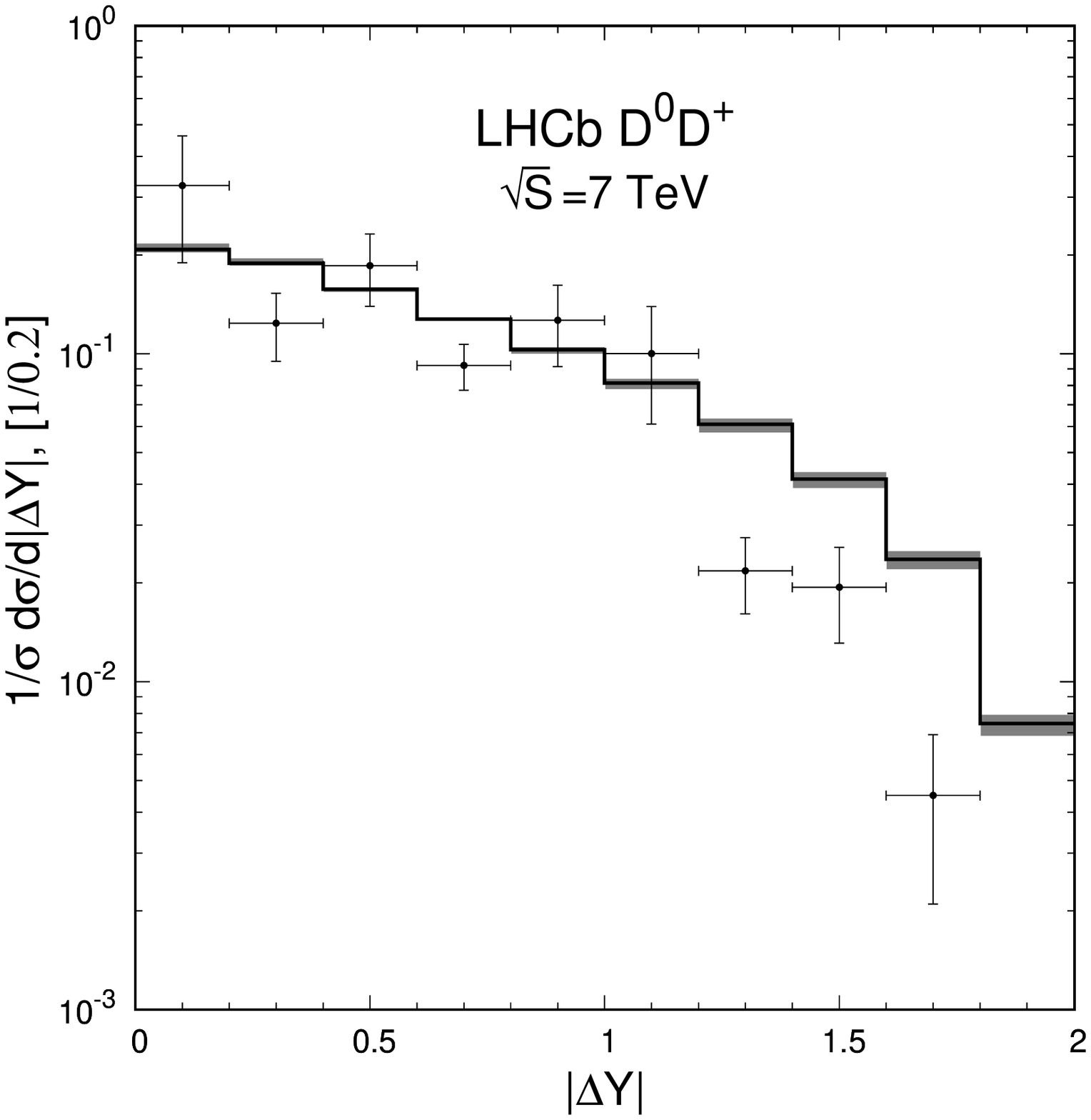}\includegraphics[width=0.5\textwidth, angle=-90,origin=c, clip=]{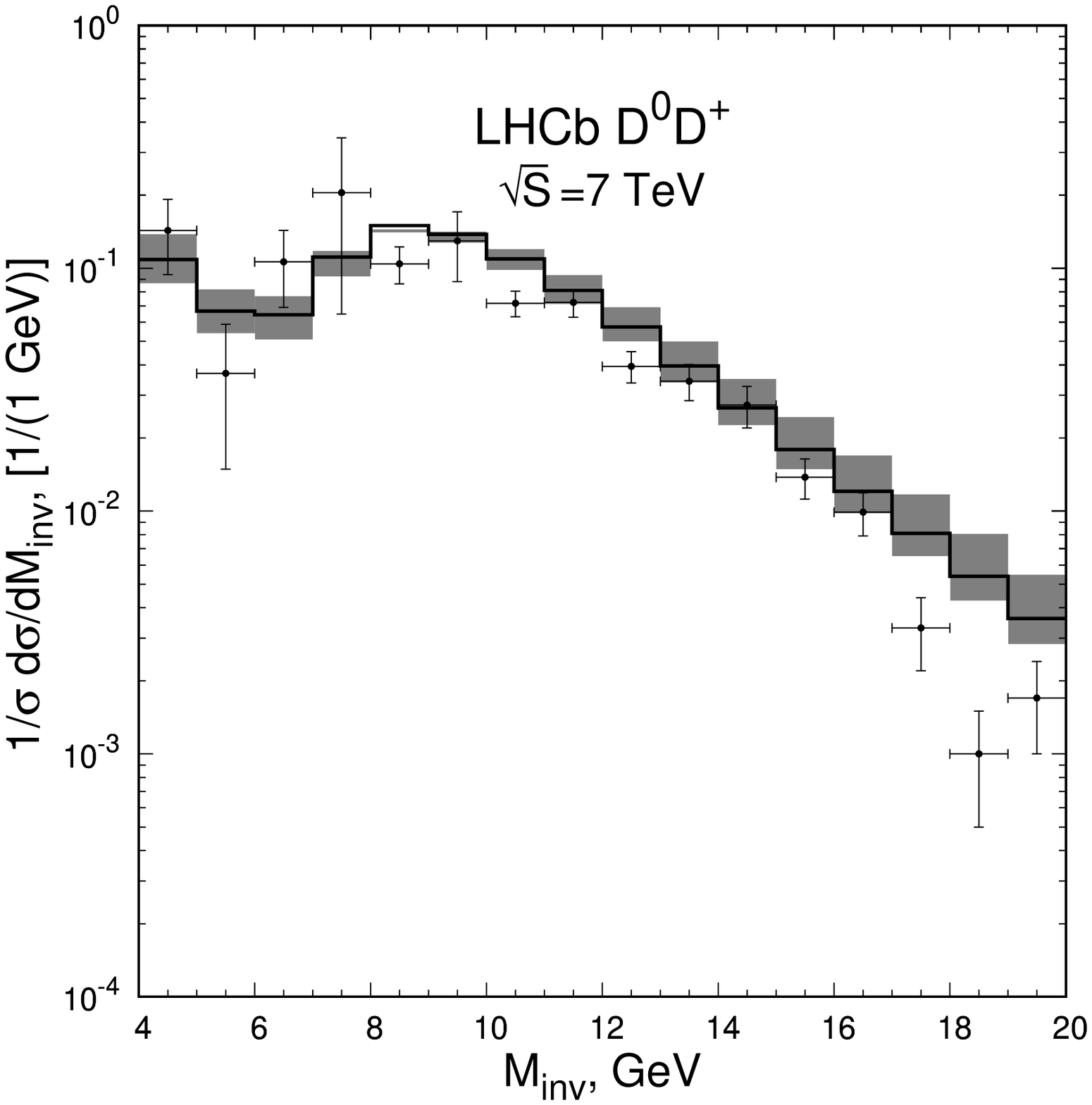}
\caption{The spectra of $D^0D^+$ pairs. The notations as in the
Fig.~\ref{fig:9} .\label{fig:10}}
\end{center}
\end{figure}

\begin{figure}[ph]
\begin{center}
\includegraphics[width=0.5\textwidth, angle=-90,origin=c, clip=]{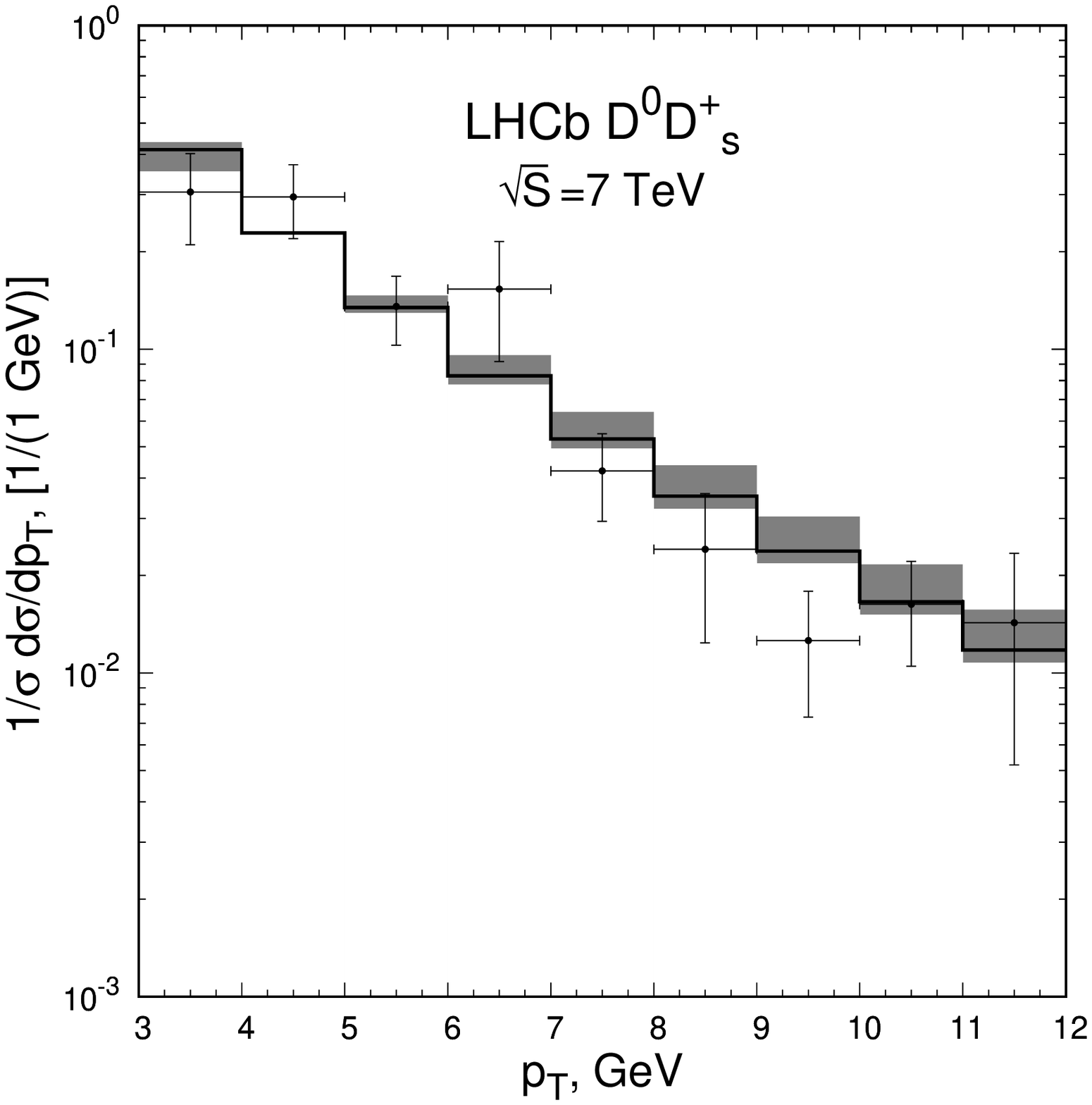}\includegraphics[width=0.5\textwidth, angle=-90,origin=c, clip=]{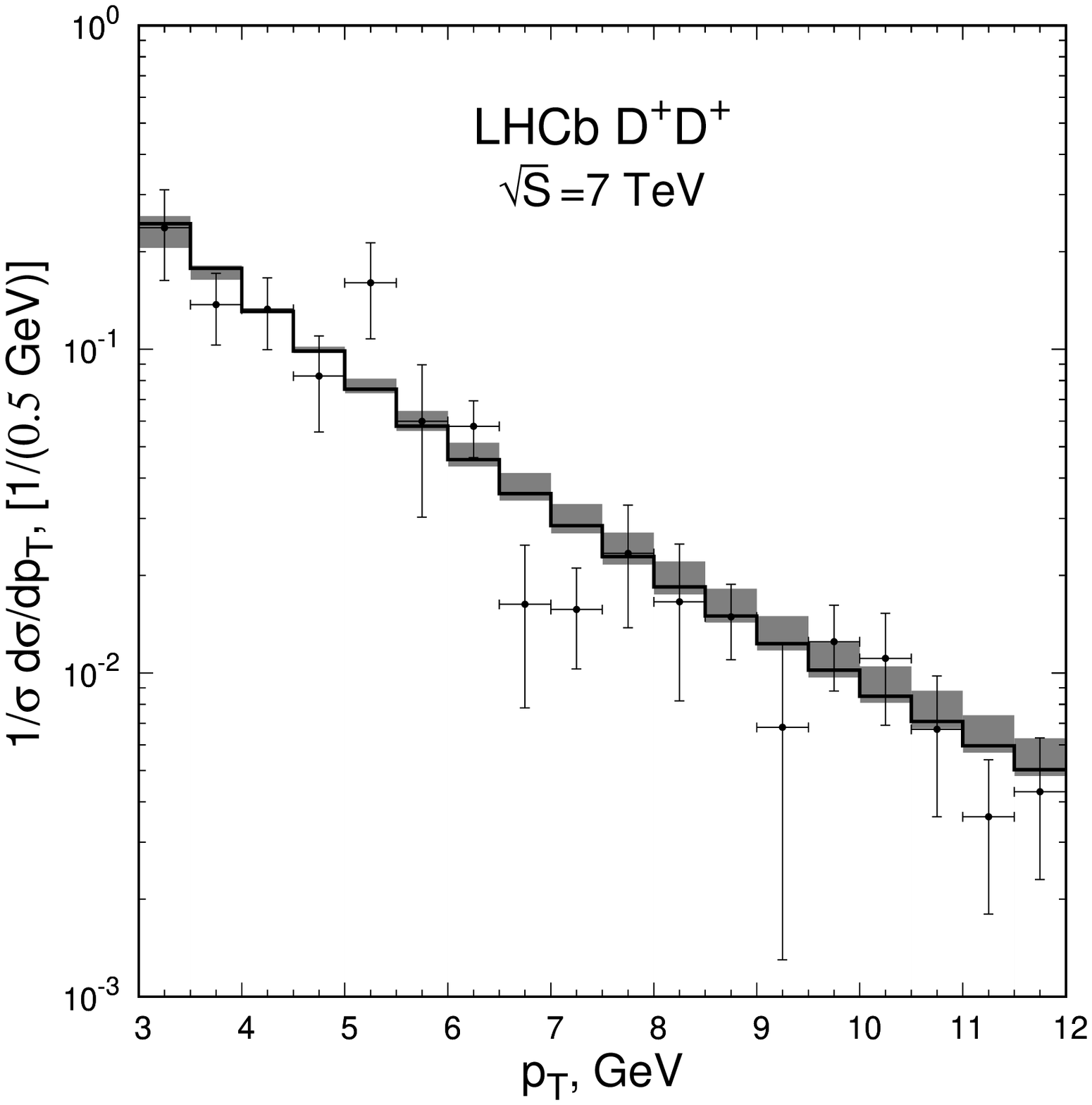}
\includegraphics[width=0.5\textwidth, angle=-90,origin=c, clip=]{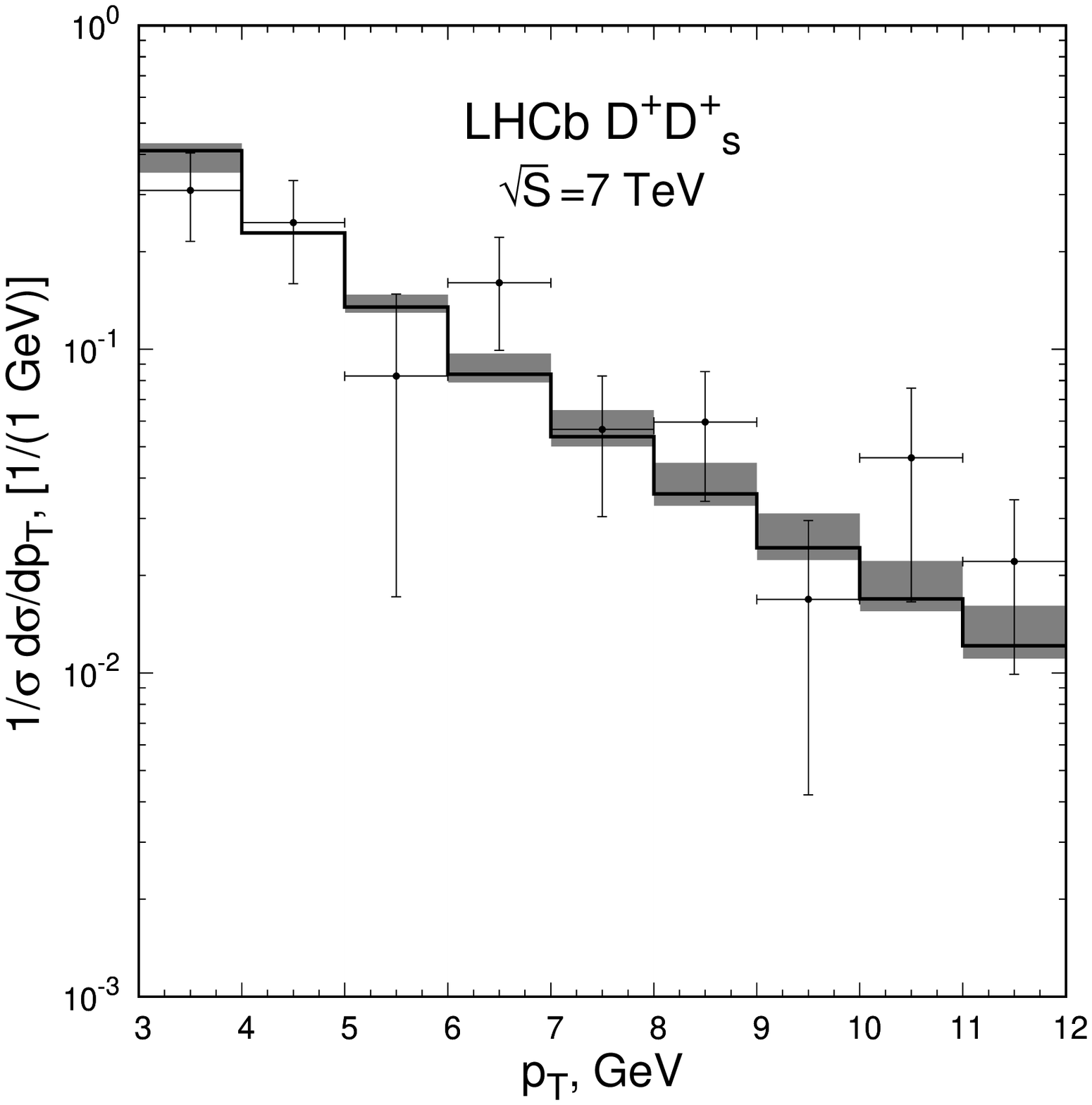}
\caption{The transverse momentum distributions of $D^0D_s^+$ (left,
top), $D^+D^+$ (right, top) and $D^+D_s^+$ (bottom) pairs at the
$2<y<4$ and  $\sqrt{S}=7$~TeV. The LHCb data at LHC are from the
Ref.~\cite{LHCb_Pair}. Solid line represents the leading
contribution of gluon fragmentation in gluon-gluon
fusion.\label{fig:11}}
\end{center}
\end{figure}

The production of $DD$ pairs is mostly generated by the gluon
fragmentation into the $D$ meson in the subprocess of gluon-gluon
fusion. The contribution of two-gluon production in the
quark-antiquark annihilation is suppressed by two orders of
magnitude in the same way as the contributions from the channel of
$c\bar c c\bar c$ double-pair production. In the Figs.
(\ref{fig:9})-(\ref{fig:10}), we compare our predictions based on
two-gluon fragmentation mechanism with the experimental data for
correlation spectra of $D^0D^0$ and $D^0D^+$, correspondingly. In
the Fig. (\ref{fig:11}), the transverse momentum spectra for
$D^0D_s^+$, $D^+D^+$ and $D^+D^+_s$ pairs are presented. We found a
good agreement between predictions obtained in the LO PRA
calculations and experimental data from LHCb Collaboration for all
$DD$-pair correlation observables and there is no place for
contribution of DPS production mechanism, as it is discussed in
literature.

\section{Conclusions}

\label{sec:six}

We introduce a comprehensive study of  single and pair fragmentation
production of $D$ mesons in proton-proton collisions at the energies
$\sqrt{S}=7$ and $\sqrt{S}=13$ TeV in the forward rapidity region at
the LHC in the framework of parton Reggeization approach. We use the
gauge invariant amplitudes of hard parton subprocesses in the LO
level of theory with Reggeized gluons and Reggeized quarks in the
initial state in a self-consistent way together with unintegrated
parton distribution functions proposed by Kimber, Martin and Ryskin.
To describe the non-perturbative transition of produced gluons and
$c$-quarks into the $D$ mesons we use the universal fragmentation
functions obtained from the fit   of $e^+ e^-$ annihilation data
from CERN LEP1.  We found our results for $D$-meson and $D\bar
D(DD)$-pair production to be in the good coincidence with
experimental data from the  LHCb Collaboration. The predictions for
the  $D^0\bar D^0(D^0D^0)$-pair production correlation spectra in
the large rapidity region for the energy $\sqrt S=13$ TeV are also
presented. We have found that in case of single $D$-meson production
at the LHC energies, both mechanisms, gluon and $c$-quark
fragmentation, are important with some preference for the gluon
fragmentation production. In the $D\bar D$-pair production we found
$c$-quark production mechanism to be the main one, while in the $DD$-pair production the gluon fragmentation is a totally dominant
source. Additionally, we arrive at the important conclusion that
$DD$-pair production can not be used for differentiation between
single and double parton scattering approaches of high-energy QCD.

\section{Acknowledgements}
We are particularly indebted to R.~Maciula and A.~Szczurek for
discussion of problems in $DD$-pair production at the LHCb and
B.~Kniehl and G.~Kramer for explaining some details of KKKS08
fragmentation functions.

The work of A.V.~Shipilova and V.A.~Saleev  was supported in part by
the Russian Foundation for Basic Research through the Grant N
14-02-00021 and by the Ministry of Education and Science of Russia
under Competitiveness Enhancement Program of Samara University for
2013-2020; the work of M.A.~Nefedov and A.V.~Karpishkov was
supported by the Ministry of Education and Science of Russia through
the Contract No. 1394. The work of A.V.~Karpishkov was supported in
part by the German Academic Exchange Service (DAAD) and the Ministry
of Science and Education of the Russian Federation through Michail
Lomonosov Grant Linie A, 2016 (57212599).

\newpage
\newpage

\newpage

\newpage

\newpage

\newpage

\newpage

\newpage

\newpage

\newpage

\newpage

\newpage

\newpage
\begin{figure}[ph]
\begin{center}
\includegraphics[width=0.5\textwidth, angle=-90,origin=c, clip=]{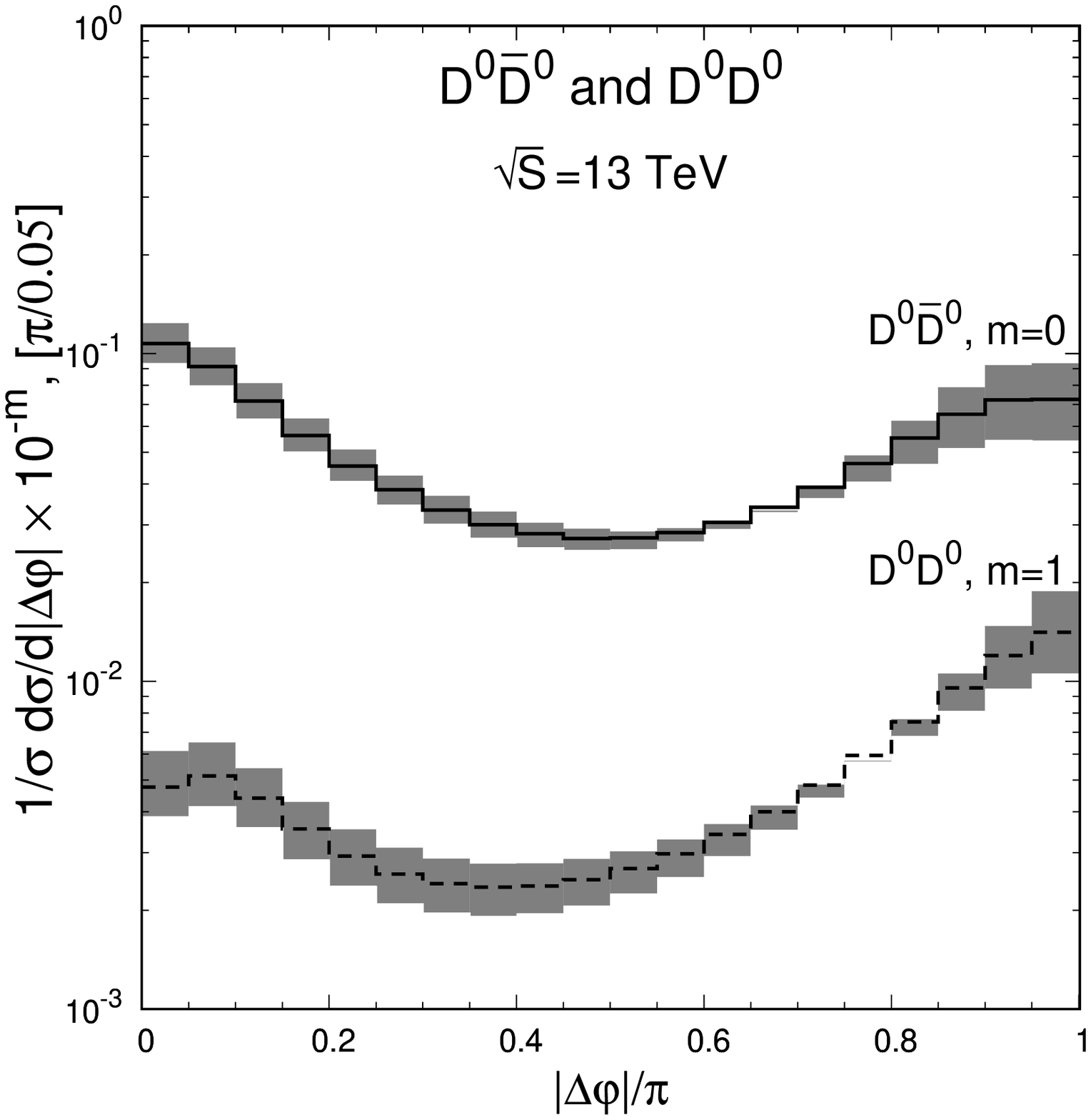}\includegraphics[width=0.5\textwidth, angle=-90,origin=c, clip=]{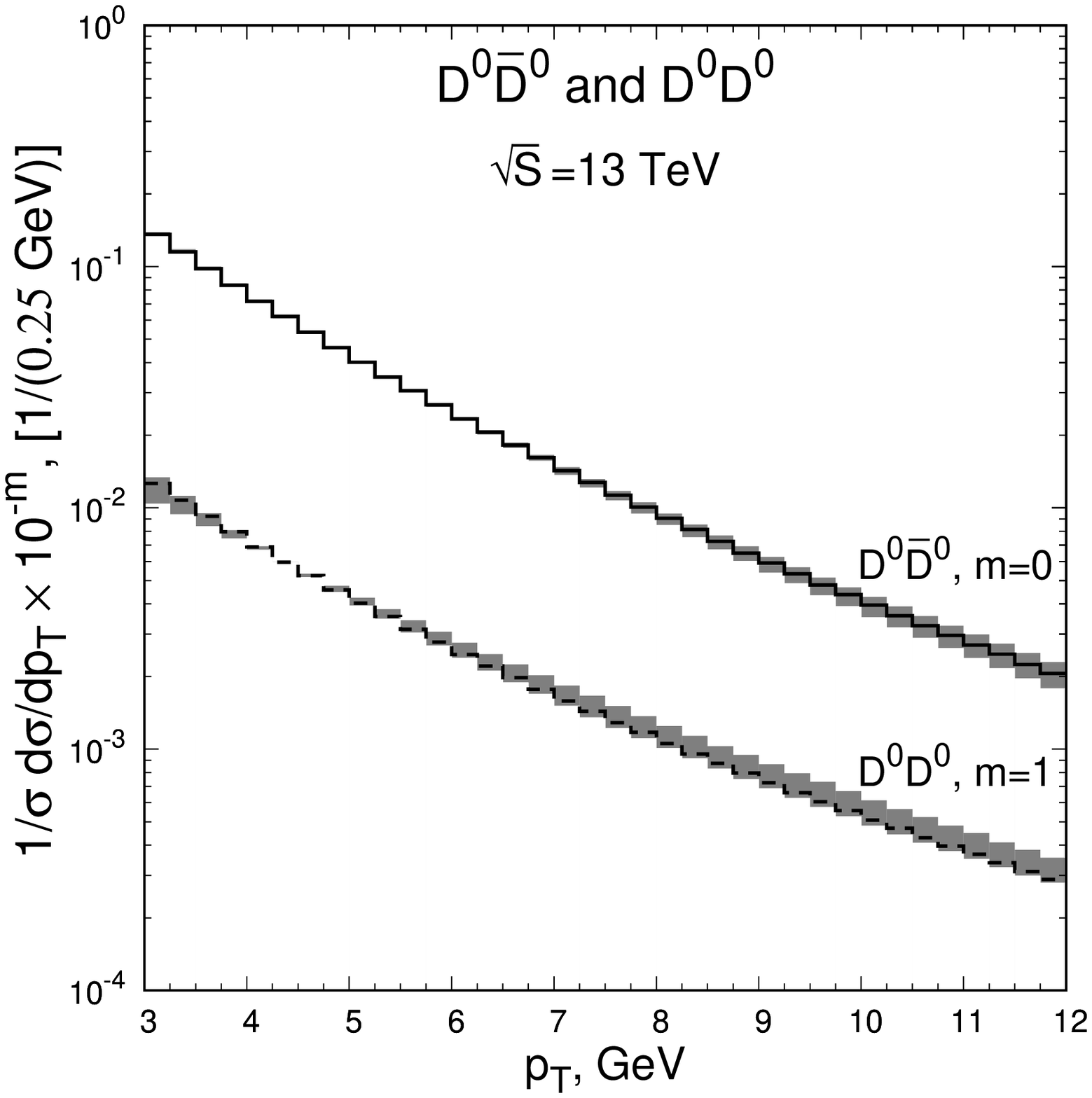}
\includegraphics[width=0.5\textwidth, angle=-90,origin=c, clip=]{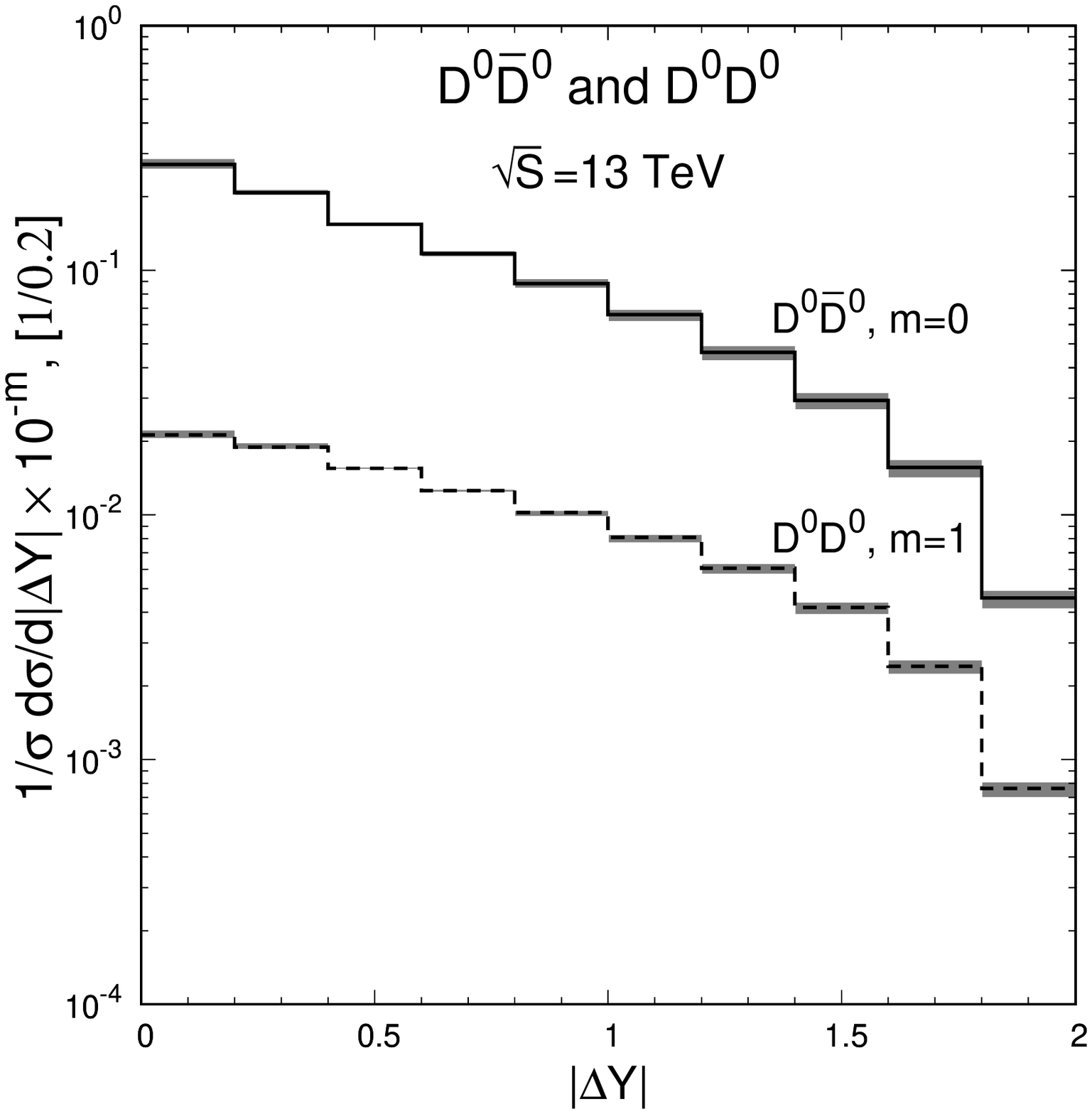}\includegraphics[width=0.5\textwidth, angle=-90,origin=c, clip=]{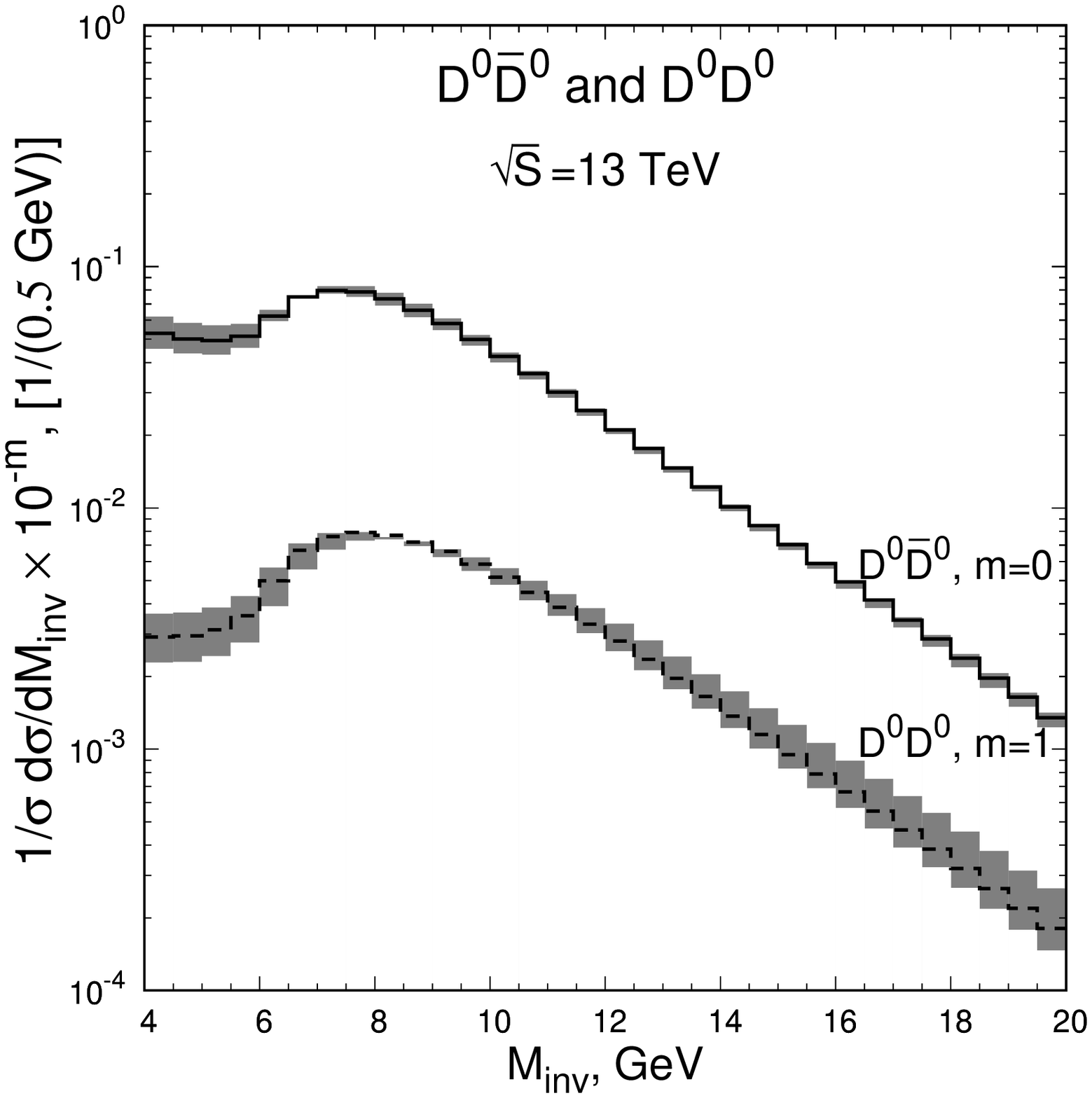}
\caption{The predicted spectra of $D^0\overline{D^0}$ (solid line,
$m=0$) and $D^0D^0$ (dashed line, $m=1$) pairs differential in
azimuthal angle (left, top), transverse momentum (right, top),
rapidity distance (left, bottom) and invariant mass of the pair
(right, bottom) at the $2<y<4$ and $\sqrt{S}=13$~TeV.\label{fig:12}}
\end{center}
\end{figure}

\end{document}